\documentstyle[epsfig,aps,prc,eqsecnum]{revtex}

\def\lixo#1{}
\def\lp {\left( }
\def\rp {\right) }
\def\lb {\left[ }
\def\rb {\right] }
\def\lc {\left\{ }
\def\rc {\right\} }

\def\nn {\nonumber}

\def\beq{\begin{equation}}
\def\eeq{\end{equation}}
\def\bea{\begin{eqnarray}}
\def\eea{\end{eqnarray}}
\def\ni{\noindent}

\def\cd {\!\cdot\!}

\def\rar {\rightarrow}

\def\sm {s\!-\!m^2}
\def\um {u\!-\!m^2}
\def\bb {\bar b}
\def\db {\bar d}
\def\ub {\bar u}

\def\Qs {\not\!\!Q}

\def\NL {N\!\!L}

\def\Pt {\tilde{\P}}

\def\V12 {\lp 1-\frac{t}{4m^2}\rp}

\def\Pb {\bar{\Pi}}
\def\Pbb {\bar{\Pb}}

\def\b{\beta}
\def\beb {\bar{\b}}
\def\cdg {\chi^\dagger}
\def\d {\delta}
\def\deb {\bar{\d}}
\def\D {\Delta}
\def\e{\epsilon}
\def\g{\gamma}
\def\D {\Delta}

\def\l {\lambda}

\def\m{\mu}
\def\n{\nu}

\def\O{\Omega}
\def\p{\pi}
\def\P{\Pi}
\def\r{\rho}
\def\s{\sigma}
\def\S{\Sigma}
\def\t{\tau}
\def\th {\theta}

\def\cI {{\cal I}}

\def\cM {{\cal M}}
\def\cO {{\cal O}}
\def\cT {{\cal T}}

\def\bsig {\mbox{\boldmath $\sigma$}}
\def\btau {\mbox{\boldmath $\tau$}}

\def\bO {\mbox{\boldmath $\O$}}
\def\bp {\mbox{\boldmath $p$}}
\def\bP {\mbox{\boldmath $P$}}

\def\btau {\mbox{\boldmath $\tau$}}
\def\hq {\mbox{\boldmath $\hat{q}$}}
\def\bq {\mbox{\boldmath $q$}}
\def\bQ {\mbox{\boldmath $Q$}}
\def\bz {\mbox{\boldmath $z$}}

\begin{document}

\title{TWO-PION EXCHANGE\\
NUCLEON-NUCLEON POTENTIAL:\\
${\cal O}(q^4)$ RELATIVISTIC CHIRAL EXPANSION}


\author{R. Higa$^{a,b}$ and  M. R. Robilotta$^{a}$}

\address{${}^{a}$Instituto de F\'{\i}sica, Universidade de S\~{a}o Paulo,\\
C.P. 66318, 05315-970, S\~ao Paulo, SP, Brazil}

\address{${}^{b}$Thomas Jefferson National Accelerator Facility,\\
12000 Jefferson Avenue, Newport News, VA 23606, USA}

\date{\today}

\maketitle
\abstract{We present a relativistic procedure for the chiral expansion 
of the two-pion exchange component of the $NN$ potential, which emphasizes 
the role of intermediate $\p N$ subamplitudes. The relationship between power 
counting in $\p N$ and $NN$ processes is discussed and results are expressed 
directly in terms of observable subthreshold coefficients. Interactions are 
determined by one- and two-loop diagrams, involving pions, nucleons, and other 
degrees of freedom, frozen into empirical subthreshold coefficients. The full 
evaluation of these diagrams produces amplitudes containing  many different 
loop integrals. Their simplification by means of relations among these 
integrals leads to a set of intermediate results. Subsequent truncation to 
$\cO(q^4)$ yields the relativistic potential, which depends on six loop 
integrals, representing bubble, triangle, crossed box, and box diagrams. 
The bubble and triangle integrals are the same as in $\p N$ scattering and 
we have shown that they also determine the chiral structures of box and 
crossed box integrals. Relativistic threshold effects make our results 
to be not equivalent with those of the heavy baryon approach. Performing a 
formal expansion of our results in inverse powers of the nucleon mass, 
even in regions where this expansion is not valid, we recover most of the 
standard heavy baryon results. The main differences are due to the 
Goldberger-Treiman discrepancy and terms of $\cO(q^3)$, possibly 
associated with the iteration of the one-pion exchange potential.}

\vspace{5mm}

\section{introduction}\label{sec:introd}

A considerable refinement in the description of nuclear interactions has 
occurred in the last decade, due to the systematic use of chiral symmetry.
As the non-Abelian character of QCD prevents low-energy calculations, one 
works with effective theories that mimic, as much as possible, the basic 
theory. In the case of nuclear processes, where interactions are dominated 
by the quarks $u$ and $d$, these theories are required to be Poincar\'e 
invariant and to have approximate $SU(2)\times SU(2)$ symmetry.
The latter is broken by the small quark masses, which give rise to the pion 
mass at the effective level.

In the 1960s, it became well established that the one-pion exchange 
potential (OPEP) provides a good description of $NN$ interactions at large 
distances. When one moves inward, the next class of contributions corresponds 
to exchanges of two uncorrelated pions \cite{TNS} and, until recently, there 
was no consensus in the literature as how to treat this component of the 
force. An important feature of the two-pion exchange potential (TPEP) is 
that it is closely related to the pion-nucleon ($\p$N) amplitude, a point 
stressed more than thirty-five years ago by Cottingham and Vinh Mau 
\cite{CVM}. This idea allowed one to overcome the early difficulties associated 
with perturbation theory \cite{PT} and led to the construction of the 
successful Paris potential \cite{P}, where the intermediate part of the 
interaction is obtained by means of dispersion relations. This has the 
advantages of minimizing the number of unnecessary hypotheses and yielding 
model independent results, but it does not help in clarifying the role of 
different dynamical processes, which are always treated in bulk. 

Field theory provides an alternative framework for the evaluation of the 
TPEP. In this case, one uses a Lagrangian, involving the degrees of freedom 
one considers to be relevant, and calculates amplitudes using Feynman 
diagrams, which are subsequently transformed into a potential. An important 
contribution along this line was given in the early 1970s by Partovi and 
Lomon, who considered box and crossed box diagrams, using a Lagrangian 
containing just pions and nucleons with pseudoscalar (PS) coupling 
\cite{PL}. A study of the same diagrams using a pseudovector (PV) coupling 
was performed later by Zuilhof and Tjon \cite{ZT}. The development of this 
line of research led to the Bonn model for the $NN$ interaction, which included 
many important degrees of freedom and proved to be effective in reproducing 
empirical data \cite{B}. On the phenomenological side, accurate potentials 
also exist, which can reproduce low-energy observables employing parametrized 
forms of the two-pion exchange component \cite{A}.

Nowadays, it is widely acknowledged that chiral symmetry provides the best 
conceptual framework for the construction of nuclear potentials. The 
importance of this symmetry was pointed out in the 1970s by 
Brown and Durso \cite{BD} and by Chemtob, Durso, and Riska \cite{CDR}, 
who stressed that it constrains the form of the intermediate $\pi$N 
amplitude present in the TPEP. 

In the early 1990s, the works by Weinberg restating the role of chiral 
symmetry in nuclear interactions \cite{WNN} were followed by an effort by 
Ord\'o\~nez and van Kolck \cite{OvK} and other authors \cite{ChNN,RR94} 
to construct the TPEP in that framework. The symmetry was then realized 
by means of non linear Lagrangians containing only pions and nucleons. 
This minimal chiral TPEP is consistent with the requirements of chiral 
symmetry and reproduces, at the nuclear level, the well known cancellations 
present in the intermediate $\pi N$ amplitude \cite{ChC}. On the other hand, 
a Lagrangian containing just pions and nucleons could not describe 
experimental $\pi N$ data \cite{H83} and the corresponding potential 
missed even the scalar-isoscalar medium range attraction \cite{RR94}.

One needed other degrees of freedom. The $\Delta$ contributions were shown 
to improve predictions by Ord\'o\~nez, 
Ray, and van Kolck \cite{ORK} and other authors \cite{KGW}. Empirical 
information about the low-energy $\pi N$ amplitude is 
normally summarized by 
means of subthreshold coefficients \cite{H83,HJS}, which can be used either 
directly in 
the construction of the TPEP or to 
determine unknown coupling constants (LECs) in chiral Lagrangians. 
The inclusion of this information allowed 
satisfactory descriptions of the asymptotic $NN$ data to be produced, with no 
need of free parameters \cite{RR97,KBW,Nij,EGM}.

As far as techniques for implementing the symmetry are concerned, recent 
calculations of 
the TPEP were performed using both heavy baryon chiral perturbation theory 
(HBChPT)  and covariant Lagrangians. In the former case 
\cite{OvK,ORK,KBW,Nij,EGM,K3}, one uses  non relativistic effective 
Lagrangians, which include unknown counterterms, and amplitudes are derived 
in which loop and counterterm contributions are organized in well defined 
powers of a typical low-energy scale. 
In this approach, relativistic corrections required by precision have to be 
added separately \cite{K4}.

QCD is a theory without formal ambiguities and the same should happen 
with effective theories designed to be used at the hadron level. In 
the case of nuclear interactions, this allows one to expect that the 
chiral TPEP should be unique, except for the iteration of the OPEP, 
which depends on the dynamical equation employed. 

In the meson sector, chiral perturbation is indeed unique and predictions at 
a given order are unambiguous. However, the problem becomes much more 
difficult for systems containing baryons. At present, the uniqueness 
problem is under scrutiny and two competing calculation procedures 
are available based on either heavy 
baryon (HBChPT) or relativistic (RBChPT) techniques. If both approaches 
are correct, they should produce fully equivalent predictions for a 
given process. Descriptions of single nucleon properties were found to 
be consistent, provided the nucleon mass is used as the dimensional 
regularization scale \cite{BKKM}. 
In the case of $\pi N$ scattering, comparison of predictions 
became possible only recently, through the works of Fettes, Mei\ss ner, 
and Steininger \cite{stein} (HBChPT) and Becher and Leutwyler 
\cite{BL1,BL2} (RBChPT). Differences were found, associated with the fact 
that some classes of diagrams cannot be fully represented by the heavy 
baryon series. Discussions of the pros and cons of these techniques may be 
found in refs. \cite{meissrev,EllisT}. 

In the $NN$ problem, all perturbative calculations produced so far were 
based on HBChPT 
\cite{OvK,ORK,KGW,KBW,Nij,EGM,K3,K4}. 
On the other hand, an indication exists that $NN$ results are approach 
dependent,
for the large distance properties of the central potential were shown to be
dominated by diagrams that cannot be expanded in the HB series \cite{R01}.
The main motivation of the present work is to extend the discussion of 
the uniqueness of chiral predictions to the $B=2$ sector. 
We do this by calculating covariantly the TPEP to order ${\cal O}(q^4)$ 
and comparing our results with the HB potential at the same order.

Our presentation is organized as follows. In section \ref{sec:tpep-form} 
we give the formal relations between the relativistic TPEP and the 
intermediate $\pi N$ amplitude, whose chiral structure is analyzed in 
section \ref{sec:inter-piN}.
We discuss how power counting in $\pi N$ is transferred to the TPEP in 
section \ref{sec:pw-count} and how it is reflected into 
subthreshold coefficients in section \ref{sec:subt-coef}.
The problem of the triangle diagram, in which heavy baryon and 
relativistic descriptions disagree, is briefly reviewed in section 
\ref{sec:rel-hb}. In section \ref{sec:dynamics} we discuss the dynamical
content of the potential
and the properties of important loop integrals used to express it.
Our TPEP is suited to Lippmann-Schwinger
dynamics and, in appendix \ref{ap:iterat}, we review the subtraction of 
the OPEP iteration, needed to avoid double counting. 
The full TPEP, which represents an extension of our earlier work 
\cite{RR94,RR97}, is derived in appendix \ref{ap:fullres}.
This potential is transformed using relations among integrals given 
in appendix \ref{ap:relat} and a new form is given in appendix 
\ref{ap:interes}, which is simpler by the neglect of short range 
contributions. The truncation of these results gives rise to our 
${\cal O}(q^4)$ invariant amplitudes and potential components, 
displayed in sections \ref{sec:covamp} and \ref{sec:tpep}.
In section \ref{sec:compare} we compare our TPEP with the standard heavy 
baryon version, using expansions for loop integrals derived 
in appendix \ref{ap:relex}.
Conclusions are presented in section \ref{sec:concl}, whereas appendixes 
\ref{ap:kin} and \ref{ap:integ} deal with kinematics and relativistic loop 
integrals.

\section{TPEP - formalism }\label{sec:tpep-form}

\begin{figure}[ht]
\begin{center}
\epsfig{figure=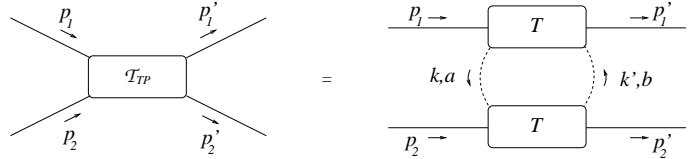, width=9cm}
\end{center}
\caption{Two-pion exchange amplitude.}\label{fig1}
\end{figure}

The TPEP is obtained from the $T$ matrix $\cT_{T\!P}$, which describes the 
on-shell process  $N(p_1)\;N(p_2)\rar N(p'_1)\;N(p'_2)$ and contains two 
intermediate pions, as represented in fig.\ref{fig1}. In order to derive the 
corresponding potential, one goes to the center of mass frame and subtracts 
the  iterated OPEP, so as to avoid double counting. The $NN$ interaction is 
thus closely associated with the off-shell $\p N$ amplitude.

The coupling of the two-pion system to a nucleon is described by $T$, the 
amplitude for the process $\pi^a(k) N(p)\rar\pi^b(k') N(p')$. It has the 
isospin structure

\beq
T_{ba} = \d_{ab} T^+ +  i\e_{bac}\t_c \; T^-
\label{2.1}
\eeq

\ni
and the evaluation of fig.\ref{fig1} yields

\beq
{\cT}_{T\!P} =  \lb 3\;{\cT}^+ + 2\; \btau^{(1)} \cd \btau^{(2)} \; 
{\cT}^- \rb 
\label{2.2}
\eeq

\ni
with

\beq
\cT^\pm = -\;\frac{i}{2!} \int \frac{d^4Q}{(2\pi)^4}\;
\frac{ [T^\pm]^{(1)}\;[T^\pm]^{(2)}}{[k^2-\mu^2]\;[k'^2-\mu^2]} \;,
\label{2.3}
\eeq

\ni 
where $\m$ is the pion mass and the factor $1/2!$ accounts for the 
exchange symmetry of the intermediate pions. The integration variable 
is $Q=(k'+k)/2$ and we also define $q=(k'-k)$, $t=q^2$, and 
$\n_i=(p_i'+p_i)\cd Q /2m$. Our kinematical variables are fully displayed 
in appendix \ref{ap:kin}.

For on-shell nucleons, the sub amplitudes $T^\pm$ may be written as

\beq
T^\pm = \ub(\bp') \lb A^\pm +  \Qs \; B^\pm \rb u(\bp)
\label{2.4}
\eeq

\ni 
and the functions $A^\pm$ and $B^\pm$ are determined dynamically. 
An alternative possibility is  

\beq
T^\pm = \ub(\bp') \lb D^\pm - \;\frac{i}{2m} \s_{\m\n} (p'-p)^\m Q^\n \; 
B^\pm \rb u(\bp) 
\label{2.5}
\eeq

\ni 
with $D^\pm=A^\pm+\n B^\pm$. This second form tends to be more convenient 
when one is interested in the chiral content of the amplitudes. The 
information needed about the pion-nucleon sub amplitudes $A^\pm$, $B^\pm$, 
and $D^\pm$ may be found in the comprehensive review by H\"ohler \cite{H83} 
and in the recent chiral analysis by Becher and Leutwyler \cite{BL2}. 

The intermediate $\p N$ subamplitudes $A^\pm$, $B^\pm$, and $D^\pm$ depend 
on the variables $k^2$, $k'^2$, $\nu$, and $t$. For 
physical processes one has $k'^2=k^2=\m^2$, $\nu\ge\mu$ and $t \le 0$. 
On the other hand, the conditions of integration in eq.(\ref{2.2}) are 
such that the pions are off-shell and the main contributions come from 
the region $\nu\approx 0$. Physical amplitudes cannot be 
directly employed in the evaluation of the TPEP and must be continued 
analytically to the region below threshold, by means of either dispersion 
relations or field theory.
In both cases one should preserve the analytic structure of the $\p N$ 
amplitude, which plays an important role in the TPEP.

The relativistic spin structure of the TPEP is obtained by using 
eq.(\ref{2.5}) into eq.(\ref{2.3}) and one has, for each isospin channel, 

\bea
\cT &=& [\ub\;u]^{(1)}[\ub\;u]^{(2)}\; \cI_{DD}
- \frac{i}{2m} [\ub\;u]^{(1)}[\ub \; \s_{\m\l} \; 
(p'-p)^\m \;u]^{(2)} \; \cI_{DB}^\l
\nn\\[2mm]
&-& \frac{i}{2m} [\ub \; \s_{\m\l} (p'-p)^\m \;u]^{(1)}[\ub\;u]^{(2)} 
\; \cI_{BD}^\l
\nn\\[2mm]
&-& \frac{1}{4m^2} [\ub \; \s_{\m \l} (p'-p)^\m \;u]^{(1)}[\ub \; 
\s_{\n\r}(p'-p)^\n \;u]^{(2)} \; \cI_{BB}^{\l\r}\;,
\label{2.6}
\eea

\ni 
where 

\bea
&& \cI_{DD} = -i/2 \int [\cdots] \;  [D]^{(1)}[D]^{(2)}\;,
\label{2.7}\\
&& \cI_{DB}^\l = -i/2 \int [\cdots] \;[D]^{(1)}[Q^\l \; B]^{(2)}\;,
\label{2.8}\\
&& \cI_{BD}^\l = -i/2 \int [\cdots] \; [Q^\l \; B]^{(1)}[D]^{(2)}\;,
\label{2.9}\\
&& \cI_{BB}^{\l\r} = -i/2 \int [\cdots] \; [Q^\l\;B]^{(1)}[Q^\r\; B]^{(2)}\;,
\label{2.10}
\eea

\ni
and

\beq
\int [\cdots]  =  \int \frac{d^4Q}{(2\pi)^4}\; 
\frac{1}{[k^2-\mu^2]\;[k'^2-\mu^2]} \;.
\label{2.11}
\eeq

The Lorentz structure of the integrals $\cI$ is realized in terms of the 
external quantities $q$, $z$, $W$, and $g^{\m\n}$, defined in appendix 
\ref{ap:kin}. Terms proportional to $q$ do not contribute and we write 

\bea
&& \cI_{DB}^\l = \frac{W^\l}{2m}\;  \cI_{DB}^{(w)} +  \frac{z^\l}{2m}\; 
\cI_{DB}^{(z)} \;,
\label{2.12}\\
&& \cI_{BD}^\l = \frac{W^\l}{2m}\;  \cI_{DB}^{(w)} -  \frac{z^\l}{2m}\;  
\cI_{DB}^{(z)} \;,
\label{2.13}\\
&& \cI_{BB}^{\l\r} =  g^{\l\r}\;\cI_{BB}^{(g)}+ \frac{W^\l W^\r}{4 m^2}\; 
\cI_{BB}^{(w)} +  \frac{z^\l z^\r}{4m^2}\;  \cI_{BB}^{(z)} \;.
\label{2.14}
\eea

These expressions and the spinor identities (\ref{a19}) and (\ref{a21}) 
yield 

\bea
&& \cT =  [\ub\;u]^{(1)}[\ub\;u]^{(2)}\; 
\lb \cI_{DD}+\frac{q^2}{2m^2}\;\cI_{DB}^{(w)} + 
\frac{q^4}{16m^4}\;\cI_{BB}^{(w)}\rb
\nn\\[2mm]
&& - \frac{i}{2m}\lc [\ub\;u]^{(1)}[\ub \; \s_{\m\l} (p'-p)^\m \;u]^{(2)}
-  [\ub \; \s_{\m\l} (p'-p)^\m \;u]^{(1)}[\ub\;u]^{(2)} \rc \frac{z^\l}{2m}
\nn\\[2mm]
&& \times \lb \cI_{DB}^{(w)} + \cI_{DB}^{(z)} + \frac{q^2}{4m^2}\; 
\cI_{BB}^{(w)}\rb 
\nn\\[2mm]
&& - \frac{1}{4m^2} [\ub \; \s_{\m \l} (p'-p)^\m \;u]^{(1)}[\ub \; 
\s_{\n\r}(p'-p)^\n \;u]^{(2)} 
\lb g^{\l\r}\; \cI_{BB}^{(g)} + \frac{z^\l z^\r}{4m^2} \lp -\cI_{BB}^{(w)} + 
\cI_{BB}^{(z)} \rp \rb \;.
\label{2.15}
\eea

In order to display the ordinary spin content of this amplitude, we go to 
the center of mass frame and use identities (\ref{a31})--(\ref{a34}), which 
allow one to rewrite $\cT_{T\!P}$, without approximations, in terms of 
the $(2\times 2)$ identity matrix and the operators\footnote{We use here 
the notation and results from Partovi and Lomon \cite{PL}, 
eqs.(4.26)--(4.28).}

\bea 
\bO_{SS}=\bq^2\; \bsig^{(1)}\cd\bsig^{(2)}\;, && \bO_T=-\bq^2\;
(3\bsig^{(1)}\cd\hq\bsig^{(2)}\cd\hq\!-\!\bsig^{(1)}\cd\bsig^{(2)})\;,
\nn\\
\bO_{LS}= i\;(\bsig^{(1)}\!+\!\bsig^{(2)}) \cd \bq\times \bz /4 \;, && 
\bO_Q = \bsig^{(1)}\cd \bq\!\times\!\bz \;\bsig^{(2)}\cd\bq\!\times\!\bz \;.
\nn
\eea

The two-component momentum space amplitude in the center of mass (CM) 
is derived by dividing ${\cT}$ by the factor $(4Em)$, present in the 
relativistic normalization, 
and introducing back the isospin coefficients as in eq.(\ref{2.2}). 
We then have the decomposition

\beq
t_{cm}^\pm \equiv \t^\pm \;\frac{{\cT_{cm}^\pm}}{4Em} 
= t_C^\pm + \frac{\bO_{SS}}{m^2}\; t_{SS}^\pm  + \frac{\bO_T}{m^2}\; t_T^\pm 
+ \frac{\bO_{LS}}{m^2}\; t_{LS}^\pm 
+ \frac{\bO_Q}{m^4}\; t_Q^\pm 
\label{2.16}
\eeq

\ni
with $\t^+ = 3$ and $\t^- = 2$.
Finally, the momentum space potential, denoted by $\hat{t}^\pm$, is obtained 
by subtracting  the iterated OPEP from this expression, so as to avoid double 
counting.

\section{intermediate $\pi N$ amplitude}\label{sec:inter-piN}

The theoretical soundness of the TPEP relies heavily on the description 
adopted for the intermediate $\p N$ amplitude. In this work we employ the 
relativistic chiral representation produced by the Bern group and 
collaborators \cite{BL1,BL2,GSS}, which incorporates the correct analytic 
structure. For the sake of completeness, in this section we summarize some 
of their results.  

At low and intermediate energies, the $\pi N$ amplitude is given by the 
nucleon pole contribution, superimposed to a smooth background. Chiral 
symmetry is realized differently in these two sectors and it is useful 
to disentangle the pseudovector Born term $(pv)$ from a remainder $(R)$.
We then write

\beq
T^\pm = T_{pv}^\pm + T_R^\pm \;.
\label{3.1}
\eeq

The $pv$ contribution involves two observables, namely, the nucleon 
mass $m$ and the $\p N$ coupling constant $g$, as prescribed by the 
Ward-Takahashi identity \cite{WTI}. 
In chiral perturbation theory, depending on the order one 
is working with, the calculation of these quantities may involve 
different numbers of loops and several coupling constants\footnote{
For instance, up to ${\cal O}(q^4)$ $T^\pm_{pv}$ receives contributions 
from tree graphs of ${\cal L}^{(1)}\cdots{\cal L}^{(4)}$ and one-loop 
graphs from ${\cal L}^{(1)}$ and ${\cal L}^{(2)}$, expressed in terms 
of its bare coupling constants.}. Nevertheless,
at the end, results must be organized in such a way as to
reproduce the physical values of both $m$ and $g$ in $ T_{pv}^\pm$ \cite{Moj}.
Following H\"ohler \cite{H83} and the Bern group, 
\cite{GSS,BL2} in their treatments of the Born term, we use the constant $g$ 
in these equations, instead of  $(g_A/f_\p)$. The motivation for this choice 
is that the $\p N$ coupling constant is indeed the observable determined by 
the residue of the nucleon pole. We write 

\bea
D_{pv}^+ &=& \frac{g^2}{2m}\; \lp \frac{k'\cd k}{s-m^2}+
\frac{k'\cd k}{u-m^2}\rp \rar \cO(q^2) \;,
\label{3.2}\\
B_{pv}^+ &=&  - g^2 \; \lp \frac{1}{s-m^2}-\;\frac{1}{u-m^2}\rp 
\rar \cO(q^{-1}) \;,
\label{3.3}\\
D_{pv}^- &=&  \frac{g^2}{2m} \; \lp \frac{k\cd k'}{s-m^2}-\;
\frac{k\cd k'}{u-m^2}-\; \frac{\n}{m} \rp \rar \cO(q) \;,
\label{3.4}\\
B_{pv}^- &=& - g^2 \; \lp \frac{1}{s-m^2}+ \frac{1}{u-m^2} +
\frac{1}{2m^2} \rp \rar \cO(q^0) \;,
\label{3.5}
\eea

\ni
where $s=(p+k)^2=(p'+k')^2$ and $u=(p-k')^2=(p'-k)^2$.
The arrows after the equations indicate their chiral orders, estimated by 
using $s-m^2 \sim W\cd Q$ and $u-m^2 \sim - W\cd Q$, with 
$W= p_1+p_2 = p'_1+p'_2$. When the relative sign between the $s$ and $u$ 
poles is negative, these contributions add up and we have 
$[1/(s-m^2)-1/(u-m^2)] \rar \cO(q^{-1})$. On the other hand, when the 
relative sign is positive, the leading contributions cancel out and we 
obtain $[1/(s-m^2)+1/(u-m^2)] \rar \cO(q^0)$.

In ChPT, the structure of the amplitudes $T_R^\pm$ involves both tree and 
loop contributions. The former can be read directly from the basic 
Lagrangians and correspond to polynomials in $\n$ and $t$, with coefficients 
given by the renormalized LECs. The calculation 
of the latter is more complex and results may be expressed in terms of 
Feynman integrals. In the description of $\p N$ processes below threshold, 
it is useful to approximate these contributions by polynomials, using

\beq
X_R = \sum x_{mn} \n^{2m}t^n \;,
\label{3.6}
\eeq

\ni where $X_R$ stands for $D_R^+$, $B_R^+/\n$, $D_R^-/\n$, or $B_R^-$.
The values of the coefficients $x_{mn}$ can be determined empirically, by 
using dispersion relations in order to extrapolate physical scattering data
to the subthreshold region \cite{H83,HJS}. As such, they acquire the status
of observables and become a rather important source of information about the
values of the LECs.

The isospin odd subthreshold coefficients include leading order 
contributions, which yield the predictions made by Weinberg \cite{W66} 
and Tomozawa \cite{T66} (WT) for $\p N$ scattering lengths, given by

\bea
&& D_{WT}^- = \frac{\n}{2 f_\p^2} \rar\cO(q) \;,
\label{3.7}\\
&& B_{WT}^- = \frac{1}{2 f_\p^2} \rar\cO(q^0)\;.
\label{3.8}
\eea

Sometime ago, we developed a chiral description of the TPEP based on 
the empirical values of the subthreshold coefficients, which could 
reproduce asymptotic $NN$ data \cite{RR97}. As we discuss in the 
sequence, that description has to be improved when one goes beyond 
$\cO(q^3)$. In nuclear interactions, the ranges of the various processes 
are associated with the variable $t$ and must be accurately described. 
In particular, the pion cloud of the nucleon gives rise to scalar and 
vector form factors \cite{GSS}, which correspond, in configuration space, 
to structures that extend well beyond 1 fm \cite{R01}. On the other hand, the 
representation of an amplitude by means of a power series, as in 
eq.(\ref{3.6}), amounts to a zero-range expansion, for its Fourier 
transform yields only $\d$ functions and its derivatives. So, this kind 
of representation is suited for large distances only.
At shorter distances, the extension of the objects begins to appear. 

\begin{figure}[ht]
\begin{center}
\epsfig{figure=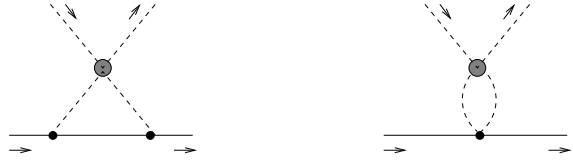, width=7.5cm}
\end{center}
\caption{Long range contributions to the scalar and vector form factors.}
\label{fig2}
\end{figure}

In the work of Becher and Leutwyler \cite{BL2} we can check that the only
sources of $NN$ medium range effects are their diagrams $k$ and $l$,
reproduced in figure \ref{fig2}, which contain two pions propagating in
the $t$ channel. Here we consider explicitly their full contributions and 
our amplitudes $A_R^\pm$ and $B_R^\pm$ are written as 

\bea
D_R^+&=&D_{mr}^+(t)+\lb\db_{00}^++d_{10}^+\n^2+\db_{01}^+t\rb_{(2)}
+\lb d_{20}^+\n^4+d_{11}^+\n^2t+\db_{02}^+t^2\rb_{(3)}\;,
\label{3.9}\\[2mm]
B_R^+&=&B_{mr}^+(t)+\lb b_{00}^+\n\rb_{(1)}\;,
\label{3.10}\\[2mm]
D_R^-&=&D_{mr}^-(t)+\lb \n/(2f_\p^2)\rb_{(1)}+\lb\db_{00}^-\n+d_{10}^-\n^3
+\db_{01}^-\n t\rb_{(3)}\;,
\label{3.11}\\[2mm]
B_R^-&=&B_{mr}^-(t)+\lb 1/(2f_\p^2)+\bb_{00}^-\rb_{(0)}+\lb b_{10}^-\n^2+
\bb_{01}^-t\rb_{(1)}\;.
\label{3.12}
\eea

In these expressions, the labels $(n)$ outside the brackets indicate the 
presence of leading terms of $\cO (q^n)$, whereas the label $mr$ denotes 
the contribution from the medium range diagrams of fig.\ref{fig2}. This 
decomposition implies the redefinition of some subthreshold coefficients, 
indicated by a bar over the appropriate symbol. Their explicit forms will 
be displayed in the sequence. 

\begin{figure}[ht]
\begin{center}
\epsfig{figure=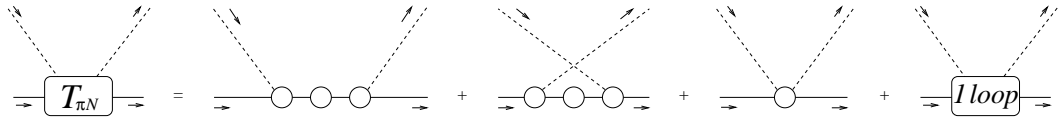, width=14cm}
\end{center}
\caption{Dynamical structure of the ${\cal O}(q^4)$ $\p N$ amplitude; 
the blobs represent terms coming directly from the effective Lagrangians.}
\label{fig3}
\end{figure}

\begin{figure}[ht]
\begin{center}
\epsfig{figure=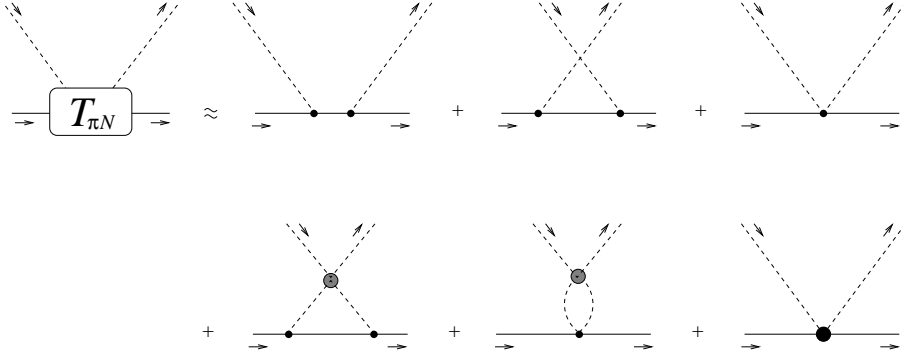, width=12cm}
\end{center}
\caption{Dynamical content of the approximate $\pi N$ amplitude.}
\label{fig4}
\end{figure}

The dynamical content of the $\cO(q^4)$ $T_{\p N}$ amplitude derived in 
\cite{BL2} is shown in fig.\ref{fig3} and our approximation in fig.\ref{fig4}.
In the latter, the first two 
diagrams correspond to the direct and crossed PV Born amplitudes, with 
physical masses and coupling constants. The third one represents the contact 
interaction associated with the Weinberg-Tomozawa vertex, whereas the next 
two describe the medium range effects associated with the scalar and vector 
form factors. Finally, the last diagram summarizes the terms within square 
brackets in eqs.(\ref{3.9})--(\ref{3.12}).

\section{power counting}\label{sec:pw-count}

One begins the expansion of the TPEP to a given chiral order by recasting 
the explicitly covariant $\cT_{T\!P}$ into the two-component form of 
eq.(\ref{2.16}). This procedure involves no approximations and one finds, 
in the CM frame,

\bea
t_C^\pm  &=& \tau^\pm \;\frac{m}{E} \lb \lp 1\!+\! \bq^2\!/\l^2 \rp^2 
\cI_{DD}^\pm-\frac{\bq^2}{2m^2}\lp 1\! +\! \bq^2\!/\l^2 \rp \lp 1 \!+ \! 
\bq^2\!/\l^2 \!+\! \bz^2\!/ \l^2 \rp \cI_{DB}^{(w)\pm}\right.
\nn\\[2mm]
&+& \left.\frac{\bq^4}{16m^4} \lp 1 \!+\! \bq^2\!/\l^2 \!+\! \bz^2\!/\l^2 
\rp^2 \cI_{BB}^{(w)\pm}+\frac{\bq^4}{16m^4} \lp 1 \!+\! 4 m^2 \bz^2\!/ \l^4 
\rp \cI_{BB}^{(g)\pm}\right.
\nn\\[2mm]
&-& \left.  \frac{\bq^2\bz^2}{2 m^2 \l^2} \lp 1 \!+\! \bq^2\!/\l^2 \rp 
\cI_{DB}^{(z)\pm}-\; \frac{\bq^4\bz^4}{16 m^4 \l^4}\; \cI_{BB}^{(z)\pm}\rb \;,
\label{4.1}\\[2mm]
t_{SS}^\pm  &=&  \tau^\pm \;\frac{m}{E} \lb -\; \frac{1}{6 }\; 
\cI_{BB}^{(g)\pm}\rb  \;,
\label{4.2}\\[2mm]
t_T^\pm  &=& \tau^\pm \;\frac{m}{E} \lb -\; \frac{1}{12} \; 
\cI_{BB}^{(g)\pm}\rb  \;,
\label{4.3}\\[2mm]
t_{LS}^\pm  &=& \tau^\pm \;\frac{m}{E} \lb -\; \frac{4m^2}{\l^2} \lp 1 
\!+\! \bq^2\!/ \l^2 \rp \cI_{DD}^\pm+ \lp 1\!+\! 2 \bq^2\!/\l^2 \rp \lp 1 
\!+\! \bq^2\!/\l^2 \!+\! \bz^2\!/\l^2 \rp \cI_{DB}^{(w)\pm}\right.
\nn\\[2mm]
&-& \left. \frac{\bq^2}{4m^2} \lp 1 \!+\! \bq^2\!/\l^2 \!+\! \bz^2\!/\l^2 
\rp^2 \cI_{BB}^{(w)\pm}-\; \frac{\bq^2}{4m^2} \lp 1 \!+\! 4m^2\!/ \l^2 \!
+\! 4 m^2 \bz^2 \! / \l^4 \rp \cI_{BB}^{(g)\pm}\right.
\nn\\[2mm]
&+& \left. \lp 1 \!+\! \bq^2\! / \l^2 \!+\! \bz^2\!/ \l^2 
+2\bq^2\bz^2 / \l^4\rp 
\cI_{DB}^{(z)\pm}+\frac{\bq^2\bz^2}{4 m^2 \l^2} \lp 1 \!+\! \bz^2\! / 
\l^2 \rp  \cI_{BB}^{(z)\pm}\rb \;,
\label{4.4}\\[2mm]
t_Q^\pm  &=&  \tau^\pm \;\frac{m}{E} \lb -\; \frac{m^4}{\l^4} \; 
\cI_{DD}^\pm
+ \frac{m^2}{2\l^2} \lp 1 \!+\! \bq^2\!/ \l^2 \!+\! \bz^2\! / \l^2 \rp 
\cI_{DB}^{(w)\pm}\right.
\nn\\[2mm]
&& \left.-\; \frac{1}{16} \lp 1 \!+\! \bq^2\! / \l^2 \!+\! \bz^2\! / \l^2 
\rp^2 \cI_{BB}^{(w)\pm}- \; \frac{1}{16} \lp 1 \!+\! 8m^2\!/ \l^2 \!+\! 
4 m^2 \bz^2\! / \l^4\rp \cI_{BB}^{(g)\pm}\right.
\nn\\[2mm]
&& \left. + \frac{m^2}{2 \l^2} \lp 1 \!+\! \bz^2 \! / \l^2 \rp 
\cI_{DB}^{(z)\pm}+ \frac{1}{16} \lp 1 \!+\! \bz^2 \! / \l^2 \rp^2 
\cI_{BB}^{(z)\pm}\rb \;,
\label{4.5}
\eea

\ni 
with $\bq = \bp'\!-\!\bp$, $\bz = \bp'\!+\!\bp$ and $\l^2 = 4m(E+m)$.

The potential to order $\cO(q^n)$ is determined by $t_C^\pm \rar \cO(q^n)$, 
$\{t_{SS}^\pm, t_T^\pm, t_{LS}^\pm\} \rar \cO(q^{n\!-\!2})$ and 
$t_Q^\pm \rar \cO(q^{n\!-\!4})$. This means that one needs 
$\cI_{DD}^\pm \rar \cO(q^n)$, $\{ \cI_{DB}^{(w)\pm}, \cI_{DB}^{(z)\pm}, 
\cI_{BB}^{(g)\pm}\} \rar \cO(q^{n\!-\!2})$ and 
$\{ \cI_{BB}^{(w) \pm}, \cI_{BB}^{(z) \pm}\} \rar \cO(q^{n\!-\!4})$. We 
now discuss how the chiral powers in these functions are related with 
those in the basic $\p N$ amplitude. This relationship involves a subtlety, 
associated with the fact that $D_{pv}^+$ and $B_{pv}^-$ contain chiral 
cancellations. 

A generic subamplitude $\cI_{XY}^\pm$ is given by the product of the 
corresponding $\p N$ contributions and we have 

\beq
\cI_{XY}^\pm = \int [\cdots] \lc \lb X_{pv}^\pm \rb^{(1)}\! \lb Y_{pv}^\pm 
\rb^{(2)}\!\!+\! \lb X_{pv}^\pm \rb^{(1)}\! \lb Y_R^\pm \rb^{(2)} 
\!\!+\! \lb X_R^\pm \rb^{(1)}\! \lb Y_{pv}^\pm \rb^{(2)}
\!\!+\! \lb X_R^\pm \rb^{(1)}\! \lb Y_R^\pm \rb^{(2)}\rc\;.
\label{4.6}
\eeq 

The loop integral and the two pion propagators, as given by eq.(\ref{2.11}), 
do not interfere with the counting of powers, since 
$\int [\cdots] \rar \cO(q^0)$. The loop integration is symmetric under the 
operation $Q\rar - Q$, which gives rise to the exchange 
$s \leftrightarrow u$ in the Born terms. In the case of 
$[X_{pv}^\pm]^{(1)}\; [Y_{pv}^\pm ]^{(2)}$, one is allowed to use 

\beq
\lp \frac{1}{s-m^2}\pm \frac{1}{u-m^2}\rp^{(i)} \lp \frac{1}{s-m^2}\pm 
\frac{1}{u-m^2}\rp^{(j)}\rar 2  \lp \frac{1}{s-m^2}\rp^{(i)} \lp 
\frac{1}{s-m^2}\pm \frac{1}{u-m^2}\rp^{(j)}
\label{4.7}
\eeq

\ni 
within the integrand. For the specific components this yields

\bea
\lb D_{pv}^+\rb^{(i)} \lb D_{pv}^+\rb^{(j)}\! \rar \cO(q^3), \;\;\;\;\;
&&  \lb D_{pv}^-\rb^{(i)} \lb D_{pv}^-\rb^{(j)}\! \rar \cO(q^2), 
\nn\\ 
\lb D_{pv}^+\rb^{(i)} \lb Q B_{pv}^+\rb^{(j)}\! \rar \cO(q), \;\;\;\;\;
&& \lb D_{pv}^-\rb^{(i)} \lb Q B_{pv}^-\rb^{(j)}\! \rar \cO(q),
\nn\\
\lb Q B_{pv}^+\rb^{(i)} \lb Q B_{pv}^+\rb^{(j)}\! \rar \cO(q^0), \;\;\;\;\;
&& \lb Q B_{pv}^-\rb^{(i)} \lb Q B_{pv}^-\rb^{(j)}\! \rar \cO(q).
\nn
\eea

These results show that, inside the integral, $D_{pv}^+$ and $B_{pv}^-$ 
cannot be always counted as $\cO(q^2)$ and $\cO(q^{-1})$, respectively. 
For the products  $[ X_{pv}^\pm ]^{(i)}\; [Y_R^\pm ]^{(j)}$ and  
$[X_R^\pm]^{(i)}\; [Y_{pv}^\pm]^{(j)}$, one uses

\beq
\lp \frac{1}{s-m^2}\pm \frac{1}{u-m^2}\rp^{(i)} \rar 2  \lp 
\frac{1}{s-m^2}\rp^{(i)} 
\label{4.8}
\eeq

\ni 
and has $D_{pv}^\pm \rar \cO(q)$ and $B_{pv}^\pm \rar \cO(q^{-1})$. 
Assuming $ [X_R^\pm]^{(i)}, [Y_R^\pm]^{(j)} \rar \cO(q^r)$, one gets 

\bea
\lb D_{pv}^\pm\rb^{(i)} \lb D_R^\pm\rb^{(j)}\! \rar \cO(q^{1+r}),\;\;\;\;\;
&& \lb D_{pv}^\pm\rb^{(i)} \lb Q B_R^\pm\rb^{(j)}\! \rar \cO(q^{2+r}),
\nn\\
\lb D_R^\pm\rb^{(i)} \lb Q B_{pv}^\pm\rb^{(j)}\! \rar \cO(q^{r}),\;\;\;\;\;
&& \lb Q B_{pv}^\pm\rb^{(i)} \lb Q B_R^\pm \rb^{(j)}\! \rar \cO(q^{1+r}).
\nn
\eea

Finally, in the case of  $ [X_R^\pm]^{(i)} [Y_R^\pm]^{(j)}$, one just adds 
the corresponding powers. 

In this work we consider the expansion of the potential to $\cO(q^4)$ and 
need $\cI_{DD}^\pm \rar \cO(q^4)$, 
$\{ \cI_{DB}^{(w)\pm}, \cI_{DB}^{(z)\pm}, \cI_{BB}^{(g)\pm} \}\rar \cO(q^2)$, 
and $\{ \cI_{BB}^{(w)\pm}, \cI_{BB}^{(z)\pm}\}\rar \cO(q^0)$. This means 
that, in the intermediate $\p N$ amplitude, we must consider 
$D_R^\pm$ to $\cO(q^3)$ and $B_R^\pm$ to $\cO(q)$.

\section{subthreshold coefficients}\label{sec:subt-coef}

The polynomial parts of the amplitudes $T_R^\pm$ to order $\cO(q^3)$, as
given by eqs.(\ref{3.7})--(\ref{3.10}), are determined by the subthreshold
coefficients of ref. \cite{BL2}, which we reproduce below

\bea
d_{00}^+ &=& - \frac{2\;(2c_1 - c_3)\; \m^2}{f_\p^2}+
\frac{8\; g_A^4\;\m^3}{64\; \p \;f_\p^4}
+ \lb \frac{3\;g_A^2 \; \m^3}{64\; \p \;f_\p^4}\rb_{mr}\;,
\label{5.1}\\
d_{10}^+ &=& \frac{2\; c_2}{f_\p^2}- \frac{ (4+ 5 \;g_A^4)\;\m}{32\; \p \;
f_\p^4}\;,
\label{5.2}\\
d_{01}^+ &=& - \frac{c_3}{f_\p^2} - \frac{48\; g_A^4 \;\m}{768\; \p\; f_\p^4}
- \lb \frac{77\; g_A^2 \; \m}{768\; \p\; f_\p^4}\rb_{mr} \;,
\label{5.3}\\
d_{20}^+ &=& \frac{12 + 5\;g_A^4}{192\; \p \;f_\p^4\;\m}\;,
\label{5.4}\\
d_{11}^+ &=& \frac{g_A^4}{64\; \p \;f_\p^4\;\m}\;,
\label{5.5}\\
d_{02}^+ &=& \lb \frac{193\; g_A^2}{15360 \; \p \;f_\p^4\;\m}\rb_{mr} \;,
\label{5.6}\\
b_{00}^+ &=& \frac{4\;m\; (\tilde{d}_{14} - \tilde{d}_{15})}{f_\p^2} -
\frac{g_A^4 \; m}{8\; \p^2 \;f_\p^4}\;,
\label{5.7}\\
d_{00}^- &=& \lb \frac{1}{2\;f_\p^2}\rb_{WT}
+ \frac{4 \; (\tilde{d}_1 + \tilde{d}_2  + 2\; \tilde{d}_5) \;\m^2}{f_\p^2}
+ \frac{g_A^2\;(-3 + g_A^2)\;\m^2}{48\;\p^2\;f_\p^4}+ \lb
\frac{3\; g_A^2\;\m^2}{48\;\p^2\;f_\p^4}\rb_{mr} \;,
\label{5.8}\\
d_{10}^- &=&  \frac{4 \; \tilde{d}_3}{f_\p^2} -
\frac{15 + 7\; g_A^4}{240 \;\p^2\;f_\p^4}\;,
\label{5.9}\\
d_{01}^- &=& - \frac{2\;(\tilde{d}_1+\tilde{d}_2)}{f_\p^2} -
\frac{ 2\; g_A^4}{192 \;\p^2\;f_\p^4}
- \lb \frac{1 + 7 \; g_A^2}{192 \;\p^2\;f_\p^4}\rb_{mr} \;,
\label{5.10}\\
b_{00}^- &=& \lb \frac{1}{2\;f_\p^2}\rb_{WT}
+ \frac{2\; c_4 \;m}{f_\p^2}- \frac{g_A^4 \; m\; \m}{8\; \p \;f_\p^4}
- \lb \frac{g_A^2 \; m\; \m}{8\; \p \;f_\p^4}\rb_{mr} \;,
\label{5.11}\\
b_{10}^- &=&  \frac{g_A^4 \; m}{32\; \p \;f_\p^4\;\m}\;,
\label{5.12}\\
b_{01}^- &=&  \lb \frac{g_A^2 \; m}{96\; \p \;f_\p^4\;\m}\rb_{mr} \;,
\label{5.13}
\eea

\ni
where the parameters $c_i$ and $\tilde{d}_i$ are the usual renormalized
coupling constants of the chiral Lagrangians of order 2 and 3, respectively 
\cite{BKKM}. The terms within square brackets labelled $(mr)$ in some of
these results are due to the medium range diagrams shown in fig.\ref{fig2}
and must be neglected\footnote{In ref. \cite{BL2}, the contribution of the
triangle diagram to $d_{00}^+$ includes both short and medium range terms
and only the latter must be excluded.}, because we already include their
contributions in $D_{mr}^\pm$ and $B_{mr}^\pm$. The terms bearing the
$(W\!T)$ label must also be excluded, for they were explicitly considered
in eqs.(\ref{3.9})--(\ref{3.12}). This corresponds to the redefinition
mentioned at the end  of section \ref{sec:inter-piN}.

\begin{table}[t]
\begin{center}
\caption{experimental values for the subthreshold coefficients and medium
range (mr) contributions in $\m^{-n}$ units; experimental results are 
taken from ref. [\ref{reftab}].\label{tab1}}
\begin{tabular} {|c|c|c|c|c|c|c|}
\hline
&$d_{00}^+$&$d_{10}^+$&$d_{01}^+$&$d_{20}^+$&$d_{11}^+$&$d_{02}^+$ \\\hline
exp   &-1.46$\pm$0.10 & 1.12$\pm$0.02  &  1.14$\pm
$0.02 &0.200$\pm$0.005  & 0.17$\pm$0.01 &  0.036$\pm$0.003  \\ \hline
mr              & 0.12 & -  &  -0.25 &  - &  - &  0.032                 \\
\hline\hline
&$b_{ 00}^+$& & & & & \\ \hline exp   & -3.54$\pm$0.06 & & & & &\\
\hline\hline
&$d_{ 00}^-$&$d_{10}^-$&$d_{01}^-$& & & \\ \hline exp & 1.53$\pm
$0.02 & -0.167$\pm$0.005 & -0.134$\pm$0.005 & & & \\ \hline
$W\!T+mr$               & 1.18  & - & -0.032 & & &  \\ \hline\hline
&$b_{ 00}^-$&$b_{10}^-$&$b_{01}^-$ & & & \\ \hline exp & 10.36$
\pm$0.10 & 1.08 $\pm$0.05 & 0.24$\pm$0.01 & & & \\ \hline
$W\!T+mr$       & -0.99 & - & 0.18 & & & \\ \hline
\end{tabular}
\end{center}
\end{table}

The values of the subthreshold coefficients are determined from $\p N$ 
scattering data and, in a chiral expansion to $\cO(q^3)$, they are used to 
fix the otherwise undetermined parameters $c_i$ and $\tilde{d}_i$. In our 
formulation of the TPEP, we bypass the use of these unknown parameters, for 
the redefined subthreshold coefficients are already the dynamical ingredients 
that determine the strength of the various interactions. This allows the 
potential to be expressed directly in terms of observable quantities. 

In table \ref{tab1} we show the experimental values of the subthreshold 
coefficients determined in ref. \cite{H83} and the sum of $(W\!T)$ and $(mr)$ 
contributions. The redefined values are obtained by just subtracting the 
latter from the former\footnote{We use $g_A=1.25$, $f_\p=93$ MeV, 
$\m=139.57$ MeV and $m=938.28$ MeV.}. It is worth noting that the values 
of $\db_{02}^+$ and $\bb_{01}^-$ are compatible with zero. 

When writing the results for the TPEP, it is very convenient to display 
explicitly the chiral scales of the various contributions. With this 
purpose in mind, we will employ the dimensionless subthreshold constants 
defined in table \ref{tab2}.

\begin{table}[h]
\begin{center}
\caption{Dimensionless subthreshold coefficients.\label{tab2}}
\begin{tabular} {|c|c|c|c||c|}
\hline
&$\deb_{00}^+$&$ \d_{10}^+$&$\deb_{01}^+$ & $ \b_{00}^+$    \\\hline
definition      & $m f_\p^2 \;d_{00}^+ /\m^2$ & $m f_\p^2 \;d_{10}^+$  &
$m f_\p^2 \;d_{01}^+ $ & $m f_\p^2 \;b_{00}^+$ \\ \hline
value   & -4.72  & 3.34 &  4.15 & -10.57            \\ \hline\hline
&$\deb_{ 00}^-$&$\d_{10}^-$&$\deb_{01}^-$& $\beb_{00}^-$   \\ \hline
definition & $m^2 f_\p^2 \;\db_{00}^- /\m^2$ & $m^2 f_\p^2 \;d_{10}^-$ &
$m^2 f_\p^2 \;\db_{01}^-$ &  $f_\p^2 \;\bb_{00}^-$ \\ \hline
value   & 7.02 & -3.35 &-2.05 & 5.04   \\ \hline
\end{tabular}
\end{center}
\end{table}

\section{relativistic and heavy baryon formulations}\label{sec:rel-hb}

In this section we review briefly the relativistic formulation of
baryon ChPT and its relationship with the widely used heavy
baryon techniques.
Chiral perturbation theory is a systematic expansion of 
low-energy amplitudes in 
powers of momenta and quark masses, generically denoted by $q$. The 
chiral Lagrangian consists of a string of terms, labeled by its 
power in $q$. To a given order, one builds the most general Lagrangian, 
consistent with Poincar\'e invariance and other 
symmetries of QCD (parity, time reversal, and approximate chiral 
symmetry). 
A Lagrangian of order $n$ produces tree graphs of the same order, 
while loop graphs are expected to contribute at higher orders, following 
a power counting scheme. This is indeed what happens in the mesonic 
sector, where loop graphs are two orders higher than tree graphs, if 
one uses dimensional regularization. 

In relativistic baryon ChPT, dimensional regularization no longer leads 
to a well defined power counting~\cite{GSS}, loops start 
at the same order as tree graphs and the 
connection between loop and momentum expansion is lost. 
A similar phenomenon is observed in the mesonic sector if one uses 
another regularization scheme, such as Pauli-Villars. 

In HBChPT, this problem is overcome by means of the 
expansion of the original Lagrangian around the infinite 
nucleon mass limit \cite{JenkMan}. One integrates out the heavy 
degrees of freedom of the nucleon field, eliminates its 
mass $m$ from the propagator, and expands the resulting vertices 
in powers of $1/m$. This formulation 
gives rise to a power counting scheme, but Lorentz invariance is no longer 
explicit. It can still be recovered, but only after a 
resummation of all terms in this expansion. 

The HB approach also has a more serious 
problem, pointed out recently by Becher and Leutwyler~\cite{BL1}, 
namely, that it fails to converge in part of the low-energy 
region. In order to avoid this, they proposed a new regularization 
scheme, the so 
called Infrared Regularization, which is manifestly Lorentz invariant 
and gives rise to a power counting. The method is based on a previous 
work by Ellis and Tang~\cite{EllisT}, where a loop integral $H$ was 
separated into ``soft", infrared ($I$) and ``hard", regular ($R$) 
pieces. The former satisfies a power counting rule and has the same 
analytic structure as $H$ in the low-energy domain. The latter may 
contain singularities only at high energies --- in the low-energy 
region, it is well behaved and can be expanded in a 
Taylor series, resulting in polynomials of the generic momentum $q$. 
Therefore the hard pieces, which are the power counting violating 
terms, can be absorbed in the appropriate coupling constants of 
the Lagrangian and one considers only $I$, the infrared-regularized 
part of $H$\footnote{This problem has been recently reviewed by 
Mei\ss ner in sects. 3.4-3.7 of ref. \cite{meissrev}.}. 

Ellis and Tang have shown that the chiral expansion of the infrared 
regularized one-loop integral $I$, with the ratio $q/\mu$ fixed, 
reproduces formally the corresponding terms in the HBChPT 
approach~\cite{EllisT}, 
even in the cases where such an expansion is not permitted. This allows 
one to assess the domain of validity of the HB series.

For the sake of completeness, in the sequence, we reproduce some of 
the results derived by Becher and Leutwyler.
They have analyzed in detail the triangle graph of fig.\ref{fig4}, 
which contributes to the nucleon scalar form factor,
and shown that the HBChPT formulation is not suited for the 
low energy-region, near $t=4\mu^2$.
Its exact spectral representation is given by~\cite{GSS} 

\begin{equation}
\gamma(t)=\frac{1}{\pi}\int_{4\mu^2}^{\infty}\frac{dt'}{(t'-t)}
\,\mbox{Im}\gamma(t')\;,\label{eq:new00}
\end{equation}

\ni
where 

\begin{eqnarray}
\mbox{Im}\gamma(t') &=& \frac{\theta(t'-4\mu^2)}{8\pi m\sqrt{t'(4m^2-t')}}
\,\tan^{-1}\frac{\sqrt{(4m^2-t')(t'-4\mu^2)}}{t'-2\mu^2}
\nn\\[2mm]
&\simeq& \frac{\theta(t'-4\mu^2)}{16\pi m\sqrt{t'}}
\,\tan^{-1}\frac{2m\sqrt{t'-4\mu^2}}{t'-2\mu^2}\,.
\label{eq:new01}
\end{eqnarray}

Formally, the argument 

\begin{equation}
x=\frac{2m\sqrt{t'-4\mu^2}}{t'-2\mu^2}\label{eq:new02}
\end{equation}

\noindent
seems to be of order $q^{-1}$, and the HB chiral expansion of
(\ref{eq:new01}) would yield $\tan^{-1}x= \pi/2 - 1/x + 1/3x^3+\cdots $.
However, this representation of $\tan^{-1}x$ is valid only in the domain 
$|x|\geq 1$. For $|x|<1$, one should use $\tan^{-1}x=x-x^3/3+\cdots$, but 
this corresponds to an expansion in {\em inverse} powers of $q$. 
From (\ref{eq:new02}) we see that the HB expansion of (\ref{eq:new00}) 
breaks down when $t'$ approaches $4\mu^2$. 

Becher and Leutwyler have shown that it is possible to write accurately 

\begin{eqnarray}
\gamma(t)-\gamma(0) &=& \frac{t}{\pi}\int_{4\mu^2}^{\infty}
\frac{dt'}{t'(t'-t)}\,\frac{1}{16\pi m\sqrt{t'}}
\lc \lb\frac{\p}{2} - \frac{(t'-2\m^2)}{2 m \sqrt{t'\!-\!4\m^2}}\rb_{\!HB}
\right.
\nn\\[2mm]
&+&\left.\lb\frac{\m\sqrt{t'}}{2m\sqrt{t'\!-\!4\mu^2}}-\frac{\sqrt{t'}}{2\m}
\tan^{-1}\frac{\mu^2}{m\sqrt{t'\!-\!4\m^2}}\rb_{\!t\!h}\rc\;.
\label{eq:new03}
\end{eqnarray}

By keeping only the first bracket in the integrand, one recovers the heavy 
baryon result.
However, the region $t\sim 4\mu^2$ is dominated by the lower end of 
integration in $t'$, where the second term becomes important.
The HB approximation is not valid there.
The integration can be performed analytically and Becher and Leutwyler found

\begin{eqnarray}
\gamma(t)-\gamma(0)&=&\frac{1}{32\pi m\m}\lc \lb \frac{1}{\sqrt{\tau}}
\ln \frac{2+\sqrt{\tau}}{2-\sqrt{\tau}}-1
+\frac{2\m (2-\tau)}{\pi m \sqrt{\tau(4-\tau)}}\sin^{-1}\frac{\sqrt{\tau}}{2}
-\frac{\mu}{\pi m}
\rb_{\!HB}\right.
\nn\\[2mm]
&+& \left.\lb \frac{\m}{m \sqrt{4-\tau}}-\frac{\mu}{2m}
-\ln\lp 1+\frac{\m}{m\sqrt{4-\tau}}\rp+\ln\lp 1+\frac{\m}{2m}\rp
\rb_{\!t\!h}\rc
\label{eq:new003}
\end{eqnarray}

\begin{figure}[ht]\begin{center}
\epsfig{figure=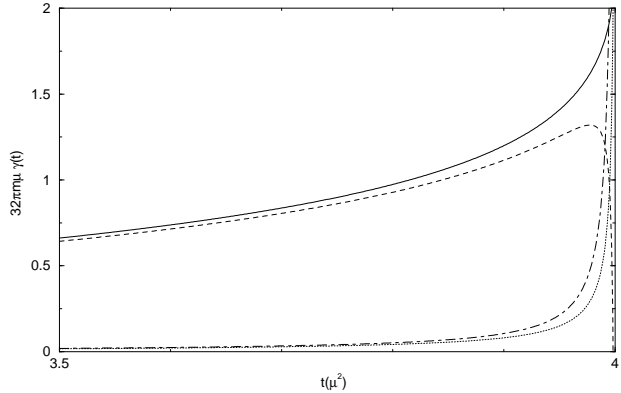, height=5.3cm} \end{center}
\caption
{Behavior of the function $\gamma(t)$ as given by eq.(\ref{eq:new003}) 
(full line) and partial contributions: HB (dashed line), $th$ 
(dotted line) and eq.(\ref{new6.11}) (dot-dashed line).}
\label{fig5a}
\end{figure}

\ni
with $\tau=t/\m^2$.
This result is interesting because it shows clearly that, for values 
of $t$ far from $4\m^2$, the contributions of the two brackets decouple 
and can be expanded in powers of $q$. The second term is then $\cO(q^2)$. 
On the other hand, when $t\sim 4\m^2$, both contributions merge, the full 
result for $\gamma(t)$ is the outcome of large cancellations between them, 
and an expansion in $q$ does not apply. In fig.\ref{fig5a}, we display the 
behavior of the various terms in eq.(\ref{eq:new003}) in the range 
$3.5\m^2 \leq t \leq 4\m^2$, where the second bracket is important. 
In this figure we also show the effect of making

\beq
\lb \frac{\m}{m\sqrt{4-\tau}}-\frac{\mu}{2m}
-\ln\lp 1+\frac{\m}{m\sqrt{4-\tau}}\rp+\ln\lp 1+\frac{\m}{2m}\rp
\rb_{\!t\!h}\rightarrow \frac{\m^2}{m^2(4 -\tau)}-\frac{\mu^2}{4m^2}\;.
\label{new6.11}
\eeq

\ni
This rough approximation is not mathematically precise, but it allows 
one to guess the order of magnitude of the threshold contribution.

The discussion of the behavior of the triangle diagram in the 
neighborhood of $t=4\m^2$ is relevant to the $NN$ potential because, in 
configuration space, this region describes its long distance properties, 
as observed numerically in our previous works 
\cite{RR97,R01}. To see this, let us take the representation 
of (\ref{eq:new00}) in configuration space: 

\begin{equation}
\Gamma(r)= \frac{1}{\pi}\int_{4\mu^2}^{\infty}dt'\int\frac{d^3q}{(2\pi)^3}\;
e^{-i{\mbox{\boldmath $q$}}\cdot{\mbox{\boldmath $r$}}}\;
\frac{\mbox{Im}\gamma(t')}{t'+{\mbox{\boldmath $q$}}^2}=
\frac{1}{4\pi^2}\int_{4\mu^2}^{\infty}dt'\;\frac{e^{-r\sqrt{t'}}}{r}\;
\,\mbox{Im}\gamma(t')\,.
\end{equation}

The exponential in the integrand shows clearly that, for large values 
of $r$, results are dominated by the lower end of the integration.
Thus, if we want to have a good
description of $\Gamma(r)$ at large distances, we need a
decent representation for $\mbox{Im}\gamma(t')$ near $t'=4\mu^2$,
which is not provided by HBChPT.

\section{dynamics}
\label{sec:dynamics}

\begin{figure}[ht]\begin{center}
\epsfig{figure=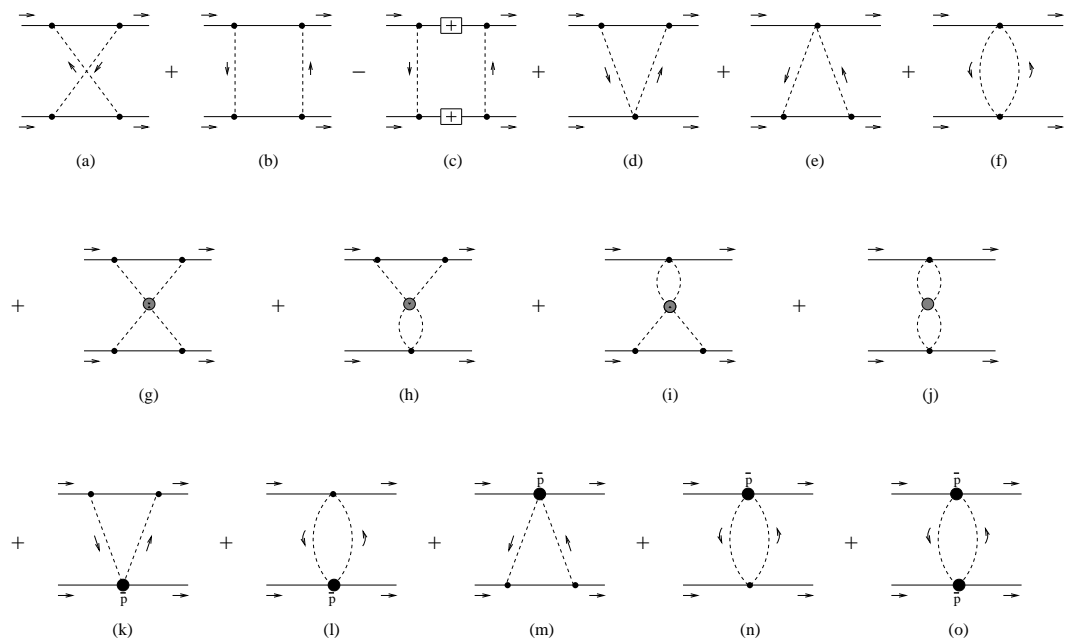, width=14cm} \end{center}
\caption
{Dynamical structure of the TPEP. The first two diagrams correspond to the 
products of Born $\p N$ amplitudes, the third one represents the iteration 
of the OPEP, whereas the next three involve contact interactions associated 
with the Weinberg-Tomozawa vertex. The diagrams on the second
line describe medium range effects associated with scalar and vector form
factors. The remaining interactions are {\em triangles} and {\em bubbles}
involving subthreshold coefficients.}
\label{fig5}
\end{figure}

The chiral two-pion exchange potential is determined by the processes
depicted in fig.\ref{fig5}, derived from the basic $\p N$ subamplitude
and organized into three different families. The first one corresponds to
the minimal realization of chiral symmetry \cite{RR94}, includes the
subtraction of the iterated OPEP, and involves only pion-nucleon
interactions with a single loop, associated with the constants $m$, $g$, 
and $f_\p$. The same constants also determine the two-loop processes of the
second family. The last family includes chiral corrections associated with
subthreshold coefficients and LECs, representing either higher order 
processes or other degrees of freedom.

The first two diagrams of fig.\ref{fig5}, known, respectively, as
{\em crossed box} and {\em box}, come from the products of the $\p N$ PV
Born amplitudes, given by eqs.(\ref{3.2})--(\ref{3.5}) and involve the
propagations of two pions and two nucleons. The third one represents the
iteration of the OPEP and gives rise to an amplitude denoted by $ \cT_{it}$,
derived after the work of Partovi and Lomon 
\cite{PL} and discussed in detail in appendix \ref{ap:iterat}. The 
remaining interactions correspond to {\em triangle} and {\em bubble} 
diagrams, which contain a single or no nucleon propagators, besides 
those of two pions.

The construction of the TPEP begins with the determination of the
relativistic profile functions, eqs.(\ref{2.7})--(\ref{2.10}), using the
$\p N$ subamplitudes $D^\pm$ and $B^\pm$ discussed in section
\ref{sec:inter-piN}. Results are then expressed in terms of the one-loop 
Feynman integrals presented in appendixes \ref{ap:integ} and \ref{ap:iterat},
which may involve two, three, or four propagators. The evaluation and
manipulation of these integrals represent an important aspect of the present
work and it is worth discussing the notation employed.

Momentum space integrals are generally denoted by $\P$ and labeled in
such a way as to recall their dynamical origins. We use lower labels,
corresponding to nucleons 1 and 2, with the following meanings: $c\rar$
contact interaction; $s\rar$ $s$-channel nucleon propagation; and $u\rar$
$u$-channel nucleon propagation. This means that functions carrying the
subscripts $(cc)$, $(sc)$, $(ss)$, and $(us)$ correspond, respectively, to
{\em bubble, triangle, crossed box}, and {\em box} diagrams. 
The last class of integrals includes the OPEP cut, which needs to be 
subtracted. This subtraction is implemented by replacing the $(us)$ 
integrals by regular ones, represented by the subscript $(reg)$ and 
given in appendix \ref{ap:iterat}.
Upper labels, on the other hand, indicate the rank of the integral in the
external kinematical variables $q, z$ and $W$. For instance, the rank 2
{\em crossed box} integral  is written as

\bea
I_{ss}^{\m\n}
&=& \int \frac{d^4 Q}{(2\p)^2}\;\lp \frac{Q^\m Q^\n}{\m^2} \rp
\frac{1}{k^2-\m^2}\;\frac{1}{k'^2-\m^2}\;
\frac{2m\m}{s_1-m^2} \;\frac{2m\m}{s_2-m^2}
\nn\\[2mm]
&=& \frac{i}{(4\p)^2}\lb \frac{q^\m q^\n}{\m^2}\; \P_{ss}^{(200)} +
\frac{z^\m z^\n}{4 m^2}\; \P_{ss}^{(020)} + \frac{W^\m W^\n}{4 m^2}\;
\P_{ss}^{(002)} + g^{\m\n}\;\Pb_{ss}^{(000)}\rb \;.
\nn
\eea

All integrals are dimensionless and include suitable powers of pion
and nucleon masses, so as to make them relatively stable upon wide
variations of the latter. We have studied these integrals numerically and,
typically, they change by 30\% when one moves the nucleon mass from its
empirical value to infinity. The fact that the integrals are 
${\cal O}(q^0)$ is rather useful in discussing
chiral scales and heavy baryon limits. At present the infrared 
regularization techniques are still being developed for the case of 
two nucleon system \cite{IRreg} and we have used dimensional regularization 
whenever appropriate.
As a consequence, our results are accurate only for distances larger 
than a typical radius. Our numerical studies in configuration space 
indicate that this radius is of about 1 fm.

The covariantly expanded TPEP, to be given in section X, is expressed in 
terms of the functions $\P_{cc}^{(000)}$, $\P_{sc}^{(000)}$, 
$\P_{ss}^{(000)}$, $\P_{reg}^{(000)}$, and $\P_{reg}^{(010)}$. 
In order to simplify the notation, in the main text we call them 
$\P_\ell$, $\P_t$, $\P_\times$, $\P_b$, and $\Pt_b$, respectively.

The function $\P_\ell$ represents the {\em bubble} diagram and is given by

\beq
I_{cc} =  \int \frac{d^4Q}{(2\pi)^4}\;
\frac{1}{[k ^2\! -\!\m^2] [k'^2 \!-\!\m^2]}
= \frac{i}{(4\p)^2} \;\P_\ell \;.
\label{6.1}
\eeq

\ni
This integral can be performed analitically\footnote{The function $\P_\ell$ 
is related to the $L(q)$ used in ref. \cite{KBW} by $\P_\ell = - 2  L(q)$ 
and to the $J(t)$ of \cite{BL2} by $\Pi_\ell=(4\pi)^2\,J-1$.} 
and its regular part may be written as

\beq
\P_\ell   =  - 2\;\frac{\sqrt{1-t/4\m^2}}{\sqrt{-t/4\m^2}}\;
\ln\lp \sqrt{1-t/4\m^2}+\sqrt{-t/4\m^2}\rp \;.
\label{6.2}
\eeq

The function $\P_t$, associated with the {\em triangle} diagram, is 
expressed by

\beq
I_{sc} = \int \frac{d^4Q}{(2\pi)^4}\;
\frac{1}{[k^2\! -\!\m^2] [k'^2 \!-\!\m^2]}
\; \frac{2m\m}{ [s_1-m^2] }
= \frac{i}{(4\p)^2} \; \P_t
\label{6.3}
\eeq

\ni
and related to the function $\gamma(t)$ discussed in the preceding section 
by $\P_t = -2m\m (4\p)^2 \g(t) $. 
The heavy-baryon representation of this function is 

\beq
\P_t \rightarrow \P_t^{\!H\!B}= \P_a + \frac{\m}{2m}\;\P_t^{\NL} 
\label{6.8}
\eeq

\ni
with\footnote{
The function $\P_a$ is related to the $A(q)$ of ref. \cite{KBW}
by $\P_a= - 4 \p \m A(q)$.}

\bea
\P_a &=& - \; \frac{\p}{\sqrt{-t/4\m^2}}\tan^{-1}\sqrt{ -t /4\m^2} \;,
\label{6.9}\\[2mm]
\P_t^{\NL} &=& (1\!-\!t / 2\m^2) \; \P'_\ell \;,
\label{6.10}
\eea

\ni
and
$ \P' = \m\; (d \P / d\m)$.

The functions $\P_\times$, $\P_b$, and $\Pt_b$ are associated with
{\em crossed box} and {\em box} diagrams and their complete expressions 
are given in appendix \ref{ap:integ}. Their heavy baryon expansions 
are derived in appendix \ref{ap:relex} and read 

\bea
&& \P_\times^{\!H\!B} = -\P'_\ell - \lb \frac{\m}{m}\rb \frac{\p/2}{(1\!-\!t/4\m^2)}
- \lb \frac{\m}{m} \rb^2 \frac{1}{4}\;\lb (1\!-\!t/2\m^2)^2 \lp 2\;
\P'_\ell -\P''_\ell  \rp + (2 z^2/3\m^2)\;\P'_\ell \rb +\cdots\;,
\label{6.14}\\[2mm]
&& \P_b^{\!H\!B} = \Pi_\times^{\!H\!B} + \lb \frac{\m}{m}\rb \frac{\p/4}{(1\!-\!t/4\m^2)}
+ \lb \frac{\m}{m}\rb^2 \frac{1}{6}\;\lb (1\!-\!t/2\m^2)^2 \lp 2\;
\P'_\ell -\P''_\ell \rp \rb +\cdots\;,
\label{6.15}\\[2mm]
&& \Pt_b^{\!H\!B} = -\;\frac{1}{2}\;\P_a - \lb \frac{\m}{m}\rb \frac{1}{3}\;
(1\!-\!t / 2\m^2) \; \P'_\ell + \cdots \;.
\label{6.16}
\eea

In the heavy baryon expansion of the potential, the following 
results are useful

\bea
&& \P'_\ell = 2+ \P_\ell /(1\!-\!t/ 4\m^2)\;,
\label{6.17}\\[2mm]
&& \P''_\ell = 2/(1\!-\!t/ 4\m^2)+\lb 2/(1\!-\!t/ 4\m^2) -
1/(1\!-\!t/ 4\m^2)^2 \rb \P_\ell \;,
\label{6.18}\\[2mm]
&& \Pi_a'=\Pi_a-\pi\,t/(1-t/4\mu^2)
\;.
\label{6.18a}
\eea

For the reasons discussed in the preceding section, all these 
heavy baryon representations are inaccurate around $t\sim 4\m^2$.

\section{covariant amplitudes}
\label{sec:covamp}

The direct reading of the Feynman diagrams of fig.\ref{fig5} gives rise to our
{\bf full results} for the relativistic profile functions, displayed in
appendix \ref{ap:fullres}. These are the functions that the chiral 
expansion must converge to and hence they allow one to assess the series 
directly. On the other hand, they do not
exhibit explicitly the chiral scales of the various components of the
potential, since their net values are the outcome of several cancellations.

In order to display these scales, in appendix \ref{ap:relat} we derive
several relations among integrals, which are used to transform the
full results of appendix \ref{ap:fullres} into the forms listed 
in appendix
\ref{ap:interes}. The relations given in appendix \ref{ap:relat} are, in 
principle, exact,
provided one keeps {\em short range} integrals that contain a single or no 
pion propagators. 
However, for the sake of simplicity, we neglect those 
contributions\footnote{It would be very easy to keep those 
terms, but this would produce longer equations.}. 
The importance of this
approximation was checked by comparing numerically the Fourier transforms
of the various amplitudes of appendixes \ref{ap:fullres} and
\ref{ap:interes}. In all cases, agreement is much better than 1\% for
distances larger than 1 fm, except for $\cI_{DD}^+ $, where the difference
is 4\% at 1.5 fm and falls below 1\% beyond 2.5 fm. This has very little
influence over the full potential.

With the purpose of allowing comparison with results produced in the HB tradition,
we write our final expressions for the potential in terms of the axial 
constant $g_A$, which is related to the $\p N$ coupling constant by
$ g = \lp 1 + \D_{GT} \rp \; g_A m / f_\p $.
Here $\D_{GT}$ is the Goldberger-Treiman (GT) discrepancy\footnote{
The GT discrepancy may be
written \cite{BL2} as $\D_{GT}=-2d_{18}\m^2/g+\cO(q^4)$.},
proportional to $\m^2$.
In applications, on the other hand, we recommend the direct use of the 
$\p N$ coupling constant $g$, by making $g_A=g f_\pi/m$ and neglecting
$\D_{GT}$ in our results.

The appropriate truncation of the expressions of appendix \ref{ap:interes}, at the 
orders in $q$ prescribed at the end of section \ref{sec:pw-count}, leads
to the following results for the profile functions:

\bea
&\bullet& \cI_{DD}^+ = \frac{m^2}{16\; \p^2 f_\p^4}\;\lb \frac{\m}{m}\rb^2
\lc \frac{g_A^4}{16} \; (1\!-\!t\!/2\m^2)^2 \lp \P_\times - \P_b \rp
+\Delta_{GT}\,\frac{g_A^4}{4} \; (1\!-\!t\!/2\m^2)^2 \lp\P_\times-\P_b\rp
\right.
\nn\\[2mm]
&& \left.
+\lb \frac{\m}{m} \rb \frac{g_A^2}{8}\; (1\!-\!t\!/2\m^2) \lb -g_A^2\; \P_a
+ 4 \lp \deb_{00}^+ + \deb_{01}^+\; t/ \m^2\rp  \P_t  \rb
+ \lb \frac{\m}{m}\rb^2 \lb - \frac{g_A^2}{4}\; \d_{10}^+ \;
(1\!-\!t\!/2\m^2)^2  \right.\right.
\nn\\[2mm]
&& \left.\left. + \frac{1}{2} \lp \deb_{00}^+ + \deb_{01}^+\; t/ \m^2 +
\frac{1}{3}\;\d_{10}^+ \; (1\!-\!t\!/4\m^2) \rp^2 + \frac{2}{45} \lp \d_{10}^+
\rp^2 (1\!-\!t\!/4\m^2)^2  \rb \P_\ell \right.
\nn\\[2mm]
&& \left. - \lb \frac{\m}{m}\rb^2 \;\frac{m^2}{256\; \p^2 f_\p^2}\;g_A^4\;
(1\!-\!2 t\!/\m^2) \lb (1\!-\!t\!/2 \m^2) \; \P_t -2\p \rb^2 \rc \;,
\label{7.1}\\[4mm]
&\bullet& \cI_{DB}^{(w)+} = \frac{m^2}{16\; \p^2 f_\p^4}\;  \lb
\frac{\m}{m}\rb \lc - \frac{g_A^4}{8}\; (1\!-\! t \! / 2\m^2) \;  \P_t
\right.\nn\\[2mm]&& \left.
+ \lb \frac{\m}{m}\rb  \lb \frac{g_A^4}{16} \;(1\!-\! t \! / 2\m^2)^2 \;
\P_\times
- \frac{g_A^2}{2}\lp \deb_{00}^+ + \deb_{01}^+\; t/\m^2 +
\frac{1}{3}\; \d_{10}^+ \; (1\!-\! t \! / 4\m^2)  \rp \P_\ell \rb \rc  \;,
\label{7.2}\\[4mm]
&\bullet& \cI_{DB}^{(z)+} = \frac{m^2}{16\; \p^2 f_\p^4}\;\lb \frac{\m}{m}\rb
\lc \frac{g_A^4}{8} \lb (1\!-\! t \! / 2\m^2) \; \Pt_b-
(3/2 \!-\! 5 t\! / 8\m^2)\; \P_a \rb \right.
\nn\\[2mm]
&& \left. + \lb \frac{\m}{m}\rb \frac{g_A^2}{2} \lp \deb_{00}^+ +
\deb_{01}^+\; t / \m^2 - \d_{10}^+ \;  (1\!-\! t \! / 4\m^2)
\rp \P_\ell \rc \; ,
\label{7.3}\\[4mm]
&\bullet& \cI_{BB}^{(g)+} = \frac{m^2}{16\; \p^2 f_\p^4} \lc
\frac{g_A^4}{4}\;  (1\!-\!t/4\m^2) \lp \P_\times+ \P_b\rp
+ \Delta_{GT}\; g_A^4\;  (1\!-\!t/4\m^2) \lp \P_\times+ \P_b\rp
\right.\nn\\[2mm]&& \left.
+ \lb \frac{\m}{m}\rb \frac{g_A^4}{8} \lb (1\!-\! t \! / 2\m^2)  \lp \P_t -
\Pt_b \rp
+ (1\!-\!t/4\m^2)\;\P_a \rb
\right.\nn\\[2mm]&& \left.
- \lb \frac{\m}{m}\rb^2  g_A^2
\lb \frac{g_A^2}{16} \;(1\!-\! t \! / 2\m^2)^2\;\P_\times
+ \frac{1}{3}\; \b_{00}^+ \; (1\!-\! t \! / 4\m^2) \;\P_\ell \rb \rc \;,
\label{7.4}\\[4mm]
&\bullet& \cI_{BB}^{(w)+}=\frac{m^2}{16\;\p^2 f_\p^4}\;g_A^4\;\P_\ell\;,
\label{7.5}\\[4mm]
&\bullet& \cI_{BB}^{(z)+} = \frac{m^2}{16 \; \p^2 f_\p^4}\;
\frac{g_A^4}{3}\; \P_\ell \;,
\label{7.6}
\eea

\ni
and

\bea
&\bullet& \cI_{DD}^- = \frac{m^2}{16 \;\p^2 f_\p^4} \lb \frac{\m}{m}\rb^2
\lc \frac{g_A^4}{16}\; (1\!-\!t/2\m^2)^2 \lp \P_\times+\P_b \rp
-\frac{g_A^2}{4}\; (g_A^2-1)\; (1\!-\!t/2\m^2)\; \P_\ell \right.
\nn\\[2mm]
&& \left. + \frac{1}{24} \; (g_A^2-1)^2 (1\!-\!t/4\m^2)\; \P_\ell
+ \lb \frac{\m}{m}\rb  \frac{g_A^2}{8}\; (1\!-\!t/2\m^2) \lb g_A^2\;\P_a
+ (g_A^2\!-\!1)\; (1\!-\!t/2\m^2) \;\P_t  \rb  \right.
\nn\\[2mm]
&& \left. + \lb \frac{\m}{m} \rb^2 \lb \frac{g_A^2}{2}\; (1\!-\!t/2\m^2)
\lp (g_A^2\!-\!1)\;z^2/4\m^2 + \deb_{00}^- + \deb_{01}^-\;t/\m^2 +
\frac{1}{3}\;\d_{10}^- \;(1\!-\!t/4\m^2) \rp\right.\right.
\nn\\[2mm]
&& \left.\left.
-\frac{(g_A^2\!-\!1)}{6}\; (1\!-\!t/4\m^2) \lp (g_A^2\!-\!1)\;(t/16\m^2\!
+\! z^2/8\m^2) + \deb_{00}^- + \deb_{01}^-\;t/\m^2
+\frac{3}{5}\;\d_{10}^-\; (1\!-\!t/4\m^2) \rp \rb \P_\ell\right.
\nn\\[2mm]
&& \left.-\lb \frac{\m}{m}\rb^2 \frac{m^2}{64 \;\p^2 f_\p^2}\lb -g_A^2 \lp
(1\!-\!t/2\m^2)\; \P_\ell + 1 - t/3\m^2\rp \right.\right.
\nn\\[2mm]
&& \left.\left.+\frac{1}{3}\;(g_A^2-1) \lp (1\!-\!t/4\m^2)\;\P_\ell + 2-
t/4\m^2 \rp \rb^2  \right.
\nn\\[2mm]
&& \left. + \lb \frac{\m}{m}\rb^2 \frac{m^2}{64 \;\p^2 f_\p^2}
(z^2/ 4\m^2)\; g_A^4\lb (1\!-\!t/4\m^2) \P_t -\pi \rb^2 \right.
\nn\\[2mm]
&& \left.
+ \Delta_{G\!T}  \lb \lp \frac{1}{6}(1\!-\!t/4\m^2)\;g_A^2(g_A^2-1) 
-(1-t/2\mu^2)\,g_A^2\,(g_A^2-1/2)
\rp \P_\ell
\right.\right. \nn\\[2mm]&& \left.\left.
+ \frac{g_A^4}{4} \; (1\!-\!t/2\m^2)^2\; \lp
\P_\times+\P_b \rp\rb \rc \;,
\label{7.7}\\[4mm]
&\bullet& \cI_{DB}^{(w)-} = \frac{m^2}{16\p^2f_\p^4}\lb\frac{\m}{m}\rb\lc
\lb - \frac{g_A^4}{8}\;  (1\!-\! t / 2 \m^2) \;\P_t \rb \right.
\nn\\[2mm]
&& \left.+ \lb \frac{\m}{m} \rb \lb  \frac{1}{24}\;(g_A^2-1)(g_A^2-1-2
\beb_{00}^-) (1\!-\! t  / 4\m^2)-\frac{g_A^2}{4}\; (g_A^2-1- \beb_{00}^-)\;
(1\!-\! t  / 2\m^2) \rb \P_\ell \right.
\nn\\[2mm]
&& \left.+ \lb\frac{\m}{m}\rb \lb \frac{g_A^4}{16}\;  (1\!-\! t/2\m^2)^2
\P_\times \rb \rc \;,
\label{7.8}\\[4mm]
&\bullet& \cI_{DB}^{(z)-}=\frac{m^2}{16\p^2f_\p^4}\lb\frac{\m}{m}\rb\lc
\lb\frac{g_A^2}{4}\;(g_A^2-1)\;(1\!-\!t/4\m^2)\;\P_t-
\frac{g_A^4}{8}\;\lp(1\!-\!t/2\m^2)\;\tilde{\Pi}_b-\right.\right.\right.
\nn\\[2mm]
&& \left.\left.\left. - (3/2\!-\!5 t / 8\m^2)\;\P_a\rp \rb+\lb \frac{\m}{m}
\rb \lb  \frac{1}{24}\; (g_A^2-1) (g_A^2-1-2\beb_{00}^-)\;
(1\!-\! t  / 4\m^2) \right.\right.
\nn\\[2mm]
&& \left.\left.+\frac{g_A^2}{8}\;  (g_A^2-1+2\,\beb_{00}^-)\;
(1\!-\! t/ 2\m^2) \rb \P_\ell
-\lb \frac{\m}{m}\rb \frac{m^2}{64\p^2f_\p^2}\;
g_A^4 \lb(1\!-\! t \! / 4\m^2) \P_t-\pi\rb^2 \rc \;,
\label{7.9}\\[4mm]
&\bullet&  \cI_{BB}^{(g)-} =\frac{m^2}{16\p^2f_\p^4}\lb\frac{\m}{m}\rb \lc
\frac{g_A^2}{4}\;(g_A^2-1-2\beb_{00}^-)\; (1\!-\! t  / 4\m^2) \;\P_t
-\frac{g_A^4}{8} \lb (1\!-\!t/2\m^2)\; \tilde{\Pi}_b \right. \right.
\nn\\[2mm]
&& \left. \left. + (1\!-\!t/4\m^2)\;\P_a \rb +\lb \frac{\m}{m} \rb \lb
\frac{1}{24}\; (g_A^2-1-2\beb_{00}^-)^2 \; (1\!-\! t  /4\m^2)\right.\right.
\nn\\[2mm]
&& \left.\left. +\frac{g_A^2}{8}\;(g_A^2-1-2\beb_{00}^-)\;(1\!-\!t/2\m^2)
\rb \;\P_\ell
-\lb \frac{\m}{m}\rb \frac{m^2}{64\p^2f_\p^2}\;
g_A^4 \lb(1\!-\! t \! / 4\m^2) \P_t-\pi\rb^2 \rc \;,
\label{7.10}\\[4mm]
&\bullet& \cI_{BB}^{(w)-} \simeq \cI_{BB}^{(z)-} \simeq 0 \;.
\label{7.11}
\eea

The results for the basic subamplitudes presented in this section are
closely related to the underlying $\p N$ dynamics and, in many cases, this
relationship can be directly perceived in the final forms of our expressions.
For instance, reorganizing the contributions proportional to $\P_t$ in
eq.(\ref{7.10}), one has

\bea
\cI_{BB}^{(g)-} &=& \frac{m^2/f_\p^4}{(4\p)^2}\lb\frac{\m}{m}\rb
\lc \lb \frac{g_A^2}{4} \lp g_A^2-1-\frac{g_A^2}{f_\p^2}\;
\frac{\m m}{(4\p)^2}\lp(1\!-\! t/4\m^2) \;\P_t -\pi\rp -2\beb_{00}^-\rp
\right.\right.\nonumber\\[2mm]&&\left.\left.
\times\lp(1\!-\! t/ 4\m^2) \;\P_t-\pi\rp+\cdots \rb\rc \;.
\label{7.15}
\eea

The terms within the parentheses represent the contributions
from fig.\ref{fig4}, which read: (a) Born terms,  proportional to $g_A^2$;
(b) Weinberg-Tomozawa term; (c) two-loop medium range interactions;
(d) other degrees of freedom plus two-loop short range interactions.
The organization of the last three terms may be better understood by
noting that, around the point $t=0$, the following expansion holds: 
$(1\!-\! t  / 4\m^2) \;\P_t  \rar -\p+t\,\pi/6\mu^2$, 
and the content of the parentheses of eq.(\ref{7.15}) may be written as

\beq
\lc \frac{g^2}{2m^2} - \lb \frac{1}{2f_\p^2} + \bb_{00}^- + \frac{1}{2}\lp
-\frac{g_A^2 \;m\;\m}{8\;\p\;f_\p^4}+ t\;\frac{g_A^2\;m}{96\;\p\;f_\p^4\;\m}
\rp \rb \rc\;.
\label{7.17}
\eeq

This shows that the structure of eq.(\ref{3.12}) is recovered, except for
the medium range contribution, which is divided by a factor 2,
characteristic of the topology of Feynman diagrams.

\section{TPEP}\label{sec:tpep}

Our final result for the relativistic $\cO(q^4)$ two-pion exchange potential 
is obtained by feeding the truncated covariant profile functions of 
the preceding section into eqs.(\ref{4.1})--(\ref{4.5}).
It is ready to be used as input in other calculations and 
is expressed in terms of five basic functions (section \ref{sec:dynamics}) 
and empirical subthreshold coefficients (section \ref{sec:subt-coef}).
If one wishes, the latter may be traded by LECs, using the results 
of section \ref{sec:subt-coef}. The various components are listed below.

\bea
&&  t_C^+ = \frac{m}{E} \;\frac{3m^2}{256\pi^2f_{\pi}^4}\;\lb
\frac{\m}{m}\rb^2 \lc g_A^4\;(1\!-\!t/2\m^2)^2 \lp \P_\times-\P_b\rp\right.
\nn\\[2mm]
&& \left. + \lb \frac{\m}{m}\rb g_A^2\;(1\!-\!t/2\m^2) \lb - g_A^2\lp2 \P_a
+\Pi_t\,t/\mu^2\rp
+ 8\lp \deb_{00}^+ + \deb_{01}^+\,t / \m^2 \rp \P_t \rb
\right.
\nn\\[2mm]
&& \left. + \lb \frac{\m}{m}\rb^2 \lb
- \frac{m^2\,g_A^4}{16\pi^2f_{\pi}^2}\; (1\!-\!2 t/ \m^2)
\lp(1\!-\!t/2\m^2) \P_t-2\pi \rp^2 +\frac{g_A^4}{4}\,\frac{t}{\mu^2}
(\Pi_\times+\Pi_b)\rb
\right.
\nn\\[2mm]
&& \left. 
+\lb \frac{\m}{m}\rb^2 \lb g_A^4\,\frac{t^2}{\mu^4}
-4 g_A^2  \lp( \deb_{00}^+ + \deb_{01}^+\,t/\mu^2)t/\mu^2+ 
\d_{10}^+(1-2\,t/3\mu^2+t^2/6\mu^4) \rp
\right.\right.\nn\\[2mm]&&\left.\left. 
+8\lp \deb_{00}^+ +\deb_{01}^+\,t/\mu^2+(\d_{10}^+/3)(1-t/4\mu^2)\rp^2
+\frac{32}{45}(\d_{10}^+)^2(1\!-\!t/4\m^2)^2\rb\Pi_\ell
\right.\nn\\[2mm]&&\left.
+4\,\Delta_{GT}\,g_A^4\,(1\!-\!t/2\m^2)^2 \lp \P_\times-\P_b\rp\rc\;,
\label{8.1}\\[5mm]
&&  t_T^+ =  t_{SS}^+/2 = \frac{m}{E}\;
\frac{m^2g_A^2}{256\pi^2f_{\pi}^4}\;
\lc - g_A^2\;(1\!-\!t/4\m^2)\lb \P_\times+\P_b\rb
\right.\nn\\[2mm]&&\left.
 - \lb\frac{\m}{m}\rb \frac{g_A^2}{2}\lb
(1-t/2\mu^2)(\Pi_t-\tilde\Pi_b)+(1-t/4\mu^2)\Pi_a\rb
\right.\nn\\[2mm]&&\left.
+\lb\frac{\m}{m}
\rb^2  \lb \frac{g_A^2}{4}\;(1\!-\!t/2\m^2)^2\;\P_\times 
+\frac{4}{3}\,\beta^+_{00}(1-t/4\mu^2)\Pi_\ell\rb
\right.\nn\\[2mm]&&\left.
-4\,\Delta_{GT}\,g_A^2\,(1\!-\!t/4\m^2)\lb \P_\times+\P_b\rb\rc \;,
\label{8.2}\\[5mm]
&& t_{LS}^+ = \frac{m}{E}\;\frac{3m^2g_A^2}{128\pi^2f_{\pi}^4}\;
\lb \frac{\m}{m }\rb\lc g_A^2\lb(1- t/2 \m^2)(\tilde\Pi_b-\Pi_t)
-(3/2-5\,t/8\mu^2)\,\Pi_a\rb
\right.\nn\\[2mm]&&\left.
+ \lb\frac{\m}{m}\rb \lb \frac{g_A^2}{4}\,(1+2t/\mu^2-t^2/2\mu^4)\;
(\Pi_\times+\Pi_b)
+\lp 2\,g_A^2t/\mu^2-\frac{16}{3}\,\delta^+_{10}\,(1\!-\!t/4\m^2)\rp
\P_\ell\rb\rc\;,
\label{8.3}\\[5mm]
&& t_Q^+ = \frac{m}{E}\frac{m^2g_A^4}{256\pi^2f_{\pi}^4}\;
\P_\ell \;.
\label{8.4}
\eea

\ni
and

\bea
&&  t_C^- =\frac{m}{E}\frac{\mu^2}{8\pi^2f_{\pi}^4}\left\{
\frac{g_A^4}{16}\left(1-\frac{t}{2\mu^2}\right)^2\left(\Pi_\times
+\Pi_b\right)-\frac{g_A^2}{4}\left(g_A^2-1\right)
\left(1-\frac{t}{2\mu^2}\right)\Pi_\ell
\right.\nonumber\\[2mm]&&\left.
+\frac{1}{24}\left(g_A^2-1\right)^2\left(1-\frac{t}{4\mu^2}\right)\Pi_\ell
\right.\nonumber\\[2mm]&&\left.
+\left[\frac{\mu}{m}\right]\frac{g_A^2}{8}\left(1-\frac{t}{2\mu^2}\right)
\left[g_A^2\left(\Pi_a-\frac{t}{2\mu^2}\,\Pi_t\right)+\left(g_A^2-1\right)
\left(1-\frac{t}{2\mu^2}\right)\Pi_t\right]
\right.\nonumber\\[2mm]&&\left.
+\left[\frac{\mu}{m}\right]^2\left\{\frac{g_A^2}{2}
\left(1-\frac{t}{2\mu^2}\right)\left[\left(g_A^2-1\right)
\left(-\frac{t}{8\mu^2}+\frac{z^2}{4\mu^2}\right)+
\bar\delta^-_{00}+\bar\delta^-_{01}\,\frac{t}{\mu^2}
+\frac{1}{3}\,\delta^-_{10}\left(1-\frac{t}{4\mu^2}\right)
+\bar\beta^-_{00}\,\frac{t}{4\mu^2}\right]
\right.\right.\nonumber\\[2mm]&&\left.\left.
-\frac{(g_A^2-1)}{6}\left(1-\frac{t}{4\mu^2}\right)\left[(g_A^2-1)\,
\frac{z^2}{8\mu^2}+\bar\delta^-_{00}+\bar\delta^-_{01}\,\frac{t}{\mu^2}
+\frac{3}{5}\,\delta^-_{10}\left(1-\frac{t}{4\mu^2}\right)
+\bar\beta^-_{00}\,\frac{t}{4\mu^2}\right]\right\}\Pi_\ell
\right.\nonumber\\[2mm]&&\left.
-\left[\frac{\mu}{m}\right]^2\frac{m^2}{64\pi^2f_{\pi}^2}\left[
-g_A^2\left(\left(1-t/2\mu^2\right)\Pi_\ell+1-\frac{t}{3\mu^2}\right)
+\frac{1}{3}(g_A^2-1)\left(\left(1-t/4\mu^2\right)
\Pi_\ell+2-\frac{t}{4\mu^2}\right)\right]^2
\right.\nonumber\\[2mm]&&\left.
+\left[\frac{\mu}{m}\right]^2\frac{m^2}{256\pi^2f_{\pi}^2}\,\frac{z^2}{\mu^2}
\,g_A^4\left[\left(1-\frac{t}{4\mu^2}\right)\Pi_a-\pi\right]^2
+\left[\frac{\mu}{m}\right]^2\frac{g_A^4}{16}\left(1-t/2\mu^2\right)^2
\frac{t}{4\mu^2}\left(\Pi_\times-\Pi_b\right)
\right.\nonumber\\[2mm]&&\left.
+ \Delta_{G\!T}  \lb \lp \frac{1}{6}(1\!-\!t/4\m^2)\;g_A^2(g_A^2-1)
-(1-t/2\mu^2)\,g_A^2\,(g_A^2-1/2)
\rp \P_\ell
\right.\right. \nn\\[2mm]&& \left.\left.
+ \frac{g_A^4}{4} \; (1\!-\!t/2\m^2)^2\; \lp
\P_\times+\P_b \rp\rb
\right\}
\label{8.5}\\[5mm]
&& t_T^- =  t_{SS}^-/2 = \frac{m}{E}\; \frac{m^2}{768\p^2f_\p^4}\;
\lb \frac{\m}{m }\rb \lc g_A^4\lb (1\!-\!t/2\m^2)\,\tilde\Pi_b
+(1\!-\!t/4\m^2)\,\P_a\rb
\right.\nn\\[2mm]&&\left.
-2\,g_A^2(g_A^2-1-2\,\bar\beta^-_{00})(1-t/4\mu^2)\,\Pi_t
\right.\nn\\[2mm]&&\left.
+ \lb\frac{\m}{m}\rb \lb -g_A^2(g_A^2-1-2\,\bar\beta^-_{00})(1-t/2\mu^2)
-\frac{1}{3}\,(g_A^2-1-2\,\bar\beta^-_{00})^2(1-t/4\mu^2)\rb\Pi_\ell
\right.\nn\\[2mm]&&\left.
+\lb\frac{\m}{m}\rb\frac{m^2g_A^4}{8\p^2f_\p^2}
\lb (1-t/4\m^2)\,\P_t-\pi\rb^2\rc \;,
\label{8.6}\\[5mm]
&& t_{LS}^- = \frac{m}{E}\; \frac{m^2}{64\p^2f_\p^4}\;\lb
\frac{\m}{m }\rb \lc g_A^4\lb \lp3/2-5\,t/8\mu^2\rp\Pi_a
-(1\!-\!t/2\m^2)(\Pi_t+\tilde\Pi_b)\rb
\right.\nn\\[2mm]&&\left.
+2g_A^2(g_A^2-1)(1-t/4\mu^2)\Pi_t
\right.\nn\\[2mm]&&\left.
+\lb\frac{\m}{m}\rb \lb \frac{1}{2}(g_A^2-1)^2 (1\!-\!t/4\m^2)
+4\,g_A^2\;\bar\beta^-_{00}\;(1\!-\!t/2\m^2)
-\frac{4}{3}\;(g_A^2-1)\bar\beta^-_{00}(1-t/4\mu^2)\rb\Pi_\ell
\right.\nn\\[2mm]&&\left.
+\left[\frac{\mu}{m}\right]\frac{g_A^4}{4}\left(1-t/2\mu^2\right)^2
\left(\Pi_\times-\Pi_b\right)
- \lb\frac{\m}{m}\rb\frac{m^2g_A^4}{8\p^2f_\p^2}
\lb(1-t/4\m^2)\P_t-\pi\rb^2\rc\;,
\label{8.7}\\[5mm]
&& t_Q^- \simeq 0\;.
\label{8.8}
\eea

This potential is the main result of this work. 
If one keeps only terms up to order ${\cal O}(q^3)$, it coincides 
numerically with that derived earlier by us \cite{RR97}. 
As far as ${\cal O}(q^4)$ terms are concerned, the only difference is due to the 
explicit treatment of medium range contributions. In our previous study 
we have shown that diagrams (k)--(o) of fig.\ref{fig5} strongly dominate 
the potential. In the above expressions, these terms are represented by 
products of $g_A^2$ by subthreshold coefficients. About 70\% of the 
isoscalar potential $t^+_C$ comes from the term proportional to 
($\bar\delta_{00}^++\bar\delta_{01}^+\,t/\mu^2$), which is related to 
the scalar form factor of the nucleon \cite{R01}, given by

\begin{equation}
\sigma(t)=\frac{3\mu^3g_A^2}{64\pi^2f_{\pi}^2}\,(1-t/2\mu^2)\,\Pi_t\,.
\end{equation}

The leading contribution to $t^+_C$ then reads

\begin{equation}
t^+_C\sim 2\;\frac{(\bar\delta_{00}^++\bar\delta_{01}^+\,t/\mu^2)}{mf_{\pi}^2}\;
\sigma(t)
\sim  \frac{4}{f_{\pi}^2}\left[-2\,c_1-c_3\,(1-t/2\mu^2)\right]
\sigma(t)\,.
\end{equation}

As the scalar form factor represents the probing of the part of the 
nucleon mass 
associated with its pion cloud, the leading term of the $NN$ potential 
corresponds to a picture in which one of the nucleons, acting as a scalar 
source, disturbs the pion cloud of the other. A rather puzzling aspect of 
this problem is that the largest term in a ${\cal O}(q^2)$ potential is 
of ${\cal O}(q^3)$.

\section{comparison with heavy baryon calculations}\label{sec:compare}

The relativistic potential of the preceding section involves
five basic functions,
representing loop integrals, and subthreshold coefficients.
The latter can be reexpressed in terms of LECs and
explicit powers of $\mu/m$, using the results of
ref. \cite{BL2}, summarized in section \ref{sec:subt-coef}.
The loop functions were derived by means of  covariant techniques and one 
uses the results of section \ref{sec:dynamics} and appendix \ref{ap:integ}.
As discussed by Ellis and Tang \cite{EllisT} and in our section 
\ref{sec:rel-hb},
if one forces an expansion of the relativistic functions in powers of
$\m/m$, even in the regions where this expansion is not valid,
one recovers {\em formally} the results of HBChPT.
This procedure amounts to replacing the relativistic functions,
which cover the neighborhood of the point $t=4\m^2$,
by the heavy baryon series, which is not valid there.

Performing such a replacement in the ${\cal O}(q^4)$ results of the 
preceding section, we find (inequivalent) expressions that 
coincide largely with those produced by means of heavy baryon techniques. 
In order to allow comparison with HBChPT calculations, 
in this section we display the full $\mu/m$ expansion of our potential, 
without including terms due to the common factor $m/E$. 

We reproduce below the results of refs. \cite{KBW,K3,K4}, which include
relativistic corrections and were elaborated further by Entem and
Machleidt \cite{EM}. The few terms that are only present in our potential
are indicated by $[\cdots]^{*}$:

\lixo{
The potential of the preceding section involves functions 
and subthreshold coefficients that can be reexpressed in terms of 
explicit powers of $\mu/m$. For the latter, one uses the results of 
ref. \cite{BL2}, summarized in section \ref{sec:subt-coef}. For the 
functions, one uses the results of section \ref{sec:dynamics}, 
derived using covariant techniques. Up to ${\cal O}(q^4)$, expressions 
coincide with those produced by means of heavy baryon techniques. In this 
section we display the full $\mu/m$ dependence of our potential, without 
including terms due to the common factor $m/E$.
We reproduce below the results of refs. \cite{KBW,K3,K4}, which include 
relativistic corrections and were elaborated further by Entem and 
Machleidt \cite{EM}. The few terms which are only present in our potential 
are indicated by $[\cdots]^{*}$. 
}

\vspace{5mm}

\bea
&\bullet& V_C = t_C^+ =\frac{3 g_A^2}{16 \p f_\p^4}
\lc -\frac{g_A^2 \;\m^5}{16 m (4\m^2\!+\! \bq^2)}
+ \lb 2\m^2(2 c_1\!-\!c_3)-\bq^2 c_3 \rb (2\m^2\!+\!\bq^2)\;A(q) \right.
\nn\\[2mm]
&& \left. + \frac{g_A^2 (2\m^2\! +\!\bq^2)\;A(q)}{16m}\lb -3\bq^2+
(4\m^2\!+\!\bq^2)^{\!*}\rb  \rc
\nn\\[2mm]
&&  +\frac{g_A^2\;L(q)}{32 \p^2 f_\p^4 m}
\lc \frac{24 \m^6}{4\m^2\!+\!\bq^2}\;(2c_1\!+\!c_3) +6\m^4(c_2\!-\!2c_3)
+ 4\m^2\bq^2(6c_1\!+\!c_2\!-\!3c_3) + \bq^4 (c_2\!-\!6c_3)\rc
\nn\\[2mm]
&& - \frac{3 L(q)}{16 \p^2 f_\p^4} \lc \lb - 4\m^2 c_1\!+\!
c_3(2\m^2\!+\!\bq^2)+ c_2  (4\m^2\!+\!\bq^2)/6 \rb^2
+ \frac{1}{45}\, (c_2)^2 (4\m^2\!+\!\bq^2)^2 \rc
\nn\\[2mm]
&& +\frac{g_A^4}{32 \p^2 f_\p^4m^2} \lc L(q) \lb
\frac{2\m^8}{(4\m^2\!+\!\bq^2)^2}+\frac{8\m^6}{(4\m^2\!+\!\bq^2)} - 2\m^4
-\bq^4\rb+\frac{\m^6 /2}{(4\m^2\!+\!\bq^2)} \rc
\nn\\[2mm]
&&- \frac{3 g_A^4 [A(q)]^2}{1024 \p^2 f_\p^6}\;(\m^2\!+\!2\bq^2)\;
(2\m^2\!+\!\bq^2)^2
\nn\\[2mm]
&&-\frac{3g_A^4(2\m^2\!+\!\bq^2)\;A(q)}{1024\p^2 f_\p^6}\lc 4\m\;g_A^2\;
(2\m^2\!+\!\bq^2) +2 \m\;(\m^2\!+\!2\bq^2)\rc\;,
\label{Vc}\\[5mm]
&\bullet& V_T=-\frac{3\;t_T^+}{m^2}=\frac{3 g_A^4 \;L(q)}{64 \p^2 f_\p^4}
-\frac{g_A^4 \;A(q)}{512 \p f_\p^4 m}\lb 9(2\m^2+ \bq^2)
+3(4\m^2+\bq^2)^{\!*} \rb
\nn\\[2mm]
&& -\frac{g_A^4\;L(q)}{32 \p^2 f_\p^4 m^2} \lb \bz^2/4
+ [5/8-(3/8)^{\!*}]\bq^2 + \frac{\m^4}{4\m^2\!+\!\bq^2}\rb
\nn\\[2mm]
&& + \frac{g_A^2\;(4  \m^2+\bq^2)\;L(q)}{32 \p^2 f_\p^4}
\lb (\tilde{d}_{14}-\tilde{d}_{15})-\lp g_A^4/32\;\p^2f_{\pi}^2\rp^{\!*}\rb
+\lb 3\,\Delta_{GT}\,\frac{g_A^4L(q)}{16\pi^2f_{\pi}^4} \rb^{\!*}\;,
\label{Vt}\\[5mm]
&\bullet& V_{LS}=-\frac{t_{LS}^+}{m^2}=-\frac{3g_A^4\;A(q)}{32\p f_\p^4 m}
\;\lb(2\m^2+\bq^2) + (\m^2+3\bq^2/8)^{\!*}\rb
\nn\\[2mm]
&& -\frac{g_A^4 \;L(q)}{4 \p^2 f_\p^4 m^2} \lb \frac{\m^4}{4\m^2\!+\!\bq^2}
+\frac{11}{32}\,\bq^2 \rb
-\frac{g_A^2\;c_2\; L(q)}{8 \p^2 f_\p^4 m}
\;(4\m^2+\bq^2)  \;,
\label{Vls}\\[5mm]
&\bullet&V_{\s L}=\frac{4\;t_Q^+}{m^4}=-\frac{g_A^4\;L(q)}{32\p^2f_\p^4m^2}\;,
\label{Vq}
\eea

\ni
and

\bea
&\bullet& W_C = t_C^- =\frac{L(q)}{384\p^2 f_\p^4} \lb 4\m^2 \lp
5g_A^4-4g_A^2-1\rp + \bq^2 \lp 23g_A^4-10g_A^2-1\rp +
\frac{48 g_A^4 \m^4}{4\m^2\!+\!\bq^2}\rb
\nn\\[2mm]
&& -\frac{g_A^2}{128 \p f_\p^4 m}\lc 3g_A^2\;\frac{\m^5}{4\m^2\!+\!\bq^2}
+ A(q)\;(2\m^2\!+\!\bq^2) \lb g_A^2 \;(4\m^2\!+\!3\bq^2)- 2\;
(2\m^2\!+\!\bq^2)\right.\right.
\nn\\[2mm]
&& \left.\left.+g_A^2\;(4\m^2 +\bq^2)^{\!*}\rb \rc+\frac{\bq^2\;c_4
\;L(q)}{192 \p^2 f_\p^4 m} \lb g_A^2(8\m^2+5\bq^2) + (4\m^2+\bq^2) \rb
\nn\\[2mm]
&&
- \frac{L(q)}{768 \p^2 f_\p^4 m^2} \lc  (4\m^2\!+\!\bq^2) \bz^2
+ g_A^2 \lb \frac{48\m^6}{4\m^2\!+\!\bq^2}- 24\m^4 -
12\,(2\m^2\!+\!\bq^2)\,\bq^2
+(16\m^2\!+\!10\bq^2) \bz^2 \rb 
\right.\nn\\[2mm]&&\left.
+ g_A^4\lb \bz^2 \lp \frac{16\m^4}{4\m^2\!+\bq^2} -7\bq^2 -20\m^2 \rp
\right.\right.
\nn\\[2mm]
&& \left.\left. -\frac{64\m^8}{(4\m^2\!+\bq^2)^2} -
\frac{48\m^6}{4\m^2\!+\bq^2}+
\frac{[16-(24)^{\!*}]\m^4 \bq^2}{4\m^2\!+\!\bq^2}
+[20-(6)^{\!*}]\bq^4+24\m^2\bq^2+24\m^4\rb \rc
\nn\\[2mm]
&& + \frac{16 g_A^4 \m^6}{768 \p^2 f_\p^4 m^2}\;\frac{1}{4\m^2\!+\!\bq^2}
\nn\\[2mm]
&& -\frac{L(q)}{18432 \p^4 f_\p^6}  \lc 192\p^2 f_\p^2 (4\m^2\!+\!\bq^2)
\;\tilde{d}_3\lb 2g_A^2 (2\m^2\!+\!\bq^2) - 3/5 (g_A^2-1) (4\m^2\!+\!\bq^2)
\rb 
\right.\nn\\[2mm]&&\left.
+\lb 6 g_A^2 (2\m^2\!+\!\bq^2)-(g_A^2-1)\;(4\m^2\!+\!\bq^2)\rb\lb
384 \p^2 f_\p^2\lp (2\m^2\!+\!\bq^2)\;(\tilde{d}_1+\tilde{d}_2)
+ 4\m^2\;\tilde{d}_5 \rp
\right.\right.\nn\\[2mm]&&\left.\left.
+L(q)\lp 4\mu^2(1+2g_A^2)+\bq^2(1+5g_A^2) \rp
-\lp \frac{\bq^2}{3}\,(5+13g_A^2)+8\mu^2(1+2g_A^2) \rp
\right.\right.\nn\\[2mm]&&\left.\left.
+\lp2g_A^4(2\mu^2+\bq^2)+\frac{2}{3}\,\bq^2(1+2g_A^2)\rp^{\!*}\rb
\right.\nn\\[2mm]&&\left.
-\frac{1}{25}\lp4\mu^2+\bq^2\rp\lp15+7g_A^4\rp
\lb 10 g_A^2 (2\m^2\!+\!\bq^2)-3(g_A^2-1)\;(4\m^2\!+\!\bq^2)\rb^{\!*}\rc
\nn\\[2mm]&&
-\frac{\bz^2g_A^4}{2048\pi^2f_{\pi}^6}\lc\lb(4\mu^2+\bq^2)A(q)
\rb^{2} +2\mu(4\mu^2+\bq^2)A(q)\rc^{\!*}
\nn\\[2mm]&&
+\Delta_{GT}\,\frac{g_A^2L(q)}{96\pi^2f_{\pi}^4}\lb
g_A^2\lp\frac{48\mu^4}{4\mu^2+\bq^2}+20\mu^2+23\bq^2\rp
-8\mu^2-5\bq^2\rb^{\!*}\;,\label{Wc}
\eea

\bea
&\bullet& W_T = - \frac{3}{m^2}\; t_T^- =  \frac{g_A^2 A(q)}{32\p f_\p^4}
\lb \lp c_4 + \frac{1}{4m} \rp (4\m^2\!+\!\bq^2) -\frac{g_A^2}{8m}\;[10\m^2
+3\bq^2 - (4\m^2\!+\!\bq^2)^{\!*}] \rb
\nn\\[2mm]
&& -\frac{c_4^2\; L(q)}{96 \p^2 f_\p^4}(4\m^2\!+\!\bq^2) +
\frac{c_4 \;L(q)}{192 \p^2 f_\p^4 m} \lb g_A^2(16\m^2+7\bq^2) -
(4\m^2\!+\!\bq^2) \rb
\nn\\[2mm]
&& - \frac{L(q)}{1536 \p^2 f_\p^4 m^2} \lb g_A^4 \lp 28\m^2+17\bq^2
+ \frac{16\; \m^4}{4\m^2\!+\!\bq^2}\rp - g_A^2 (32\m^2+14\bq^2) +
(4\m^2\!+\!\bq^2) \rb
\nn\\[2mm]
&& -\frac{[A(q)]^2 g_A^4 \; (4\m^2\!+\!\bq^2)^2 }{2048 \p^2 f_\p^6}
- \frac{A(q)  g_A^4 \; (4\m^2\!+\!\bq^2)}{1024 \p^2 f_\p^6}
\;\m(1+2g_A^2)\;,
\label{Wt}\\[5mm]
&\bullet& W_{LS} = - \frac{1}{m^2}\; t_{LS}^- =
\frac{A(q)}{32\p f_\p^4m} \lb g_A^2 \, (g_A^2-1)\;(4\m^2+\bq^2)
+ g_A^4\; (2\m^2+3\bq^2/4)^{\!*} \rb
\nn\\[2mm]&&
+\frac{c_4 \; L(q)}{48\p^2 m f_\p^4} \lb g_A^2 \lp 8\m^2+5\bq^2\rp
+ (4\m^2\! + \!\bq^2) \rb
\nn\\[2mm]
&& + \frac{L(q)}{256 \p^2 m^2 f_\p^4}
\lb (4\m^2\! + \!\bq^2) - 16 g_A^2\;(\m^2+3\bq^2/8)
+ \frac{4 g_A^4}{3} \lp 9\m^2 + 11\bq^2/4
-\frac{4\m^4}{4\m^2+\bq^2}\rp \rb
\nn\\[2mm]&&
+\frac{g_A^4}{512\pi^2f_{\pi}^6}\lc\lb(4\mu^2+\bq^2)A(q)\rb
\lb(4\mu^2+\bq^2)A(q)+2\mu\rb\rc^{\!*}
\;,
\label{Wls}\\[5mm]
&\bullet& W_{\s L}^R \simeq W_{\s L}^{HB} \simeq 0  \;.
\label{Wq}
\eea

\section{summary and conclusions}\label{sec:concl}

We have presented a $\cO(q^4)$ relativistic chiral expansion of the
two-pion exchange component of the $NN$ potential, based on that derived by
Becher and Leutwyler \cite{BL1,BL2} for elastic $\p N$ scattering.
The dynamical content of the potential is given by three families of diagrams,
corresponding to the minimal
realization of chiral symmetry, two-loop interactions in the $t$ channel, 
and processes involving $\p N$ subthreshold coefficients, which represent 
frozen degrees of freedom. 

The calculation begins with the full evaluation of these
diagrams. Results are then projected into a relativistic spin basis and
expressed in terms of many different loop integrals (appendix
\ref{ap:fullres}). At this stage, the chiral structure of the problem is not
yet evident. However, chiral scales emerge when these first amplitudes are
simplified by means of relations among loop integrals. This gives rise to
our intermediate results (appendix \ref{ap:interes}), which involve no
truncations and preserve the numerical content of the various subamplitudes
for distances larger than 1 fm.
The truncation of these intermediate results to $\cO(q^4)$ yields directly
the relativistic potential (section \ref{sec:tpep}), which is ready to be
used in momentum space calculations of $NN$ observables.

Our treatment of the $NN$ interaction emphasizes the role of the intermediate
$\p N$ subamplitudes and, in this sense, it is akin to that used in the Paris
potential. We discuss how power countings in $\p N$ and $NN$ processes are
related (section \ref{sec:pw-count}) and results are expressed directly in
terms of observable subthreshold coefficients.
The LECs $c_i$ and $d_i$ are implicitly kept within these coefficients,
grouped together with two-loop short range contributions.

If the potential presented here were truncated at order $\cO(q^3)$, one
would recover numerically the results derived by us sometime ago 
\cite{RR97}. However, processes involving two loops in the $t$ channel do
show up at $\cO(q^4)$ and results begin to depart at this order.

The dependence of the potential on the external variables is incorporated
into five loop integrals, associated with bubble, triangle, crossed box, and 
box diagrams. The triangle integral
is the same entering the scalar form factor of the nucleon and can be
represented accurately by means of elementary functions 
(section \ref{sec:dynamics})
and has the correct analytic behavior at the important
point $t=4\m^2$.
We have shown that
this kind of representation can also be used to disclose the chiral
structures of box and crossed box integrals (appendix \ref{ap:relex}).
The effects associated with the correct analytic structure of relativistic
integrals are important because they dominate the long distance 
behavior of the potential.

The expansion of the functions entering the relativistic potential
in powers of $\m/m$ is not mathematically defined around $t=4\mu^2$. 
Nevertheless, in order to compare our results with those produced
by means of HBChPT, we have assumed that such an expansion could be 
made for all low-energy values of $t$. 
This expansion then reproduces most of the standard HBChPT results.
We find, however, two systematic differences, apart
from some minor scattered ones. The first one is due to the
Goldberger-Treiman discrepancy. 
The other one concerns terms of $\cO(q^3)$, whose origin is
less certain. However, the fact that they occur at the same order as the
iteration of the OPEP suggests that there may be an important dependence
on the procedure adopted for subtracting this contribution. This aspect
of the problem is rather relevant in numerical applications of the
potential and deserves being clarified.

The numerical implications of the various approximations required to 
derive the ${\cal O}(q^4)$ potential in configuration space will be 
presented in a forthcoming paper.

\section*{Acknowledgements}

We thank C. A. da Rocha for supplying his numerical profile 
functions for the $NN$ potential and J. L. Goity for useful discussions. 
R. H. also acknowledges helpful communications with T.~Becher and 
M.~Moj\v zi\v s, the kind hospitality of the Theory Group 
of Thomas Jefferson National Accelerator Facility, and the financial 
support by FAPESP (Funda\c c\~ao de Amparo \`a Pesquisa do Estado de 
S\~ao Paulo). This work was partially supported by DOE Contract No. 
DE-AC05-84ER40150 under which SURA operates the Thomas Jefferson 
National Accelerator Facility.

\appendix

\section{kinematics}\label{ap:kin}

The initial and final nucleon momenta are denoted by $p$ and $p'$, whereas 
$k$ and $k'$ are the momenta of the exchanged pions, as in fig.\ref{fig1}. 
We define the variables

\bea
&& W =  p_1+p_2 = p'_1+p'_2 \;,
\label{a1}\\
&& z  = [(p_1+p'_1) - (p_2+p'_2) ]/2 \;,
\label{a2}\\
&& q = k'-k = p'_1-p_1= p_2-p'_2 \;,
\label{a3}\\
&& Q = (k+k')/2 \;.
\label{a4}
\eea

The external nucleons are on shell and the following constraints hold

\bea
&& m^2 =  (W^2+z^2+q^2)/4\;,
\label{a5}\\
&&W\cd z = W\cd q = z\cd q = 0 \;.
\label{a6}
\eea

For the Mandelstam variables, one has

\bea
&& t = q^2 \;,
\label{a7}\\
&& s_1 = [ Q^2\! +\! Q\cd (W\!+\!z)\! -\! t / 4\! +\! m^2 ] \; ,
\label{a8}\\
&& u_1 = [ Q^2 \!-\! Q\cd (W\!+\!z)\! -\! t/4\! + \!m^2 ] \; ,
\label{a9}\\
&& \n_1= (W\!+\!z)\cd Q/2m\;,
\label{a10}\\
&& s_2 = [ Q^2\! +\! Q\cd (W\!-\!z)\! -\! t/4\! +\! m^2 ] \; ,
\label{a11}\\
&& u_2 = [ Q^2 \!-\! Q\cd (W\!-\!z)\! -\! t/4  \!+\! m^2 ] \;,
\label{a12}\\
&& \n_2= (W\!-\!z)\cd Q/2m\;.
\label{a13}
\eea

Sometimes it is useful to write

\bea
Q^2 &=& (k^2-\m^2)/2 + (k'^2-\m^2)/2 + \lp \m^2 -t / 4 \rp \;,
\label{a14}\\
Q\cdot q&=&(k'^2-\mu^2)/2-(k^2-\mu^2)/2\;.
\label{a14b}
\eea

For free spinors, the following results hold: 

\bea
&& [\ub(\bp') \not{\!q}\; u(\bp)]^{(1)} = 
[\ub(\bp') \not{\!q}\; u(\bp)]^{(2)}= 0\;,
\label{a15}\\
&& [\ub(\bp')\; (\not{\!W}+\not{\!z})\; u(\bp)]^{(1)} = 
2m \; [\ub(\bp') \; u(\bp)]^{(1)}\;,
\label{a16}\\
&& [\ub(\bp')\; (\not{\!W}- \not{\!z})\; u(\bp)]^{(2)} = 
2m \; [\ub(\bp') \; u(\bp)]^{(2)}\;,
\label{a17}
\eea

\ni
and also

\bea
&&  \{\ub \g_\l  u \}^{(1)}=
\{(W+z)_\l/2m [\ub\;u]  -i/2m [\ub\;\s_{\m\l}(p'-p)^\m\;u ]\}^{(1)} \;,
\label{a18}\\
&&  (q^2/4m^2) \{\ub\; u\}^{(1)} = 
\{-i/2m [\ub\;\s_{\m\l}(p'-p)^\m\;u] (W+z)^\l/2m \}^{(1)} \;,
\label{a19}\\
&&  \{\ub \g_\r  u \}^{(2)}=
\{(W-z)_\r/2m  [\ub\;u]  -i/2m [\ub\;\s_{\n\r}(p'-p)^\n\;u ]\}^{(2)} \;,
\label{a20}\\
&&  (q^2/4m^2) \{\ub\; u\}^{(2)}=
\{-i/2m [\ub\;\s_{\n\r}(p'-p)^\n\;u ] (W-z)^\r/2m\}^{(2)} \;.
\label{a21}
\eea

In the CM one has

\bea
&& p_1=(E;\bp)\;\; , \;\; p'_1=(E;\bp')\;,
\label{a22}\\
&& p_2=(E;-\bp)\;\; , \;\; p'_2=(E; -\bp')\;,
\label{a23}\\
&& W= (2E;0)\;,
\label{a24}\\
&& q = (0;\bp'-\bp)\;,
\label{a25}\\
&& z = (0; \bp'+\bp)
\label{a26}
\eea

\ni
and the on shell condition for nucleons reads

\beq
E^2 = m^2 +\bq^2/4 + \bz^2/4 \;.
\label{a27}
\eeq

In the CM frame, the nucleon spin functions may be expressed in terms of 
two component matrices as

\bea
&& \lc\ub(\bp')\;u(\bp)\rc^{(i)} 
= \cdg \lb 2m  + \frac{1}{2(E\!+\!m)}\lp \bq^2 - 
i \; \bsig\cd\bq\!\times\! \bz \rp \rb \chi \;,
\label{a28}\\[2mm]
&&  \lc\frac{i}{2m} \ub(\bp')\;\s_{\m 0}(p'\!-\!p)^\m\; u(\bp)\rc^{(i)}
= \cdg \lb \frac{1}{2m}\lp \bq^2 -  i\; \bsig\cd\bq\!\times\! 
\bz \rp \rb \chi \;,
\label{a29}\\[2mm]
&& \lc\frac{i}{2m} \ub(\bp')\;\s_{\m j}(p'\!-\!p)^\m\; u(\bp)\rc^{(i)}
\nonumber\\[2mm]
&&= s(i)\;\cdg \lb -i \;\bsig\!\times\!(\bp'\!-\!\bp)
+ \lp \bq^2 - i \;\bsig\cd\bq\!\times\! \bz\rp 
\frac{(\bp'\!+\!\bp)}{4m(E\!+\!m)}\rb_j \chi \;,
\label{a30}
\eea

\ni
where $s(i)=(1,-1)$ for $i=(1,2)$.
These results, which contain no approximations, allow one to write the 
identities

\bea
&&  [\ub\;u]^{(1)}[\ub\;u]^{(2)}
= 4m^2 \lb \lp 1\!+\! \bq^2\! / \l^2 \rp^2
- 4 \lp 1 \!+\! \bq^2\! / \l^2 \rp \frac{\bO_{LS}}{\l^2}
-\frac{\bO_Q}{\l^4}\rb  \;,
\label{a31}\\[2mm]
&& - \frac{i}{2m}\lc [\ub\;u]^{(1)}[\ub \; \s_{\m\l} \; (p'-p)^\m \;u]^{(2)}
- (1 \leftrightarrow 2) \rc \frac{z^\l}{2m}
= 4m^2 \lb - \lp 1 \!+\! \bq^2\!/ \l^2 \rp \frac{\bz^2\bq^2}{2m^2\l^2} \right.
\nn\\[2mm]
&& \left. + \lp 1 \!+\! \bq^2\!/ \l^2 \!+\! \bz^2\!/ \l^2 + 
2\bq^2\bz^2\!/ \l^4 \rp \frac{\bO_{LS}}{m^2}
+\lp 1 \!+\! \bz^2\!/ \l^2 \rp \frac{\bO_Q}{2 m^2\l^2}\rb \;,
\label{a32}\\[2mm]
&& - \frac{1}{4m^2} [\ub \; \s_{\m \l} (p'-p)^\m \;u]^{(1)}
[\ub \; \s_{\n\r}(p'-p)^\n \;u]^{(2)} g^{\l\r}
= 4m^2 \lb \lp 1 \!+\! 4m^2 \bz^2\!/ \l^4 \rp \frac{\bq^4}{16m^4}
- \frac{\bO_{SS}}{6m^2}\right.
\nn\\[2mm]
&& \left.  -\frac{\bO_T}{12m^2} - \lp 1 \!+\! 4m^2\!/ \l^2 \!+\! 
4m^2\bz^2\!/ \l^4 \rp \frac{\bq^2\; \bO_{LS}}{4m^4}
- \lp 1 \!+\! 8m^2\!/ \l^2 \!+\! 4m^2\bz^2\!/ \l^4 \rp 
\frac{\bO_Q}{16 m^4} \rb \;,
\label{a33}\\[2mm]
&& - \frac{1}{4m^2} [\ub \; \s_{\m \l} (p'-p)^\m \;u]^{(1)}
[\ub \; \s_{\n\r}(p'-p)^\n \;u]^{(2)}\; \frac{z^\l z^\r}{4m^2}
= 4m^2 \lb -\frac{\bq^4 \bz^4}{16 m^4 \l^4}\right.
\nn\\[2mm]
&& \left. + \lp 1 \!+\! \bz^2\!/ \l^2 \rp \frac{\bq^2\bz^2 \; 
\bO_{LS}}{4 m^4 \l^2}+\lp 1 \!+\!\bz^2\!/\l^2 \rp^2\frac{\bO_Q}{16 m^4}\rb \;,
\label{a34}
\eea

\ni
where the two-component spin operators $\bO$ were defined in section 
\ref{sec:tpep-form} and $\l^2 = 4m (E+m)$.

\section{loop integrals}\label{ap:integ}

The basic loop integrals needed in this work are 

\bea
&& I_{cc}^{\m\cdots} =  \int [\cdots] \; \lp\frac{Q^\m}{\m}\cdots \rp \;,
\label{b1}\\
&& I_{sc}^{\m\cdots} = \int [\cdots] \; \lp  \frac{Q^\m}{\m}\cdots \rp \;
\frac{2m\m}{ [ Q^2\! + \! Q\cd (W\!+\!z)\! -\! t / 4 ] } \;,
\label{b2}\\
&& I_{ss}^{\m\cdots} = \int [\cdots] \; \lp \frac{Q^\m}{\m}\cdots \rp \;
\frac{2m\m}{ [ Q^2\! + \! Q\cd (W\!+\!z)\! -\! t / 4 ] }\; \frac{2m\m}
{ [ Q^2\! + \! Q\cd (W\!-\!z)\! -\! t / 4 ] } \;,
\label{b3}
\eea

\ni
with 

\beq
\int [\cdots] = \int \frac{d^4Q}{(2\pi)^4}\; 
\frac{1}{[ (Q\!-\!q/2) ^2\! -\!\m^2] [(Q\!+\!q/2)^2 \!-\!\m^2]} \;.
\label{b4}
\eeq

All denominators are symmetric under $q \rightarrow - q$ and results 
cannot contain odd powers of this variable. The integrals are dimensionless 
and have the following tensor structure:

\bea
&& I_{cc} =  \frac{i}{(4\pi)^2} \lc \P_{cc}^{(000)}\rc \;,
\label{b5}\\
&& I_{cc}^{\m\n} = \frac{i}{(4\pi)^2}\lc \frac{1}{\m^2}\lb q^\m q^\n \;
\P_{cc}^{(200)}\rb+ g^{\m\n}\;\Pb_{cc}^{(000)}\rc \;,
\label{b6}\\
&& I_{cc}^{\m\n\l\r} = \frac{i}{(4\pi)^2}\lc 
\frac{1}{\m^4}\lb q^\m q^\n q^\l q^\r \; \Pi_{cc}^{(400)} \rb
+ \frac{1}{\m^2} \lb\lp g^{\m\n}q^\l q^\r + g^{\n\l}q^\r q^\m + 
g^{\l\r}q^\m q^\n + g^{\m\l}q^\n q^\r \right.\right.\right.
\nn\\
&& \left.\left.\left.+ g^{\m\r}q^\l q^\n + g^{\n\r}q^\m q^\l \rp 
\Pb_{cc}^{(200)}\rb+ \lb \lp g^{\m\n} g^{\l\r} + g^{\m\l} g^{\n\r} + 
g^{\m\r} g^{\n\l} \rp \Pbb_{cc}^{(000)}\rb \rc \;,
\label{b7}\\
&& I_{sc} =  \frac{i}{(4\pi)^2}\lc \P_{sc}^{(000)} \rc\;,
\label{b8}\\
&& I_{sc}^{\m} =  \frac{i}{(4\pi)^2} \lc \frac{1}{2 m}
\lb (z^\m + W^\m)\; \P_{sc}^{(001)}\rb \rc \;,
\label{b9}\\
&& I_{sc}^{\m\n} =  \frac{i}{(4\pi)^2}\lc \frac{1}{\m^2} \lb q^\m q^\n\;
\P_{sc}^{(200)}\rb+ \frac{1}{4m^2} \lb (W+z)^\m (W+z)^\n \;
\P_{sc}^{(002)}\rb + g^{\m\n}\; \Pb_{sc}^{(000)} \rc \;,
\label{b10}\\
&& I_{ss} =  \frac{i}{(4\pi)^2}\lc \P_{ss}^{(000)} \rc\;,
\label{b11}\\
&& I_{ss}^{\m} =  \frac{i}{(4\pi)^2} \lc \frac{1}{2 m}
\lb z^\m\;\P_{ss}^{(010)}+ W^\m \; \P_{ss}^{(001)}\rb \rc \;,
\label{b12}\\
&& I_{ss}^{\m\n} =  \frac{i}{(4\pi)^2} \lc \frac{1}{\m^2}\lb q^\m q^\n\;
\P_{ss}^{(200)}\rb
+ \frac{1}{4m^2} \lb z^\m z^\n \;\P_{ss}^{(020)}+ W^\m W^\n \;
\P_{ss}^{(002)}\rb + g^{\m\n}\; \Pb_{ss}^{(000)} \rc \;.
\label{b13}
\eea

The usual Feynman techniques for loop integration allow us to write

\bea
&& \P_{cc}^{(k00)} =  \int_0^1d a\;  (- C_a)^k \lb \r_0 - \ln \lp 
\frac{D_{cc}}{\m^2}\rp \rb  \;,
\label{b14}\\[2mm]
&& \Pb_{cc}^{(k00)}= - \;\frac{1}{2} \int_0^1 d a \; (-C_a)^k\; 
\frac{D_{cc}}{\m^2}\lb-\r_1 + \ln \lp \frac{D_{cc}}{\m^2}\rp \rb  \;,
\label{b15}\\[2mm]
&& \Pbb_{cc}^{(000)}= \frac{1}{8} \int_0^1 d a \; \frac{D_{cc}^2}{\m^4}
\lb \r_2 - \ln \lp \frac{D_{cc}}{\m^2}\rp \rb  \;,
\label{b16}\\[2mm]
&& \P_{sc}^{(kmn)}=  \lp\!-\; \frac{2m}{\m}\rp^{m+n+1} \int_0^1 d a\;a 
\int_0^1 d b \;\frac{\m^2\; (-C_q)^k  (C_b)^{m+n}}{D_{sc}}\;,
\label{b17}\\[2mm]
&& \Pb_{sc}^{(000)}= - \lp\frac{2m}{\m}\rp \frac{1}{2} \int_0^1 d a\;a 
\int_0^1 d b \; \lb- \r_0+\ln\lp \frac{D_{sc}}{\m^2} \rp \rb\;,
\label{b18}\\
&& \P_{ss}^{(kmn)}= \lp\!-\;\frac{2m}{\m}\rp^{m+n+2}  \int_0^1 d a\;a^2 
\int_0^1 d b\;b\int_0^1 d c \;
\frac{\m^4\;(-C_q)^k\; (C_c)^m\;(C_b)^n}{D_{ss}^{\;2}}\; ,
\label{b19}\\[2mm]
&& \Pb_{ss}^{(000)}= - \lp\frac{2m}{\m}\rp^2 \frac{1}{2} \int_0^1 d a\;a^2 
\int_0^1  d b\;b\int_0^1 d c \; \frac{\m^2}{D_{ss}} 
\label{b20}
\eea

\ni
with

\bea
&& C_a = a - 1/2 \;,
\label{b21}\\[2mm]
&& \S_{cc}^2 = -q^2/4 + \m^2 \;,
\label{b22}\\[2mm]
&& D_{cc} = C_a^2 \;q^2+\S_{cc}^2 \;,
\label{b23}\\[2mm]
&& C_b=ab/2 \;,
\label{b24}\\
&& C_q = C_a- C_b \;,
\label{b25}\\[2mm]
&& \S_{mc}^2 = -(1-2 ab)\; q^2/4 + (1-ab) \;\mu^2 \;,
\label{b26}\\[2mm]
&& D_{sc} = C_q^2 \;q^2+ C_b^2 \;\lp  z^2 + 
W^2 \rp + \S_{mc}^2 \;,
\label{b27}\\[2mm]
&& C_c= abc/2 \;,
\label{b28}\\[2mm]
&& \S_{mm}^2 = \S_{mc}^2\;,
\label{b29}\\[2mm]
&&  D_{ss} = C_q^2\; q^2 + C_c^2\; z^2 + 
C_b^2 \; W^2 + \S_{mm}^2 \;.
\label{b30}
\eea

The case $(cs)$ is obtained from $(sc)$ by making $z^\m \rightarrow - z^\m $.
The case $(us)$ is obtained from $(ss)$ by making $C_b \leftrightarrow - C_c$.

\section{OPEP iteration}\label{ap:iterat}

The iteration of the OPEP has to be subtracted from the elastic scattering 
amplitude, in order to avoid double counting in the potential. In this work 
we adopt the procedure used by Partovi and Lomon \cite{PL}, based on a 
prescription developed by Blankenbecler and Sugar \cite{BbS}. In this appendix 
we adapt their expressions to our relativistic notation and also simplify 
some of the results.

The iterated OPEP is contained in the box diagram, corresponding to the 
amplitude

\beq
{\cT}_{box} =  \lb 3 - 2\; \btau^{(1)} \cd \btau^{(2)}\rb \; {\cT}_{us} \;,
\label{c1}
\eeq

\ni
where

\beq
{\cT}_{us} =  i  \lb \frac{g}{m} \rb^4 \frac{m^2}{4} \int [\cdots]  
\frac{Q^\m Q^\n}{\m^2}\; \lb \frac{2m\m}{\um}\; \ub \g_ \m u\rb^{(1)}
\lb \frac{2m\m}{\sm} \; \ub \g_ \n u\rb^{(2)}\;.
\label{c2}
\eeq

Evaluating this integral using the results of appendix \ref{ap:integ}, one 
recovers the spin structure  of  eq.(\ref{2.6}) with

\bea
&& \left. \cI_{DD}\rb_{us} = -\frac{m^2/4}{(4\pi)^2} \lb \frac{g}{m} \rb ^4
\lb -\;\frac{z^4}{16m^4}\;\P_{us}^{(020)} + \frac{W^4}{16m^4}\;\P_{us}^{(002)}
+\frac{W^2\!-\!z^2}{4m^2}\;\Pb_{us}^{(000)} \rb \;,
\label{c3}\\[2mm]
&&  \left. \cI_{DB}^{(w)} \rb_{us}=  -\frac{m^2/4}{(4\pi)^2} \lb \frac{g}{m} 
\rb ^4\lb \frac{W^2}{4m^2}\;\P_{us}^{(002)}+ \Pb_{us}^{(000)} \rb \;,
\label{c4}\\[2mm]
&&  \left. \cI_{DB}^{(z)} \rb_{us}=  -\frac{m^2/4}{(4\pi)^2} \lb \frac{g}{m} 
\rb ^4\lb \frac{z^2}{4m^2}\;\P_{us}^{(020)}+ \Pb_{us}^{(000)} \rb \;,
\label{c5}\\[2mm]
&&  \left. \cI_{BB}^{(g)} \rb_{us} = -\frac{m^2/4}{(4\pi)^2} \lb \frac{g}{m} 
\rb ^4\lb \Pb_{us}^{(000)} \rb \;,
\label{c6}\\[2mm]
&& \left. \cI_{BB}^{(w)} \rb_{us} = -\frac{m^2/4}{(4\pi)^2} \lb \frac{g}{m} 
\rb ^4\lb \P_{us}^{(002)} \rb \;,
\label{c7}\\[2mm]
&&  \left. \cI_{BB}^{(z)} \rb_{us} = -\frac{m^2/4}{(4\pi)^2} \lb \frac{g}{m} 
\rb ^4\lb \P_{us}^{(020)} \rb \;.
\label{c8}
\eea

The iterated amplitude is denoted by ${\cT}_\p$ and given by

\beq
{\cT}_\p = \lb 3 - 2 \; \btau^{(1)} \cd \btau^{(2)} \rb \; {\cT}_{it}
\label{c9}
\eeq

\ni
with

\bea
&&{\cT}_{it}= -i \lb \frac{g}{m} \rb^4 \frac{m^2}{4}
\lc (\ub u )^{(1)}  (\ub u)^{(2)}\lp I_B - 2\; I_C \rp \right.
\nn\\[2mm]
&&\left.-\lb(\ub u)^{(1)}(\ub \;\g_i \; u)^{(2)}-
(\ub \;\g_i \; u)^{(1)}(\ub u)^{(2)}\rb\lb \frac{\m}{m}\;I_C^i+\frac{z^i}{2m}
\;\lp I_B -2\; I_C \rp \rb \right.
\nn\\[2mm]
&& -  \left. (\ub \;\g_i \; u)^{(1)} (\ub \;\g_j \; u)^{(2)} \lb
\frac{\m^2}{m^2}\lp I_A^{ij}-I_C^{ij} \rp+\frac{\m}{m} \lp \frac{z^i}{2m}\;
I_C^j  + I_C^i\; \frac{z^j}{2m}\rp \right.\right.
\nn\\[2mm]
&& \left.\left. + \frac{z^i z^j}{4m^2}\;\lp I_B - 2\; I_C \rp \rb \rc \;.
\label{c10}
\eea

The functions $I_i$ are three-dimensional loop integrals, defined as

\bea
&& I_A^{i \cdots} = i \int (\cdots) \; \lp  \frac{Q^i}{\m}\cdots \rp \;
\frac{m^3}{E [E_Q^2 - E^2] } \;,
\label{c11}\\
&& I_B = i \int (\cdots) \; \frac{m^3}{E^2 E_Q}\;,
\label{c12}\\
&& I_C^{i \cdots} =  I_A^{i \cdots}- \; I_D^{i \cdots}\;,
\label{c13}\\
&& I_D^{i \cdots} = i \int (\cdots) \; \lp \frac{Q^i}{\m}\cdots \rp \; 
\frac{m^3}{ E_Q [E_Q^2 - E^2]} \;,
\label{c14}
\eea

\ni
where $E_Q = \sqrt{m^2 + (\bQ\!-\!\bz/2)^2}$ and

\beq
\int (\cdots) = \int \frac{d^3Q}{(2\pi)^3}\; 
\frac{m}{[(\bQ-\bq/2)^2+\m^2] [(\bQ+\bq/2)^2+\m^2]} \;.
\label{c15}
\eeq

The usual Feynman parametrization techniques, the representation

\beq
\frac{m}{ E_Q} = \frac{1}{\p} \int_{-\infty}^{\infty} 
\frac{m E\; d\e}{[(\bQ-\bz/2)^2+ m^2 +\e^2 E^2]}\;,
\label{c16}
\eeq

\ni
and the tensor decomposition

\bea
&& I_x =  \frac{i}{(4\pi)^2}\lc \P_x^{(000)} \rc\;,
\label{c26}\\
&& I_x^i =  \frac{i}{(4\pi)^2} \lc \frac{z^i}{2 m} \; \P_x^{(010)}\rc \;,
\label{c27}\\
&& I_x^{ij} =  \frac{i}{(4\pi)^2} \lc \frac{q^i q^j}{\m^2}\;\P_x^{(200)}
+\frac{z^iz^j}{4 m^2}\; \P_x^{(020)}+ g^{ij}\; \Pb_x^{(000)} \rc 
\label{c28}
\eea

\ni
(for $x=A,B,C$) yield

\bea
&& \left. \cI_{DD}\rb_{it} =  \frac{m^2/4}{(4\pi)^2} \lb \frac{g}{m} \rb ^4
\lc \frac{\m^2}{m^2} \lp \frac{z^4}{16m^4}\;\P_A^{(020)}+ \frac{z^2}{4m^2}\;
\Pb_A^{(000)}\rp+\lp 1-\frac{z^2}{4m^2}\rp^2 \lp \P_B^{(000)}\! -\! 2\;
\P_C^{(000)}\rp \right.
\nn\\[2mm]
&& \left.-\;\frac{z^2}{4m^2} \lp \frac{\m^2}{m^2}\;\Pb_C^{(000)}\! + 
\!\frac{2\m}{m}\;\P_C^{(010)}\rp- \frac{z^4}{16m^4} \lp \frac{\m^2}{m^2}\;
\P_C^{(020)} -  \frac{2\m}{m}\; \P_C^{(010)}\rp \rc\;,
\label{c29}\\[2mm]
&& \left. \cI_{DB}^{(z)} \rb_{it} =  \frac{m^2/4}{(4\pi)^2} \lb \frac{g}{m} 
\rb ^4\lc -\;\frac{\m^2}{m^2}\lp \frac{z^2}{4m^2}\; \P_A^{(020)}+ 
\Pb_A^{(000)} \rp+ \frac{\m^2}{m^2} \;\Pb_C^{(000)}  + \frac{\m}{m}\;
\P_C^{(010)}\right.
\nn\\[2mm]
&& \left.+\lp 1-\frac{z^2}{4m^2}\rp \lp \P_B^{(000)}\! -\! 2\; \P_C^{(000)}\rp
+\frac{z^2}{4m^2} \lp \frac{\m^2}{m^2}\; \P_C^{(020)}- \frac{2\m}{m}\; 
\P_C^{(010)}\rp \rc\;,
\label{c30}\\[2mm]
&& \left. \cI_{BB}^{(g)} \rb_{it} = - \frac{W^2}{4m^2}
 \left. \cI_{BB}^{(w)} \rb_{it} =
\frac{m^2/4}{(4\pi)^2} \lb \frac{g}{m} \rb ^4\lc\frac{\m^2}{m^2}\lb - 
\Pb_A^{(000)} + \Pb_C^{(000)}\rb \rc\;,
\label{c31}\\[2mm]
&& \left. \cI_{BB}^{(z)} \rb_{it} = \frac{m^2/4}{(4\pi)^2} \lb \frac{g}{m} 
\rb ^4\lc - \frac{\m^2}{m^2}\; \P_A^{(020)}+ \frac{\m^2}{m^2}\; \P_C^{(020)}\!
-\frac{2\m}{m}\; \P_C^{(010)}\!-\P_B^{(000)}\!+\! 2\;\P_C^{(000)}\rc\;.
\label{c32}
\eea

The functions $\Pi$ and $\bar\Pi$ are written as 

\bea
\P_A^{(020)} &=&  \lp \frac{2m}{\m}\rp^2 \;\frac{4 m^4}{E} 
\int_0^1d a\;a\; \int_0^1 db\;\int_0^\infty dQ \; 
\frac{(C_b)^2} {[Q^2 +\S_A^2 -\bP_I^2]^2} \;,
\label{c36}\\[2mm]
\Pb_A^{(000)} &=&  -\;\frac{2 m^4}{\m^2 E} \int_0^1d a\;a\; \int_0^1 db\;
\int_0^\infty dQ \; \frac{1} {[Q^2 +\S_A^2 -\bP_I^2]} \;,
\label{c37}\\[2mm]
\P_B^{(000)} &=& \frac{4 m^4}{\p E} \int_0^1d a\;a\; \int_0^1 db\; 
\int_{-\infty}^\infty d\e\;\int_0^\infty dQ \; \frac{1} {[Q^2 +
\S_B^2 -\bP_I^2]^2} \;,
\label{c38}\\[2mm]
\P_C^{(0n0)} &=& \lp \frac{2m}{\m}\rp^n\; \frac{4 m^4}{\p E} 
\int_0^1 d a\;a \int_0^1 d b \int_{-\infty}^\infty \frac{d\e}{1+\e^2}\;
\int_0^\infty dQ \;  \frac{(C_b)^n} {[Q^2 +\S_B^2 -\bP_I^2]^2} \;,
\label{c39}\\[2mm]
\Pb_C^{(000)} &=& -\; \frac{2 m^4}{\p \m^2 E} \int_0^1 d a\;a 
\int_0^1 d b\; \int_{-\infty}^\infty \frac{d\e}{1+\e^2}\;
\int_0^\infty dQ \; \frac{1} {[Q^2 +\S_B^2 -\bP_I^2]}\;,
\label{c40}
\eea

\ni
where

\bea
&& \bP_I = C_q\;\bq - C_b\;\bz \;,
\label{c20}\\
&& \S_A^2= \S_{mc}^2 \;,
\label{c21}\\
&& \S_B^2=  \S_{mc}^2 + ab\;(1+\e^2) \;E^2 \;,
\label{c22}\\
&& \S_D^2= \S_{mc}^2 + ab \l\;(1+\e^2) \; E^2  \;.
\label{c23}
\eea

The contribution from the OPEP cut in the functions $\P_{us}$ 
is canceled by the integrals $\P_A$. We parametrize the loop momentum 
in those integrals as $Q=(abc\;W/2)=(-C_c\;W)$, and have 
$[Q^2 +\S_A^2 -\bP_I^2]=  D_{us}$ and write

\beq
\frac{\m^2}{m^2}\;\P_A^{(020)} \equiv  \P_{it}^{(020)}\;,
\;\;\;\;\;\;\;\;\;\;\;\;\;\;
\frac{\m^2}{m^2}\;\Pb_A^{(000)} \equiv \Pb_{it}^{(000)} 
\label{c41}
\eeq

\ni
with

\bea
&&  \P_{it}^{(kmn)}
= \lp- \frac{2m}{\m}\rp^{m+n+2}\int_0^1 da\;a^2\int_0^1 d b\;b\int_0^\infty 
dc\;\frac{\m^4\; (C_q)^k (-C_b)^m (-C_c)^n}{D_{us}^{\;2}}\;,
\label{c42}\\[2mm]
&& \Pb_{it}^{(000)}= - \lp\! \frac{2m}{\m}\rp^2 \frac{1}{2}\int_0^1 d a\;a^2 
\int_0^1 d b\;b \int_0^\infty dc\;\frac{\m^2}{D_{us}}\;.
\label{c43}
\eea

The integrals $\P_B$ and $\P_C$ can also be simplified, by adopting the new 
variables $c$ and $\th$, defined by the relations $\e = \sqrt{a^2 b^2 c^2-a b}
\;\cos\th /\sqrt{a b}$, $Q = E \sqrt{a^2 b^2 c^2-a b}\;\sin\th$. Performing 
the angular integrations, we have

\bea
&&\P_B^{(000)} = \lp \frac{2m}{\m}\rp^4 \frac{1}{4}\int_0^1 da\;a^2 
\int_0^1 db\;b
\int_{1/\sqrt{ab}}^\infty dc\;\frac{\m^4 \sqrt{ab}\;c}{D_{us}^{\;2}}\;,
\label{c45}\\[2mm]
&& \P_C^{(0m0)} = \lp \frac{2m}{\m}\rp^{m+4}\frac{1}{4}\int_0^1 da\;a^2 
\int_0^1 db\;b
\int_{1/\sqrt{ab}}^\infty dc\;\frac{\m^4 (C_b)^m}{D_{us}^{\;2}}\;,
\label{c46}\\
&& \Pb_C^{(000)} =  -\;\lp \frac{2m}{\m}\rp^4 \frac{1}{16} \int_0^1 da\;a^2
\int_0^1 db\;b\int_{1/\sqrt{ab}}^\infty dc\;\frac{\m^2}{D_{us}}\;.
\label{c47}
\eea

The results presented so far in this appendix correspond just to a 
reorganization of those obtained by Partovi and Lomon \cite{PL}. They may 
be further simplified by noting that 

\bea
I_B &=& i \int(\cdots)\; \frac{m^3}{E^2 E_Q}
\nn\\[2mm]
& \simeq & i\;\frac{m^3}{E^3}\int(\cdots) \lb 1- (\bQ^2\!-\!\bq^2/4\! -\! 
\bQ\cd\bz )/ 2E^2+ 3 (\bQ^2\!-\!\bq^2/4\! -\! \bQ\cd\bz )^2 / 8E^4 \rb \;,
\label{c48}\\[2mm]
I_C^{i\cdots} &=& i \int(\cdots) \lp \frac{Q^i}{\m}\cdots\rp \;
\frac{m^3}{E E_Q (E+ E_Q)}
\nn\\[2mm]
& \simeq & i\;\frac{m^3}{2 E^3}\int(\cdots) \lb 1- 3(\bQ^2\!-\!\bq^2/4\!
-\! \bQ\cd\bz )/4E^2+ 5 (\bQ^2\!-\!\bq^2/4\! -\! \bQ\cd\bz )^2 / 8E^4 \rb \;.
\label{c49}
\eea

The integrals $\int(\cdots)$ can be performed analytically and we have

\bea
&& \int(\cdots) = -\;\frac{1}{(4\p)^2}\;\frac{2m}{\m}\; \P_a \;,
\label{c50}\\[2mm]
&& \int(\cdots) \lp \frac{Q_i Q_j}{\m^2}\rp  = \frac{1}{(4\p)^2}\;\frac{m}{\m}
\lp \delta_{ij}- \frac{q_i q_j}{\bq^2}\rp \lp 1+\frac{\bq^2}{4\m^2}\rp
\P_a \;,
\label{c51}
\eea

\ni
where $\P_a$ is the function given in eq.(\ref{6.9}).

Thus, 

\bea
&& \P_B^{(000)} =  -\;\frac{2m}{\m}\;\frac{m^3}{E^3} \lb 1+\frac{1}{2m^2}
(\m^2 - q^2/2)+\frac{1}{8m^4} (q^2+z^2)(\m^2- q^2/2)  \right.
\nn\\[2mm]
&& \left. + \frac{3}{8m^4} (\m^2 - q^2/2)^2 + \frac{3}{16m^4}\;
z^2(\m^2- q^2/4) \rb  \P_a\;,
\label{c52}\\[2mm]
&&\P_C^{(000)} = -\;\frac{m}{\m}\;\frac{m^3}{E^3} \lb 1+\frac{3}{4m^2}
(\m^2 - q^2/2)+\frac{3}{16 m^4} (q^2+ z^2)(\m^2 - q^2/2)  \right.
\nn\\[2mm]
&& \left. + \frac{5}{8m^4} (\m^2 - q^2/2)^2 + \frac{5}{16m^4}\;
z^2(\m^2 - q^2/4) \rb  \P_a\;,
\label{c53}\\[2mm]
&&\P_C^{(010)} = \frac{3}{4} \lp 1 - \;\frac{q^2}{4\m^2}\rp \P_a\;,
\label{c54}\\[2mm]
&&\P_C^{(020)} = 0 \;,
\label{c55}\\[2mm]
&&\Pb_C^{(000)} = -\;\frac{m}{2\m} \lp 1-\; \frac{q^2}{4\m^2}\rp 
\P_a\;.
\label{c56}
\eea

The results presented in eqs.(\ref{c3})--(\ref{c8}), 
(\ref{c29})--(\ref{c32}), (\ref{c41})--(\ref{c43}), and 
(\ref{c52})--(\ref{c56}) allow one to write

\bea
&& \left. \cI_{DD}\rb_{us}- \left. \cI_{DD}\rb_{it} =
\frac{m^2/4}{(4\pi)^2} \lb \frac{g}{m} \rb ^4
\lc - \frac{z^4}{16m^4}\;\P_{reg}^{(020)} + \frac{W^4}{16m^4}\;
\P_{reg}^{(002)}+ \frac{W^2\!-\!z^2}{4m^2}\;\Pb_{reg}^{(000)} \right.
\nn\\[2mm]
&& \left.-\frac{\m}{2m} \lp1- \frac{q^2}{2\m^2}\rp  \lp 1 + \frac{\m^2}{m^2}
+ \frac{q^2}{8 m^2}+ \frac{z^2}{8m^2} \rp \P_a \rc\;,
\label{c57}\\[2mm]
&&  \left. \cI_{DB}^{(w)} \rb_{us}- \left. \cI_{DB}^{(w)} \rb_{it} =
\frac{m^2/4}{(4\pi)^2} \lb \frac{g}{m} \rb ^4\lc\frac{W^2}{4m^2}\;
\P_{reg}^{(002)}+\Pb_{reg}^{(000)} \rc\;,
\label{c58}\\[2mm]
&&  \left. \cI_{DB}^{(z)} \rb_{us}- \left. \cI_{DB}^{(z)} \rb_{it} =
\frac{m^2/4}{(4\pi)^2} \lb \frac{g}{m} \rb^4\lc\frac{z^2}{4m^2}\; 
\P_{reg}^{(020)}+ \Pb_{reg}^{(000)}-\;\frac{\m}{2m}\lp \frac{3}{2}-
\frac{5 q^2}{8\m^2} \rp \P_a \rc\;,
\label{c59}\\[2mm]
&& \left. \cI_{BB}^{(g)} \rb_{us}- \left. \cI_{BB}^{(g)} \rb_{it}
= \frac{m^2/4}{(4\pi)^2} \lb \frac{g}{m} \rb^4\lc \Pb_{reg}^{(000)}
+ \frac{\m}{2m} \lp 1- \frac{q^2}{4\m^2} \rp \P_a \rc\;,
\label{c60}\\[2mm]
&& \left. \cI_{BB}^{(w)} \rb_{us}- \left. \cI_{BB}^{(w)} \rb_{it}
= \frac{m^2/4}{(4\pi)^2} \lb \frac{g}{m} \rb ^4\lc \P_{reg}^{(002)} \rc\;,
\label{c61}\\[2mm]
&& \left. \cI_{BB}^{(z)} \rb_{us}- \left. \cI_{BB}^{(z)} \rb_{it}
= \frac{m^2/4}{(4\pi)^2} \lb \frac{g}{m} \rb ^4\lc \P_{reg}^{(020)}  \rc\;,
\label{c62}
\eea

\ni
where the integrals $\P_{reg}\equiv \P_{it}-\P_{us}$  are regular and given 
by

\bea
&& \P_{reg}^{(kmn)}= \lp - \frac{2m}{\m}\rp^{m+n+2}\int_0^1 d a\;a^2
\int_0^1 d b\;b \int_1^\infty dc\;
\frac{\m^4\;(C_q)^k(-C_b)^m(-C_c)^n}{D_{us}^{\;2}}\;,
\label{c63}\\[2mm]
&& \Pb_{reg}^{(000)}= - \lp\! \frac{2m}{\m}\rp^2\frac{1}{2}\int_0^1 da\;a^2
\int_0^1 db\;b\int_1^\infty dc\;\frac{\m^2}{D_{us}}\;.
\label{c64}
\eea

\section{full results}\label{ap:fullres}

In this appendix we list the results for the amplitudes that enter 
eq.(\ref{2.15}), obtained by reading the diagrams of fig.\ref{fig5} 
and representing loop integrals by means of the functions displayed in 
appendices \ref{ap:integ} and \ref{ap:iterat}.

\vspace{5mm}
\ni
\underline{\bf family 1} (diagrams a+b+c+d+e+f)

\bea
&\bullet& \cI_{DD}^+ = \frac{m^2/4}{(4\p)^2}\;\frac{g^4}{m^4}
\lc 2\;\P_{cc}^{(000)}- 4\; \frac{W^2\!+\!z^2}{4m^2}\; \P_{sc}^{(001)}
-\frac{z^4}{16m^4} \P_{ss}^{(020)}+ \frac{W^4}{16m^4} \P_{ss}^{(002)} \right.
\nn\\[2mm]
&& \left. + \frac{W^2\!-\!z^2}{4m^2} \Pb_{ss}^{(000)}
-\frac{z^4}{16m^4} \P_{reg}^{(020)}+ \frac{W^4}{16m^4} \P_{reg}^{(002)}
+ \frac{W^2\!-\!z^2}{4m^2} \Pb_{reg}^{(000)}\right.
\nn\\[2mm]
&& \left. - \frac{\m}{2m} \lp1- \frac{q^2}{2\m^2}\rp  \lp 1 + \frac{\m^2}{m^2}
+ \frac{q^2}{8 m^2}+ \frac{z^2}{8m^2} \rp \P_a \rc \;,
\label{d1}\\[4mm]
&\bullet& \cI_{DB}^{(w)+} = \frac{m^2/4}{(4\p)^2} \;\frac{g^4}{m^4}
\lc -2\;\P_{sc}^{(001)} + \frac{W^2}{4m^2}\P_{ss}^{(002)}+\Pb_{ss}^{(000)}
+ \frac{W^2}{4m^2} \P_{reg}^{(002)} + \Pb_{reg}^{(000)}  \rc \;,
\label{d2}\\[4mm]
&\bullet& \cI_{DB}^{(z)+} = \frac{m^2/4}{(4\p)^2} \;\frac{g^4}{m^4}
\lc 2\;\P_{sc}^{(001)}+ \frac{z^2}{4m^2} \P_{ss}^{(020)}+ \Pb_{ss}^{(000)}
+\frac{z^2}{4m^2}\P_{reg}^{(020)} + \Pb_{reg}^{(000)}
\right.
\nn\\[2mm]
&& \left. - \frac{\m}{2m} \lp \frac{3}{2}- \frac{5 q^2}{8\m^2}\rp \P_a 
\rc \;,
\label{d3}\\[4mm]
&\bullet& \cI_{BB}^{(g)+} = \frac{m^2/4}{(4\p)^2} \;\frac{g^4}{m^4}
\lc \Pb_{ss}^{(000)} + \Pb_{reg}^{(000)}+\frac{\m}{2m}\lp 1-
\frac{q^2}{4\m^2}\rp \P_a \rc \;,
\label{d4}\\[4mm]
&\bullet& \cI_{BB}^{(w)+} = \frac{m^2/4}{(4\p)^2} \;\frac{g^4}{m^4}\lc 
\P_{ss}^{(002)} + \P_{reg}^{(002)} \rc \;,
\label{d5}\\[4mm]
&\bullet&  \cI_{BB}^{(z)+} = \frac{m^2/4}{(4\p)^2} \;\frac{g^4}{m^4}
\lc \P_{ss}^{(020)} + \P_{reg}^{(020)} \rc \;,
\label{d6}
\eea

\ni
and

\bea
&\bullet&  \cI_{DD}^-  =\frac{m^2/4}{(4\p)^2}
\lc \frac{\m^2}{2m^2} \lp \frac{g^2}{m^2} - \frac{1}{f_\p^2}\rp^2
\frac{W^2\!-\!z^2}{4m^2}\; \Pb_{cc}^{(000)}\right.
\nn\\[2mm]
&&+ \left. \frac{2\m}{m}\;\frac{g^2}{m^2} \lp \frac{g^2}{m^2}-
\frac{1}{f_\p^2} \rp\frac{W^2\!-\!z^2}{4m^2}\lb \frac{W^2\!+\!z^2}{4m^2}\; 
\P_{sc}^{(002)}+  \Pb_{sc}^{(000)}\rb \right.
\nn\\[2mm]
&& \left. + \frac{g^4}{m^4}\;\lb -\frac{z^4}{16m^4} \P_{ss}^{(020)}+ 
\frac{W^4}{16m^4} \P_{ss}^{(002)}+\frac{W^2\!-\!z^2}{4m^2} 
\Pb_{ss}^{(000)}\right.\right.
\nn\\[2mm]
&& \left.\left. + \frac{z^4}{16m^4} \P_{reg}^{(020)} - \frac{W^4}{16m^4}
\P_{reg}^{(002)}-\frac{W^2\!-\!z^2}{4m^2} \Pb_{reg}^{(000)} \right.\right.
\nn\\[2mm]
&& \left.\left. + \frac{\m}{2m} \lp1- \frac{q^2}{2\m^2}\rp  \lp 1+
\frac{\m^2}{m^2}+\frac{q^2}{8 m^2}+ \frac{z^2}{8m^2} \rp \P_a\rb \rc \;,
\label{d7}\\[4mm]
&\bullet& \cI_{DB}^{(w)-} = \frac{m^2/4}{(4\p)^2}\lc \frac{\m^2}{2m^2}\lp
\frac{g^2}{m^2} - \frac{1}{f_\p^2}\rp^2  \Pb_{cc}^{(000)}+ \frac{2\m}{m}\;
\frac{g^2}{m^2} \lp \frac{g^2}{m^2}- \frac{1}{f_\p^2}\rp
\lb \frac{W^2}{4m^2}\; \P_{sc}^{(002)}+  \Pb_{sc}^{(000)}\rb \right.
\nn\\[2mm]
&& \left. +\frac{g^4}{m^4} \lb \frac{W^2}{4m^2} \P_{ss}^{(002)} + 
\Pb_{ss}^{(000)}-\frac{W^2}{4m^2} \P_{reg}^{(002)} - \Pb_{reg}^{(000)}\rb 
\rc \;,
\label{d8}\\[4mm]
&\bullet& \cI_{DB}^{(z)-}  = \frac{m^2/4}{(4\p)^2}
\lc \frac{\m^2}{2m^2} \lp \frac{g^2}{m^2} - \frac{1}{f_\p^2}\rp^2 
\Pb_{cc}^{(000)}+\frac{2\m}{m}\;\frac{g^2}{m^2}  \lp \frac{g^2}{m^2}-
\frac{1}{f_\p^2} \rp\lb \frac{z^2}{4m^2}\; \P_{sc}^{(002)}+
\Pb_{sc}^{(000)}\rb \right.
\nn\\[2mm]
&& \left.+\frac{g^4}{m^4}\lb \frac{z^2}{4m^2} \P_{ss}^{(020)}+
\Pb_{ss}^{(000)}-\frac{z^2}{4m^2} \P_{reg}^{(020)} - \Pb_{reg}^{(000)}
+ \frac{\m}{2m} \lp \frac{3}{2}- \frac{5q^2}{8\m^2}\rp \P_a \rb \rc \;,
\label{d9}\\[4mm]
&\bullet& \cI_{BB}^{(g)-} = \frac{m^2/4}{(4\p)^2}
\lc \frac{\m^2}{2m^2} \lp \frac{g^2}{m^2} - \frac{1}{f_\p^2}\rp^2
\Pb_{cc}^{(000)}+\frac{2\m}{m}  \;\frac{g^2}{m^2} \lp \frac{g^2}{m^2}-
\frac{1}{f_\p^2} \rp \Pb_{sc}^{(000)}\right.
\nn\\[2mm]
&& \left. + \frac{g^4}{m^4}\lb \Pb_{ss}^{(000)}-\Pb_{reg}^{(000)}
-\frac{\m}{2m}\lp 1-\frac{q^2}{4\m^2}\rp \P_a\rb \rc \;,
\label{d10}\\[4mm]
&\bullet& \cI_{BB}^{(w)-} = \frac{m^2/4}{(4\p)^2}  \; \frac{g^2}{m^2}
\lc \frac{2\m}{m}  \lp \frac{g^2}{m^2}- \frac{1}{f_\p^2} \rp \P_{sc}^{(002)}
+ \frac{g^2}{m^2} \lb  \P_{ss}^{(002)}-\P_{reg}^{(002)}\rb \rc \;,
\label{d11}\\[4mm]
&\bullet& \cI_{BB}^{(z)-} = \frac{m^2/4}{(4\p)^2}  \; \frac{g^2}{m^2}
\lc \frac{2\m}{m}  \lp \frac{g^2}{m^2}- \frac{1}{f_\p^2} \rp \P_{sc}^{(020)}
+ \frac{g^2}{m^2} \lb \P_{ss}^{(020)}-\P_{reg}^{(020)}\rb  \rc \;.
\label{d12}
\eea

\vspace{5mm}
\ni
\underline{\bf family 2} (diagrams g+h+i+j)

\bea
&\bullet& \cI_{DD}^+ = - \frac{\m^2 /4f_\p^2}{(4\p)^4} \;\frac{g^4}{m^2}\;
(1- 2 q^2/\m^2) \lb \P_{cc}^{(000)}- \P_{sc}^{(001)}+1\rb^2 \;,
\label{d13}
\eea

\ni
and

\bea
&\bullet& \cI_{DD}^- = - \frac{\m^2/ f_\p^2}{(4\p)^4}  \; m^2
\lc \frac{W^2}{4m^2}\lb \frac{g^2}{m^2} \lp \frac{W^2}{4m^2} \P_{sc}^{(002)}
+\Pb_{sc}^{(000)} \rp +\frac{\m}{2m}\lp \frac{g^2}{m^2}-\frac{1}{f_\p^2}\rp
\Pb_{cc}^{(000)}\rb^2 \right.
\nn\\[2mm]
&& \left. -\; \frac{z^2}{4m^2}\lb \frac{g^2}{m^2}\lp \frac{z^2}{4m^2}
\P_{sc}^{(002)} +\Pb_{sc}^{(000)}\rp+\frac{\m}{2m}\lp \frac{g^2}{m^2}
-\frac{1}{f_\p^2}\rp \Pb_{cc}^{(000)}\rb^2  \rc \;,
\label{d14}\\[2mm]
&\bullet& \cI_{DB}^{(w)-} = - \frac{\m^2/f_\p^2}{(4\p)^4} \; \frac{g^4}{m^2}
\lb \frac{W^2}{4m^2} \P_{sc}^{(002)}+ \Pb_{sc}^{(000)}
+ \frac{\m}{2m}\lp 1-\frac{m^2}{g^2 f_\p^2}\rp \Pb_{cc}^{(000)}\rb^2 \;,
\label{d15}\\[2mm]
&\bullet& \cI_{DB}^{(z)-} = - \frac{\m^2/f_\p^2}{(4\p)^4} \; \frac{g^4}{m^2}
\lb \frac{z^2}{4m^2} \P_{sc}^{(002)}+ \Pb_{sc}^{(000)}
+ \frac{\m}{2m}\lp 1-\frac{m^2}{g^2 f_\p^2}\rp \Pb_{cc}^{(000)}\rb^2 \;,
\label{d16}\\[2mm]
&\bullet& \cI_{BB}^{(g)-} = - \frac{\m^2/f_\p^2}{(4\p)^4} \; \frac{g^4}{m^2}
\lb \Pb_{sc}^{(000)} + \frac{\m}{2m}\lp 1-\frac{m^2}{g^2 f_\p^2}\rp
\Pb_{cc}^{(000)}\rb^2 \;,
\label{d17}\\[2mm]
&\bullet& \cI_{BB}^{(w)-} = - \frac{\m^2/f_\p^2}{(4\p)^4} \; \frac{g^4}{m^2}
\lc \frac{W^2}{4m^2}\lb \P_{sc}^{(002)}\rb^2 \right.
\nn\\[2mm]
&& \left.+ 2\;\P_{sc}^{(002)}\lb \Pb_{sc}^{(000)}+\frac{\m}{2m}\lp 1
-\frac{m^2}{g^2 f_\p^2}\rp \Pb_{cc}^{(000)}\rb \rc \;,
\label{d18}\\[2mm]
&\bullet& \cI_{BB}^{(z)-} = - \frac{\m^2/f_\p^2}{(4\p)^4} \; \frac{g^4}{m^2}
\lc \frac{z^2}{4m^2}\lb \P_{sc}^{(002)}\rb^2 \right.
\nn\\[2mm]
&& \left. + 2\;\P_{sc}^{(002)}\lb \Pb_{sc}^{(000)}+\frac{\m}{2m}\lp 1-
\frac{m^2}{g^2 f_\p^2}\rp \Pb_{cc}^{(000)}\rb \rc \;,
\label{d19}
\eea

\vspace{5mm}
\ni
\underline{\bf family 3} (diagrams (k+l+m+n+o)

\bea
&\bullet& \cI_{DD}^+ =  \frac{1}{(4\p)^2} \; \frac{g^2}{m^2}\lc m \lp 
\db_{00}^+ + q^2 d_{01}^+ \rp \lb \P_{cc}^{(000)}- \frac{W^2\!+\!z^2}{4m^2}\; 
\P_{sc}^{(001)}\rb\right.
\nn\\[2mm]
&& \left. + \frac{\m^3}{2} \lp1-q^2/2\m^2 \rp \lp d_{10}^+ + q^2 d_{11}^+ 
\rp\lb \frac{(W^2\!+\!z^2)^2}{16 m^4} \P_{sc}^{(002)} + \frac{W^2\!+\!z^2}{4m^2}
\Pb_{sc}^{(000)}\rb \rc
\nn\\[2mm]
&& +\frac{1/2}{(4\p)^2} \lc \lp \db_{00}^+ + q^2 d_{01}^+ \rp^2
\P_{cc}^{(000)}+ 2 \m^2  \lp \db_{00}^+ + q^2 d_{01}^+ \rp  d_{10}^+ \; 
\Pb_{cc}^{(000)}+ 3 \m^4 \lp d_{10}^+ \rp^2 \Pbb_{cc}^{(000)} \rc\;,
\label{d20}\\[4mm]
&\bullet& \cI_{DB}^{(w)+} = - \frac{m/2}{(4\p)^2} \; \frac{g^2}{m^2}
\lc \lp \db_{00}^+ + q^2 d_{01}^+ \rp \P_{sc}^{(001)}
+ \m^2 \lp d_{10}^+ + q^2 d_{11}^+ \rp \Pb_{cc}^{(000)} \rc \;,
\label{d21}\\[4mm]
&\bullet& \cI_{DB}^{(z)+} =   \frac{m/2}{(4\p)^2} \; \frac{g^2}{m^2}
\lc \lp \db_{00}^+ + q^2 d_{01}^+ \rp \P_{sc}^{(001)}
- 3\,\m^2  \lp d_{10}^+ + q^2 d_{11}^+ \rp \Pb_{cc}^{(000)} \rc \;,
\label{d22}\\[4mm]
&\bullet& \cI_{BB}^{(g)+} =  - \frac{\m^2 m}{(4\p)^2}\; \frac{g^2}{m^2}
 \; b_{00}^+ \;\Pb_{cc}^{(000)} \;,
\label{d23}
\eea

\ni
and

\bea
&\bullet&  \cI_{DD}^-  = - \frac{\m m}{(4\p)^2} \; \frac{W^2\!-\!z^2}{4m^2}\; 
\lp \db_{00}^- + q^2 \db_{01}^- \rp\lc \frac{\m}{2m} \lp \frac{g^2}{m^2} - 
\frac{1}{f_\p^2}-\db_{00}^- - q^2 \db_{01}^- \rp  \Pb_{cc}^{(000)} \right.
\nn\\[2mm]
&& \left. + \frac{g^2}{m^2} \; \lb \frac{W^2\!+\!z^2}{4m^2}\; 
\P_{sc}^{(002)}+  \Pb_{sc}^{(000)}\rb \rc
\nn\\[2mm]
&&  + \frac{\m^4/2}{(4\p)^2}\lc  d_{10}^- \lb - 3 \lp \frac{g^2}{m^2} - 
\frac{1}{f_\p^2}\rp \Pbb_{cc}^{(000)}+ \frac{g^2}{m^2} \; (1-q^2/2\m^2) \;
\Pb_{cc}^{(000)} \rb \rc \;,
\label{d24}\\[4mm]
&\bullet& \cI_{DB}^{(w)-} = - \frac{\m m/2}{(4\p)^2}\lc \frac{\m}{2m} \lp 
\frac{g^2}{m^2} - \frac{1}{f_\p^2}\rp \bb_{00}^- \; \Pb_{cc}^{(000)}
+ \frac{g^2}{m^2}\;\bb_{00}^-\lb \frac{W^2}{4m^2}\; \P_{sc}^{(002)}+
\Pb_{sc}^{(000)}\rb \rc \;,
\label{d25}\\[4mm]
&\bullet& \cI_{DB}^{(z)-}  = \cI_{DB}^{(w)-} \;,
\label{d26}\\[4mm]
&\bullet& \cI_{BB}^{(g)-} = - \frac{\m m}{(4\p)^2}\; \bb_{00}^- \;\lc 
\frac{\m}{2m} \lp \frac{g^2}{m^2} - \frac{1}{f_\p^2}- \bb_{00}^- \rp 
\Pb_{cc}^{(000)}+\frac{g^2}{m^2}\;\Pb_{sc}^{(000)} \rc \;.
\label{d27}
\eea

\section{relations among integrals}\label{ap:relat}

We display here the relations among integrals needed for the
chiral expansion of the potential. The derivation of these relations
is based on the fact that the numerators of some integrands can be
simplified. For instance, a result for $I_{sc}^{\mu}$ may be
obtained through

\bea
\frac{(W\!+\!z)_\m}{2m} \; I_{sc}^\m
&=& \int [\cdots]\; \frac{Q\cd(W\!+\!z)}{(s_1\!-\!m^2)}
= \int [\cdots] \lb 1 - \frac{(Q^2\!-\!q^2/4)}{(s_1\!-\!m^2)}\rb
\nn\\[2mm]
=  \frac{(W\!-\!z)_\m}{2m} \; I_{cs}^\m
&=& I_{cc}-\frac{\m}{2m}\lp1-\frac{t}{2\m^2}\rp I_{sc}+\cdots\;,
\label{e12}
\eea

\ni
where the ellipsis indicates that short range contributions were discarded.
The combination of both results produces

\bea
&\bullet&\;\;\;
\frac{W^2\!+\!z^2}{4m^2} \;\P_{sc}^{(001)}
= \P_{cc}^{(000)} - \frac{\m}{2m} \lp 1 -
\;\frac{t}{2\m^2}\rp \P_{sc}^{(000)}+\cdots\;.
\label{e13}
\eea

The repetition of this procedure yields

\bea
&\bullet&\;\;\;
\Pb_{cc}^{(000)} = \frac{1}{3}\lp 1-\frac{t}{4\m^2}\rp
\P_{cc}^{(000)} + \cdots\;,
\label{e10}
\\[2mm]
&\bullet&\;\;\;
\Pbb_{cc}^{(000)} = \frac{1}{15} \lp
1-\frac{t}{4\m^2}\rp^2 \P_{cc}^{(000)} + \cdots\;,
\label{e11}
\\[2mm]
&\bullet&\;\;\;
\frac{W^2\!+\!z^2}{4m^2}\;\P_{sc}^{(002)}+\Pb_{sc}^{(000)}
= - \frac{\m}{2m} \lp 1 -\;\frac{t}{2\m^2}\rp \P_{sc}^{(001)}+ \cdots \;,
\label{e15}
\\[2mm]
&\bullet&\;\;\;
\Pb_{sc}^{(000)} = \frac{1}{2}\lp 1-\frac{t}{4\m^2}\rp
\P_{sc}^{(000)}+\frac{\m}{4m}\lp 1-\frac{t}{2\m^2}\rp \P_{sc}^{(001)}+\cdots\;,
\label{e17}
\\[2mm]
&\bullet&\;\;\;
\frac{W^2}{4m^2} \;\P_{ss}^{(001)} =\P_{sc}^{(000)}
-\frac{\m}{2m}\lp 1-\frac{t}{2\m^2}\rp \P_{ss}^{(000)}+\cdots \;,
\label{e18}
\\[2mm]
&\bullet&\;\;\;
\frac{z^2}{4m^2} \;\P_{ss}^{(020)} +
\Pb_{ss}^{(000)}= - \P_{sc}^{(001)}+\cdots \;,
\label{e20}
\\[2mm]
&\bullet&\;\;\;
\frac{W^2}{4m^2} \;\P_{ss}^{(002)} + \Pb_{ss}^{(000)}=
\P_{sc}^{(001)} -\frac{\m}{2m}\lp 1-\frac{t}{2\m^2}\rp \P_{ss}^{(001)}
+\cdots \;,
\label{e21}
\\[2mm]
&\bullet&\;\;\;
\frac{W^4}{16m^4} \;\P_{ss}^{(002)} + \frac{W^2}{4m^2} \Pb_{ss}^{(000)}
\nn\\[2mm]&&
=\frac{\m^2}{4m^2} \lp 1-\frac{t}{2\m^2}\rp^2 \;\P_{ss}^{(000)}
+\frac{W^2}{4m^2}\;\P_{sc}^{(001)}-\frac{\m}{2m} \lp 1-\frac{t}{2\m^2}\rp \;
\P_{sc}^{(000)} + \cdots \;,
\label{e23}
\\[2mm]
&\bullet&\;\;\;
\Pb_{ss}^{(000)} = \lp 1-\frac{t}{4\m^2}\rp \P_{ss}^{(000)}
+\frac{\m}{2 m} \lp 1-\frac{t}{2\m^2} \rp \P_{ss}^{(001)}+\cdots\;.
\label{e24}
\\[2mm]
&\bullet&\;\;\;
\frac{z^2}{4m^2} \;\P_{reg}^{(010)} =  \frac{\m}{2m}\lp
1-\frac{t}{2\m^2}\rp \P_{reg}^{(000)}+\P_{sc}^{(000)}-\frac{2m}{W}\;
\P_a +\cdots \;,
\label{e32}
\\[2mm]
&\bullet&\;\;\;
\frac{W^2}{4m^2} \;\P_{reg}^{(002)} + \Pb_{reg}^{(000)}=
\P_{sc}^{(001)} \;,
\label{e34}
\\[2mm]
&\bullet&\;\;\;
\frac{z^2}{4m^2} \;\P_{reg}^{(020)} + \Pb_{reg}^{(000)}=
\frac{\m}{2m}\lp 1-\frac{t}{2\m^2}\rp \P_{reg}^{(010)}-
\P_{sc}^{(001)}+\cdots \;,
\label{e40}
\\[2mm]
&\bullet&\;\;\;
\frac{z^4}{16m^4} \;\P_{reg}^{(020)}+\frac{z^2}{4m^2}\;
\Pb_{reg}^{(000)}=\frac{\m^2}{4m^2}\lp 1-\frac{t}{2\m^2}\rp^2
\P_{reg}^{(000)} -\frac{z^2}{4m^2}\;\P_{sc}^{(001)}
\nn\\[2mm]
&& +  \frac{\m}{2m} \lp 1-\frac{t}{2\m^2}\rp \lp \P_{sc}^{(000)}-
\frac{2m}{W}\;\P_a \rp +\cdots \;,
\label{e42}
\\[2mm]
&\bullet&\;\;\;
\Pb_{reg}^{(000)} = \lp 1-\frac{t}{4\m^2}\rp \P_{reg}^{(000)}
- \frac{\m}{2 m} \lp 1-\frac{t}{2\m^2} \rp \P_{reg}^{(010)}\;.
\label{e43}
\eea

\vspace{5mm}

Other two relations involving $\P_{ss}^{(000)}$ and $\P_{reg}^{(000)}$
are obtained by deriving eq.(\ref{b20}) and eq.(\ref{c64}) with respect
to $\m$, 

\beq
\bullet\;\;\;\m\;\frac{d \;\Pb_{ss}^{(000)}}{d\m}= \P_{ss}^{(000)}
+\frac{\m}{m}\;\P_{ss}^{(001)}\;,
\label{e25}
\eeq

\beq
\bullet\;\;\;\m\;\frac{d \;\Pb_{reg}^{(000)}}{d\m}= \P_{reg}^{(000)}-
\frac{\m}{m}\;\P_{reg}^{(010)}\;.
\label{e44}
\eeq

\section{intermediate results}\label{ap:interes}

The results presented here for the TPEP were obtained by using the relations 
among integrals of the preceding appendix into the full expressions of 
appendix \ref{ap:fullres}. In this procedure we just neglected short range  
integrals and both sets of 
equations are equivalent for distances larger than 1 fm. In 
family 3, we did not keep 
contributions larger than $\cO(q^4)$, in order to avoid unnecessarily long 
equations.

\vspace{5mm}
\ni
\underline{\bf family 1} (diagrams  a+b+c+d+e+f)

\bea
&\bullet& \cI_{DD}^+ = \frac{\m^2/8}{(4\p)^2} \;\frac{g^4}{m^4}
\lc\frac{1}{2}\; (1\!-\!t/2\m^2)^2\;\lb \P_{ss}^{(000)}-\P_{reg}^{(000)}\rb
- \frac{\m}{m}\; (1\!-\!t/2\m^2)\;\P_a \rc \;,
\label{f1}\\[4mm]
&\bullet& \cI_{DB}^{(w)+} = -\;\frac{\m m/8}{(4\p)^2}\; \frac{g^4}{m^4}\;
(1\!-\!t/ 2\m^2)\;\P_{ss}^{(001)}\;,
\label{f2}\\[4mm]
&\bullet& \cI_{DB}^{(z)+} = \frac{\m m/8}{(4\p)^2}\; \frac{g^4}{m^4}
\lc (1\!-\!t/2\m^2)\; \P_{reg}^{(010)} - (3/2 - 5 t / 8\m^2)\; 
\P_a \rc \;,
\label{f3}\\[4mm]
&\bullet& \cI_{BB}^{(g)+} = \frac{m^2/4}{(4\p)^2}\; \frac{g^4}{m^4}
\lc (1\!-\!t/4\m^2) \lp \P_{ss}^{(000)} + \P_{reg}^{(000)} \rp \right.
\nn\\[2mm]
&& \left. +\frac{\m}{2m}  \lb (1\!-\!t/2\m^2) \lp \P_{ss}^{(001)} - 
\P_{reg}^{(010)} \rp+(1- t / 4\m^2) \; \P_a\rb \rc \;,
\label{f4}\\[4mm]
&\bullet& \cI_{BB}^{(w)+} = \frac{m^2/4}{(4\p)^2}\; \frac{g^4}{m^4}
\lc \P_{ss}^{(002)} + \P_{reg}^{(002)} \rc \;,
\label{f5}\\[4mm]
&\bullet&  \cI_{BB}^{(z)+} = \frac{m^2/4}{(4\p)^2}\; \frac{g^4}{m^4}
\lc \P_{ss}^{(020)} + \P_{reg}^{(020)} \rc \;,
\label{f6}
\eea

\ni
and

\bea
&\bullet&  \cI_{DD}^-  = \frac{\m^2/8}{(4\p)^2}\lc \frac{1}{3} \lp
\frac{g^2}{m^2}-\frac{1}{f_\p^2} \rp^2 \lb1-\frac{\m^2}{m^2}(t/4\m^2
+z^2/2\m^2)\rb \;(1\!-\!t/4\m^2)\;\P_{cc}^{(000)}\right.
\nn\\[2mm]
&&- \left. 2\:\frac{g^2}{m^2} \lp \frac{g^2}{m^2} - \frac{1}{f_\p^2} \rp 
\lb1-\frac{\m^2}{m^2}(t/4\m^2 + z^2/2\m^2)\rb(1\!-\!t/ 2\m^2) \;
\P_{sc}^{(001)}\right.
\nn\\[2mm]
&& \left. + \frac{g^4}{m^4}\;\lb \frac{1}{2} \;(1\!-\!t/ 2\m^2)^2 \lp 
\P_{ss}^{(000)}+ \P_{reg}^{(000)}\rp+\frac{\m}{m} (1\!-\!t/ 2\m^2) \;
\P_a\rb  \rc \;,
\label{f7}\\[4mm]
&\bullet& \cI_{DB}^{(w)-} = \frac{\m m/8}{(4\p)^2}\lc\frac{\m}{3m} \lp 
\frac{g^2}{m^2} - \frac{1}{f_\p^2}\rp^2 (1\!-\!t/ 4\m^2)\; \P_{cc}^{(000)}
-\frac{g^2}{m^2} \lp \frac{g^2}{m^2} - \frac{1}{f_\p^2}\rp \right.
\nn\\[2mm]
&& \left. \times
\frac{\m}{m} \lb 2\; (1\!-\!t/ 2\m^2) \;\P_{sc}^{(001)}+\frac{\m}{m}\;
(z^2/ \m^2)\; \P_{sc}^{(002)}\rb-\;\frac{g^4}{m^4}\;(1\!-\!t/ 2\m^2)\;
\P_{ss}^{(001)} \rc \;,
\label{f8}\\[4mm]
&\bullet& \cI_{DB}^{(z)-}  = \frac{\m m/8}{(4\p)^2}\lc\frac{\m}{3m} \lp 
\frac{g^2}{m^2} - \frac{1}{f_\p^2}\rp^2 (1\!-\!t/4\m^2)\; \P_{cc}^{(000)}
+ \frac{g^2}{m^2}  \lp \frac{g^2}{m^2} - \frac{1}{f_\p^2} \rp \right.
\nn\\[2mm]
&& \left. \times \lb 2\;(1\!-\!t/4\m^2)\;\P_{sc}^{(000)} + 
\frac{\m}{m}\;(1\!-\!t/2\m^2)\;\P_{sc}^{(001)}+\frac{\m^2}{m^2}\;(z^2/\m^2)\; 
\P_{sc}^{(002)}\rb \right.
\nn\\[2mm]
&& \left. -  \frac{g^4}{m^4}\lb (1\!-\!t/ 2\m^2)\; \P_{reg}^{(010)}
- (3/2 - 5t /8 \m^2) \P_a\rb  \rc \;,
\label{f9}\\[4mm]
&\bullet& \cI_{BB}^{(g)-} = \frac{\m m/8}{(4\p)^2}\lc \frac{\m}{3 m} \lp 
\frac{g^2}{m^2} - \frac{1}{f_\p^2}\rp^2 (1\!-\!t/ 4\m^2)\; 
\P_{cc}^{(000)}\right.
\nn\\[2mm]
&& \left. + \frac{g^2}{m^2}  \lp \frac{g^2}{m^2}- \frac{1}{f_\p^2} \rp
\lb 2\; (1\!-\!t/4\m^2)\; \P_{sc}^{(000)}+ \frac{\m}{m}\;(1\!-\!t/2\m^2)\;
\P_{sc}^{(001)}\rb \right.
\nn\\[2mm]
&& \left. - \;\frac{g^4}{m^4} \lb (1\!-\!t / 2\m^2) \; \P_{reg}^{(010)}
+ (1\!-\!t/4\m^2) \;\P_a+\frac{\m}{m}\;(z^2/ 2\m^2) \; \lp 
\P_{ss}^{(020)}-\P_{reg}^{(020)}\rp \rb \rc \;,
\label{f10}\\[4mm]
&\bullet& \cI_{BB}^{(w)-} = \frac{m^2/4}{(4\p)^2}  \; \frac{g^2}{m^2}
\lc \frac{2\m}{m}  \lp \frac{g^2}{f_\p^2}- \frac{1}{f_\p^2} \rp 
\P_{sc}^{(002)}+\frac{g^2}{m^2} \lb \P_{ss}^{(002)}-\P_{reg}^{(002)}\rb\rc\;,
\label{f11}\\[4mm]
&\bullet& \cI_{BB}^{(z)-} = \frac{m^2/4}{(4\p)^2}  \; \frac{g^2}{m^2}
\lc \frac{2\m}{m}  \lp \frac{g^2}{m^2}- \frac{1}{f_\p^2} \rp \P_{sc}^{(020)}
+ \frac{g^2}{m^2}\lb \P_{ss}^{(020)}-\P_{reg}^{(020)}\rb  \rc \;.
\label{f12}
\eea

\vspace{5mm}
\ni
\underline{\bf family 2} (diagrams  g+h+i+j)

\bea
&\bullet& \cI_{DD}^+ = - \frac{\m^4 /16f_\p^2}{(4\p)^4} \;\frac{g^4}{m^4}\;
(1- 2 t/\m^2) \lb (1-t/2\mu^2) \P_{sc}^{(000)}-2\pi\rb^2 \;,
\label{f13}
\eea

\ni
and

\bea
&\bullet& \cI_{DD}^- =  - \frac{\m^4/4 f_\p^2}{(4\p)^4}
\lc \frac{W^2}{4m^2}\lb -\frac{g^2}{m^2} \lp (1\!-\!t/2\m^2)\;\P_{sc}^{(001)}
+\frac{\m}{m}\;(z^2/2\m^2)\;\P_{sc}^{(002)}+1-t/3\mu^2\rp \right.\right.
\nn\\[2mm]
&& \left.\left. +\frac{1}{3}\lp \frac{g^2}{m^2}-\frac{1}{f_\p^2}\rp 
\lp (1\!-\!t/4\m^2)\;\P_{cc}^{(000)} +2-t/4\mu^2\rp\rb^2\!\!\! 
\right.\nn\\[2mm]&&\left.
- \frac{z^2}{4\m^2}\lb 
\frac{g^2}{m^2}\lp (1\!-\!t/4\m^2)\;\P_{sc}^{(000)}+\frac{\m}{2m}\;
(1\!-\!t/2\m^2)\; \P_{sc}^{(001)}
 +\frac{\m^2}{m^2}\;(z^2/2\m^2)\;\P_{sc}^{(002)}-\pi\rp
\right.\right.\nn\\[2mm]&& \left.\left.
+\frac{\m}{3m}\lp \frac{g^2}{m^2}-\frac{1}{f_\p^2}\rp (1\!-\!t/4\m^2)\; \P_{cc}^{(000)}\rb^2  \rc \;,
\label{f14}\\[2mm]
&\bullet& \cI_{DB}^{(w)-} = - \frac{\m^4/4 f_\p^2}{(4\p)^4}\lb -
\frac{g^2}{m^2} \lp (1\!-\!t/2\m^2)\;\P_{sc}^{(001)} +\frac{\m}{m}\;
(z^2/2\m^2)\;\P_{sc}^{(002)}+1-t/3\mu^2\rp \right.
\nn\\[2mm]
&& \left. +\frac{1}{3}\lp \frac{g^2}{m^2}-\frac{1}{f_\p^2}\rp 
\lp (1\!-\!t/4\m^2)\;\P_{cc}^{(000)}+2-t/4\mu^2\rp\rb^2 \;,
\label{f15}\\[2mm]
&\bullet& \cI_{DB}^{(z)-} = - \frac{\m^2  m^2/4 f_\p^2}{(4\p)^4}\lb 
\frac{g^2}{m^2}\lp (1\!-\!t/4\m^2)\;\P_{sc}^{(000)}+\frac{\m}{2m}\;
(1\!-\!t/2\m^2)\; \P_{sc}^{(001)}\right.\right.
\nn\\[2mm]
&& \left.\left.+\frac{\m^2}{m^2}\;(z^2/2\m^2)\;\P_{sc}^{(002)}
-\pi\rp+
\frac{\m}{3m}\lp \frac{g^2}{m^2}-\frac{1}{f_\p^2}\rp (1\!-\!t/4\m^2)\;
\P_{cc}^{(000)}\rb^2  \;,
\label{f16}\\[2mm]
&\bullet& \cI_{BB}^{(g)-} = - \frac{\m^2  m^2/4 f_\p^2}{(4\p)^4}
\lb \frac{g^2}{m^2}\lp (1\!-\!t/4\m^2)\;\P_{sc}^{(000)}
+\frac{\m}{2m}\;(1\!-\!t/2\m^2)\; \P_{sc}^{(001)}-\pi\rp\right.
\nn\\[2mm]
&& \left. +\frac{\m}{3m}\lp \frac{g^2}{m^2}-\frac{1}{f_\p^2}\rp
(1\!-\!t/4\m^2)\; \P_{cc}^{(000)}\rb^2  \;,
\label{f17}\\[2mm]
&\bullet& \cI_{BB}^{(w)-} = - \frac{\m^2  m^2/2 f_\p^2}{(4\p)^4}\;
\frac{g^2}{m^2}\;\P_{sc}^{(002)}\lb \frac{g^2}{m^2}\lp (1\!-\!t/4\m^2)\;
\P_{sc}^{(000)}-\frac{\m}{2m}\;(1\!-\!t/2\m^2)\; \P_{sc}^{(001)}\right.\right.
\nn\\[2mm]
&& \left.\left.- \frac{\m^2}{m^2}\;(z^2/\m^2)\;\P_{sc}^{(002)}\rp+
\frac{2 \m}{3m}\lp \frac{g^2}{m^2}-\frac{1}{f_\p^2}\rp (1\!-\!t/4\m^2)\;
\P_{cc}^{(000)}\rb\;,
\label{f18}\\[2mm]
&\bullet& \cI_{BB}^{(z)-} =  - \frac{\m^2  m^2/ f_\p^2}{(4\p)^4}\;
\frac{g^2}{m^2}\;\P_{sc}^{(002)}\lb \frac{g^2}{m^2}\lp (1\!-\!t/4\m^2)\;
\P_{sc}^{(000)}+\frac{\m}{2m}\;(1\!-\!t/2\m^2)\; \P_{sc}^{(001)}\right.\right.
\nn\\[2mm]
&& \left.\left.+ \frac{\m^2}{m^2}\;(z^2/4\m^2)\;\P_{sc}^{(002)}\rp
+\frac{\m}{3m}\lp \frac{g^2}{m^2}-\frac{1}{f_\p^2}\rp (1\!-\!t/4\m^2)\;
\P_{cc}^{(000)}\rb\;.
\label{f19}
\eea

\vspace{5mm}
\ni
\underline{\bf family 3} (diagrams  k+l+m+n+o)

\bea
&\bullet& \cI_{DD}^+ =  \frac{\m^3/2m f_\p^2}{(4\p)^2} \; \frac{g^2}{m^2}
\lc \lp \deb_{00}^+ + t/\m^2 \;\d_{01}^+ \rp \;(1\!-\!t/2\m^2)\; 
\P_{sc}^{(000)} \right.
\nn\\[2mm]
&& \left. - \frac{\m}{2m}\; \d_{10}^+ \;(1\!-\!t/2\m^2)^2  \; 
\P_{sc}^{(001)} \rc +\frac{\m^4/2m^2}{(4\p)^2}\; \frac{1}{f_\p^4}
\lc  \lp \deb_{00}^+ + t/\m^2\; \d_{01}^+ \rp^2  \right.
\nn\\[2mm]
&& \left. + \frac{2}{3} \lp \deb_{00}^+ + t/ \m^2\; \d_{01}^+ \rp
\d_{10}^+ \;(1\!-\!t/4\m^2)+ \frac{1}{5} \lp \d_{10}^+ \rp^2 \;
(1\!-\!t/4\m^2)^2 \rc \P_{cc}^{(000)} \;,
\label{f20}\\[4mm]
&\bullet& \cI_{DB}^{(w)+} = - \frac{\m^2/2 f_\p^2}{(4\p)^2} \; \frac{g^2}{m^2}
\lc \lb \deb_{00}^+ + (t/\m^2) \;\d_{01}^+\rb \P_{sc}^{(001)}
+ \frac{1}{3}\; \d_{10}^+ \; (1\!-\!t/4\m^2)\;\P_{cc}^{(000)}\rc \;,
\label{f21}\\[4mm]
&\bullet& \cI_{DB}^{(z)+} =   \frac{\m^2/2 f_\p^2}{(4\p)^2} \; \frac{g^2}{m^2}
\lc \lb \deb_{00}^+ + (t/\m^2) \;\d_{01}^+ \rb \P_{sc}^{(001)}
- \d_{10}^+ \; (1\!-\!t/4\m^2)\; \P_{cc}^{(000)} \rc\;,
\label{f22}\\[4mm]
&\bullet& \cI_{BB}^{(g)+} =-\frac{\m^2/f_\p^2}{(4\p)^2}\;\frac{g^2}{m^2}\;
\frac{1}{3}\; \b_{00}^+ \;(1-\!\!t/4\m^2)\;\P_{cc}^{(000)} \;,
\label{f23}
\eea

\ni
and

\bea
&\bullet&  \cI_{DD}^-  = - \frac{\m^4/2 m^2 f_\p^2}{(4\p)^2}
\lc \frac{1}{3}\lp \frac{g^2}{m^2}-\frac{1}{f_\p^2}\rp \!
\lb \deb_{00}^- + (t/\m^2)\; \deb_{01}^- + \frac{3}{5} \;
(1\!-\!t/4\m^2)\; \d_{10}^- \rb \!(1\!-\!t/4\m^2)\;\P_{cc}^{(000)}\right.
\nn\\[2mm]
&& \left. - \frac{g^2}{m^2} \; (1\!-\! t/2\m^2)\lb \lp \deb_{00}^- 
+(t/\m^2)\; \deb_{01}^- \rp \P_{sc}^{(001)}+ \frac{1}{3}\;\d_{10}^-\;
(1\!-\!t/4\m^2)\;\P_{cc}^{(000)}\rb \rc \;,
\label{f24}\\[4mm]
&\bullet& \cI_{DB}^{(w)-} = - \frac{\m^2/4 f_\p^2}{(4\p)^2} \; \beb_{00}^-
\lc \frac{1}{3} \lp \frac{g^2}{m^2} - \frac{1}{f_\p^2}\rp (1\!-\!t/4\m^2)\;
\P_{cc}^{(000)}\right.
\nn\\[2mm]
&& \left. -  \frac{g^2}{m^2} \lb (1\!-\!t/2\m^2)\; \P_{sc}^{(001)}
+ \frac{\m}{2m}\;(z^2/\m^2)\; \P_{sc}^{(002)}\rb \rc \;,
\label{f25}\\[4mm]
&\bullet& \cI_{DB}^{(z)-}  = \cI_{DB}^{(w)-}\;,
\label{f26}\\[4mm]
&\bullet& \cI_{BB}^{(g)-} = - \frac{\m m/2 f_\p^2}{(4\p)^2}  \; \beb_{00}^-
\lc \frac{\m}{3m} \lp \frac{g^2}{m^2} - \frac{1+ \beb_{00}^-}{f_\p^2}\rp 
(1\!-\!t/4\m^2)\; \P_{cc}^{(000)} \right.
\nn\\[2mm]
&& \left. + \frac{g^2}{m^2}\lb (1\!-\!t/4\m^2)\; \P_{sc}^{(000)}
+ \frac{\m}{2m}\;(1\!-\!t/2\m^2)\;\P_{sc}^{(001)} \rb \rc \;.
\label{f27}
\eea

\section{relativistic expansions}\label{ap:relex}

In section \ref{sec:dynamics} we have discussed the relativistic expansion 
of the function $\gamma(t)$ derived by Becher and Leutwyler, which does 
not coincide with the usual heavy baryon expansion. 
In this appendix we show how their results can be used to produce 
relativistic expansions for box and crossed box integrals.

The triangle, crossed box, and regularized box integrals given, 
respectively, by eqs.(\ref{b17}), (\ref{b19}) and (\ref{c63}) can 
be written as

\bea
&& \Pb_{ss}^{(000)}= - \int_0^1 d c\; \P_{sc} ^{(001)}(\cM_{ss}) \;,
\label{g1}\\[2mm]
&& \Pb_{reg}^{(000)} = - \int_1^\infty  d c \; \P_{sc} ^{(001)}(\cM_{us}) \;,
\label{g2}
\eea

\ni
where $\P_{sc} ^{(00n)}(\cM)$ is a generalized triangle integral, given by

\beq
\P_{sc}^{(00n)}(\cM)=  \lp\!-\; \frac{2m}{\m}\rp^{n+1} \int_0^1 d a\;a 
\int_0^1 d b \;\frac{\m^2\;  (ab/2)^{n}}{D(\cM)}
\label{g3}
\eeq

\ni
and the denominator $D(\cM)$ is

\beq
D(\cM) = \cM^2 a^2 b^2 - a(1\!-\!a)(1\!-\!b) \;q^2 + (1-ab)\;\m^2 \;.
\label{g4}
\eeq

\ni
When $\cM=m$, one recovers the triangle integral defined in 
eq.(\ref{b17}). On the other hand, the values $\cM^2= (W^2 + q^2 + 
c^2\;z^2)/4$ and $\cM^2= (c^2\;W^2 + q^2 + z^2)/4$ yield eqs.(\ref{g1}) 
and (\ref{g2}).

Performing explicitly the $b$ integration in eq.(\ref{g4}), we obtain the 
generalization of eq.(\ref{e13}), which reads

\beq
\lp1-t/4\cM^2 \rp \;\P_{sc}^{(001)}(\cM) = \frac{m^2}{\cM^2}\lb \P_{cc}^{(000)}
-\frac{\m}{2m}\;(1- t/2\m^2)\;\P_{sc}^{(000)}(\cM)+\P_L(\cM) \rb
\label{g5}
\eeq

\ni
with

\beq
\P_L(\cM) = \lp 1-\frac{\m^2}{\cM^2}\rp \lp \ln \frac{\cM^2}{\m^2} -2 \rp
+2 \;\frac{\m}{\cM} \sqrt{1-\frac{\m^2}{4 \cM^2}}\;
\tan^{-1}\lp \frac{\cM}{\m}\;\sqrt{\frac{1-\m/2\cM}{1+\m/2\cM}} \rp\;.
\label{g6}
\eeq

In all cases $\cM$ is a large parameter and we can use the relativistic 
expansion of $\gamma(t)$, which 
is related to our triangle integral by $\P_{sc}^{(000)}=-2m\m(4\pi)^2 \;\gamma(t)$.
We have 

\beq
\P_{sc}^{(000)}(\cM) = \frac{m}{\cM} \lc \P_a+
\frac{\m}{2 \cM}\;\P_t^{\NL}
+ \frac{\p}{2}\lb -
\frac{\m}{\cM \sqrt{1-t/4\m^2}}+2 \;\ln\lp 1+
\frac{\m}{2 \cM\sqrt{1-t/4\m^2}}\rp\rb
\rc\;,
\label{g7}
\eeq

\ni
with $\P_a$ and $\P_t^{\NL}$ given by eqs.(\ref{6.9}) and (\ref{6.10}).
Recalling that $\P_{cc}^{(000)}=\P_\ell$ and inserting these results into 
eqs.(\ref{g1}) and (\ref{g2}), we obtain

\bea
&& \Pb^{(000)} =  - \int d c\;
\frac{m^2}{\cM^2 -t/4 }\lc \P_\ell +\P_L(\cM) -\frac{\m}{2\cM}\;(1- t/2\m^2) \right.
\nn\\[2mm]
&& \left. \times
\lb \P_a+
\frac{\m}{2 \cM}\;\P_t^{\NL}
+ \frac{\p}{2}\lp -
\frac{\m}{\cM \sqrt{1-t/4\m^2}}+2 \;\ln\lp 1+
\frac{\m}{2 \cM\sqrt{1-t/4\m^2}}\rp\rp
\rb
\rc \;.
\label{g9}
\eea

In the chiral limit $\m\rar 0$, we have

\beq
\Pb_{ss}^{(000)}=\Pb_{reg}^{(000)}\rar - (1+ z^2/ 6 m^2)\;\P_\ell
\label{g10}
\eeq

\ni
and, using eqs.(\ref{e18}), (\ref{e20}), (\ref{e21}), and (\ref{e25}), and 
(\ref{e32}), (\ref{e34}), (\ref{e40}), and (\ref{e44}), one finds the 
following relationships valid in that limit

\bea
&&\P_{ss}^{(001)} = -2\;\P_{reg}^{(001)} \rar \P_a\;,
\label{g11}\\[2mm]
&&\P_{ss}^{(020)} = \P_{reg}^{(020)} \rar 2\;\P_\ell/3 \;,
\label{g12}\\[2mm]
&&\P_{ss}^{(002)} = \P_{reg}^{(002)} \rar 2\;\P_\ell \;,
\label{g13}\\[2mm]
&&\P_{ss}^{(000)} = \P_{reg}^{(000)} \rar -\; \P'_\ell \;.
\label{g14}
\eea

These results may also be combined with those presented in appendix 
\ref{ap:relat}, in order to produce relativistic $\cO(q^2)$ expansions 
for box and crossed box integrals. Eqs.(\ref{e20}), (\ref{g14}), and 
(\ref{e13}) yield

\beq
\bullet \;\;\; \Pb_{ss}^{(000)}= - \lp 1 +\frac{t}{4m^2} + 
\frac{z^2}{6m^2} \rp \P_\ell+\frac{\m}{2m} \lp 1 - \frac{t}{2\m^2}\rp \P_t
\label{g15}
\eeq

\ni
and, using eqs.(\ref{e25}) and (\ref{e18}), one has

\beq
\P_{ss}^{(000)} = - \lp 1 + \frac{\m^2}{2m^2} + \frac{z^2}{6m^2} \rp
\;\P'_\ell- \frac{\m}{2m} \lp 1- \frac{t}{2\m^2}\rp\lb \P_t - \P'_t  \rb \;.
\label{g16}
\eeq

Recalling that $\ \P_{ss}^{(000)}= \P_\times$ and using the results of 
section \ref{sec:dynamics},
we find the heavy baryon expansion

\beq
\bullet\;\;\; \P_\times^{HB}= -\P'_\ell - \frac{\m}{m}\;
\frac{\p/2}{(1\!-\!t/4\m^2)}-\frac{\m^2}{4m^2}\lb (1\!-\!t/2\m^2)^2 \lp 2\;
\P'_\ell -\P''_\ell \rp + (2 z^2/3\m^2)\;\P'_\ell \rb +\cdots\;, 
\label{g17}
\eeq

\ni
where the ellipsis represents polynomials in $t$.

For the box integrals we evaluate eq.(\ref{g9}) directly and obtain

\beq
\bullet \;\;\; \lb \Pb_{reg}^{(000)}\rb^{HB} =  - \lp 1 +\frac{t}{4m^2} + \frac{z^2}{6m^2} \rp \P_\ell
+ \frac{\m}{2m} \lp 1- \frac{t}{2\m^2}\rp\lb \frac{1}{2}\;\P_a + 
\frac{\m}{6m}\;\P_t^{\NL} \rb \;.
\label{g18}
\eeq

Comparing with eq.(\ref{e40}) and using eq.(\ref{e13}), we find

\beq
\bullet \;\;\; \Pt_b^{HB} = -\;\frac{1}{2}
\lb \P_a + \frac{2\m}{3m}\;\P_t^{\NL} \rb \;,
\label{g19}
\eeq

\ni
where $\Pt_b =  \P_{reg}^{(010)}$.
Finally, evaluating eq.(\ref{e44}), we have

\beq
\bullet\;\;\; \P_b^{HB} = -\P'_\ell - \frac{\m}{m}\;
\frac{\p/4}{(1\!-\!t/4\m^2)}-\frac{\m^2}{12m^2}\lb (1\!-\!t/2\m^2)^2 
\lp 2\;\P'_\ell -\P''_\ell \rp + (2 z^2/\m^2)\;\P'_\ell \rb +\cdots
\label{g20}
\eeq

\ni 
with $\P_b = \P_{reg}^{(000)}$.


\end{document}